\documentclass[fleqn,usenatbib]{mnras}
\usepackage{newtxtext,newtxmath}
\usepackage[T1]{fontenc}
\usepackage{graphicx}
\usepackage{xspace}
\usepackage{subcaption}
\usepackage{amsmath}
\usepackage{bm}
\usepackage{enumitem}
\usepackage{multicol}
\usepackage{multirow}
\usepackage{pdflscape}
\usepackage{cuted}
\usepackage{fontawesome}
\usepackage[para]{threeparttable}
\usepackage[pagewise]{lineno}
\usepackage[normalem]{ulem}
\DeclareRobustCommand{\VAN}[3]{#2}
\let\VANthebibliography\thebibliography
\def\thebibliography{\DeclareRobustCommand{\VAN}[3]{##3}\VANthebibliography}

\newcommand{\multiplicity}[0]{{\Large\rotatebox[origin=c]{35}{$\bm{:}\bm{:}$}}}
\newcommand{\magnification}[0]{\faSearch}

\def\LSSTunresolvedmedianTD{6.59\xspace}
\def\LSSTunresolvedminTD{1.36\xspace}
\def\LSSTunresolvedAmedianTD{8.85\xspace}
\def\LSSTunresolvedBmedianTD{7.78\xspace}
\def\LSSTunresolvedCmedianTD{2.41\xspace}
\def\LSSTunresolvedDDmedianTD{8.01\xspace}
\def\LSSTunresolvedDDDmedianTD{6.41\xspace}
\def\LSSTunresolvedDDDDmedianTD{5.84\xspace}
\def\LSSTresolvedmedianTD{1.39\xspace}
\def\LSSTresolvedAmedianTD{1.97\xspace}
\def\LSSTresolvedBmedianTD{1.52\xspace}
\def\LSSTresolvedCmedianTD{1.00\xspace}
\def\LSSTresolvedDDmedianTD{1.14\xspace}
\def\LSSTresolvedDDDmedianTD{1.37\xspace}
\def\LSSTresolvedDDDDmedianTD{1.82\xspace}

\def\LSSTunresolvedmedianMU{0.71\xspace}
\def\LSSTunresolvedstdMU{1.73\xspace}
\def\LSSTunresolvedbiasMU{-0.29\xspace}
\def\LSSTresolvedmedianMU{0.12\xspace}
\def\LSSTresolvedstdMU{0.14\xspace}
\def\LSSTresolvedbiasMU{0.05\xspace}

\def\LSSTHSTIIunresolvedAmedianTD{3.58\xspace}
\def\LSSTHSTIIunresolvedBmedianTD{3.42\xspace}
\def\LSSTHSTIIunresolvedCmedianTD{1.40\xspace}
\def\LSSTHSTIIunresolvedDDmedianTD{2.68\xspace}
\def\LSSTHSTIIunresolvedDDDmedianTD{3.06\xspace}
\def\LSSTHSTIIunresolvedDDDDmedianTD{2.67\xspace}
\def\LSSTHSTIIunresolvedArangeTD{8.15\xspace}
\def\LSSTHSTVIunresolvedAmedianTD{1.49\xspace}
\def\LSSTHSTVIunresolvedBmedianTD{1.51\xspace}
\def\LSSTHSTVIunresolvedCmedianTD{0.87\xspace}
\def\LSSTHSTVIunresolvedDDmedianTD{1.34\xspace}
\def\LSSTHSTVIunresolvedDDDmedianTD{1.39\xspace}
\def\LSSTHSTVIunresolvedDDDDmedianTD{1.23\xspace}
\def\LSSTHSTVIunresolvedAchangeTD{7.30\xspace}
\def\LSSTHSTVIunresolvedBchangeTD{5.94\xspace}
\def\LSSTHSTVIunresolvedCchangeTD{1.54\xspace}
\def\LSSTHSTVIunresolvedDDchangeTD{6.64\xspace}
\def\LSSTHSTVIunresolvedDDDchangeTD{5.21\xspace}
\def\LSSTHSTVIunresolvedDDDDchangeTD{4.75\xspace}
\def\LSSTHSToptunresolvedmedianTD{3.00\xspace}
\def\LSSTHSTnirunresolvedmedianTD{1.82\xspace}
\def\LSSTHSTnirunresolvedchangeHSTIVTD{0.4\xspace}
\def\LSSTHSToptunresolvedchangeHSTVITD{1.6\xspace}
\def\LSSTHSTIIresolvedmedianTD{1.11\xspace}
\def\LSSTHSTIVresolvedmedianTD{0.92\xspace}
\def\LSSTHSTVIresolvedmedianTD{0.82\xspace}
\def\LSSTHSTIIresolvedCrangeTD{0.93\xspace}
\def\LSSTHSTVIresolvedCrangeTD{0.17\xspace}
\def\LSSTHSToptresolvedmedianTD{1.03\xspace}
\def\LSSTHSTnirresolvedmedianTD{1.08\xspace}

\def\LSSTHSTIIunresolvedmedianMU{0.18\xspace}
\def\LSSTHSTVIunresolvedmedianMU{0.11\xspace}
\def\LSSTHSTIVunresolvedstdMU{0.13\xspace}
\def\LSSTHSTVIunresolvedbiasMU{0.05\xspace}
\def\LSSTHSToptunresolvedmedianMU{0.17\xspace}
\def\LSSTHSTnirunresolvedmedianMU{0.22\xspace}
\def\LSSTHSTIIresolvedstdMU{0.09\xspace}
\def\LSSTHSTVIresolvedmedianMU{0.11\xspace}
\def\LSSTHSTVIresolvedstdMU{0.07\xspace}
\def\LSSTHSToptresolvedmedianMU{0.10\xspace}
\def\LSSTHSToptresolvedstdMU{0.08\xspace}
\def\LSSTHSTnirresolvedmedianMU{0.11\xspace}
\def\LSSTHSTnirresolvedstdMU{0.12\xspace}

\def\LSSTHSTIIunresolvedmedianTD{2.71\xspace}
\def\LSSTJWSTunresolvedmedianTD{1.69\xspace}
\def\LSSTJWSTunresolvedAmedianTD{2.04\xspace}
\def\LSSTJWSTunresolvedAchangeTD{6.65\xspace}
\def\LSSTJWSTresolvedmedianTD{1.07\xspace}
\def\LSSTJWSTresolvedchangeTD{0.28\xspace}
\def\LSSTHSTJWSTunresolvedmedianTD{1.04\xspace}
\def\LSSTHSTJWSTunresolvedAmedianTD{1.20\xspace}
\def\LSSTHSTJWSTunresolvedBmedianTD{1.08\xspace}
\def\LSSTHSTJWSTunresolvedCmedianTD{0.64\xspace}
\def\LSSTHSTJWSTunresolvedDDmedianTD{0.97\xspace}
\def\LSSTHSTJWSTunresolvedDDDmedianTD{1.02\xspace}
\def\LSSTHSTJWSTunresolvedDDDDmedianTD{1.09\xspace}
\def\LSSTHSTJWSTunresolvedchangeTD{5.50\xspace}
\def\LSSTHSTJWSTresolvedmedianTD{0.77\xspace}
\def\LSSTHSTJWSTresolvedCmedianTD{0.60\xspace}

\def\LSSTJWSTunresolvedmedianMU{0.14\xspace}
\def\LSSTJWSTunresolvedstdMU{0.17\xspace}
\def\LSSTJWSTresolvedmedianMU{0.11\xspace}
\def\LSSTJWSTresolvedstdMU{0.11\xspace}
\def\LSSTHSTJWSTunresolvedmedianMU{0.09\xspace}
\def\LSSTHSTJWSTunresolvedstdMU{0.14\xspace}
\def\LSSTHSTJWSTresolvedmedianMU{0.09\xspace}
\def\LSSTHSTJWSTresolvedstdMU{0.09\xspace}

\def\LSSTshalunresolvedmedianTD{5.35\xspace}
\def\LSSTmediunresolvedmedianTD{4.33\xspace}
\def\LSSTdeepunresolvedmedianTD{3.13\xspace}
\def\LSSTdeepunresolvedDDmedianTD{2.58\xspace}
\def\LSSTdeepunresolvedDDDmedianTD{1.28\xspace}
\def\LSSTdeepunresolvedDDDDmedianTD{1.28\xspace}
\def\LSSTdeepresolvedmedianTD{0.81\xspace}
\def\LSSTdeepresolvedCmedianTD{0.46\xspace}
\def\LSSTdeepresolvedCrangeTD{0.34\xspace}
\def\LSSTshalresolvedmedianTD{1.49\xspace}
\def\LSSTmediresolvedmedianTD{1.21\xspace}
\def\LSSTHSTGBunresolvedmedianTD{1.59\xspace}
\def\LSSTHSTGBunresolvedchangeTD{4.57\xspace}
\def\LSSTHSTGBresolvedmedianTD{0.66\xspace}
\def\LSSTHSTGBresolvedchangeTD{0.68\xspace}

\def\LSSTshalunresolvedmedianMU{0.55\xspace}
\def\LSSTmediunresolvedmedianMU{0.46\xspace}
\def\LSSTdeepunresolvedmedianMU{0.35\xspace}
\def\LSSTmediresolvedmedianMU{0.12\xspace}
\def\LSSTdeepresolvedmedianMU{0.09\xspace}
\def\LSSTmediresolvedstdMU{0.10\xspace}
\def\LSSTdeepresolvedstdMU{0.06\xspace}
\def\LSSTdeepHSTunresolvedmedianMU{0.13\xspace}
\def\LSSTdeepHSTresolvedmedianMU{0.08\xspace}

\def\LSSTHSTresolvedABCmedianTD{0.99\xspace}
\def\LSSTHSTIVresolvedBCmedianTD{0.93\xspace}
\def\LSSTHSTresolvedABCmedianCTD{0.94\xspace}
\def\LSSTHSTIVresolvedBCmedianCTD{1.25\xspace}
\def\LSSTHSTIVresolvedAmedianTD{1.01\xspace}
\def\LSSTHSTIVresolvedAmedianCTD{0.77\xspace}
\def\saltLSSTHSTresolvedABCmedianTD{0.76\xspace}
\def\saltLSSTHSTresolvedABCminTD{0.38\xspace}
\def\saltLSSTHSTresolvedABCmedianCTD{0.57\xspace}
\def\saltLSSTHSTresolvedABCminCTD{0.35\xspace}
\def\saltLSSTHSTresolvedABCchangeCTD{0.18\xspace}

\def\LSSTHSTresolvedABCmedianMU{0.12\xspace}

\def\LSSTHSTIIresolvedABCstdMU{0.10\xspace}
\def\LSSTHSTVIresolvedABCstdMU{0.08\xspace}
\def\LSSTHSTresolvedABCmedianCMU{0.13\xspace}
\def\LSSTHSTresolvedABCstdCMU{0.09\xspace}
\def\saltLSSTHSTIIresolvedABCmedianMU{0.08\xspace}
\def\saltLSSTHSTVIresolvedABCmedianMU{0.06\xspace}
\def\saltLSSTHSTVIresolvedABCstdMU{0.04\xspace}
\def\saltLSSTHSTresolvedABCmedianCMU{0.06\xspace}
\def\saltLSSTHSTresolvedABCstdCMU{0.04\xspace}
\def\saltLSSTHSTVIresolvedABCstdCMU{0.03\xspace}

\def\IIPmatchLSSTunresolvedmedianTD{0.73\xspace}
\def\IIPmatchLSSTresolvedmedianTD{0.44\xspace}
\def\IIPLSSTunresolvedmedianTD{1.12\xspace}
\def\IIPLSSTresolvedmedianTD{0.58\xspace}

\title[Precise Time Delays for glSNe with Glimpse]{The Case for Space: Estimating Precise Time Delays from Ground- and Space-Based Observations of Lensed Supernovae with Glimpse}

\author[E.E. Hayes et al.]{
Erin~E.~Hayes,$^{1}$\thanks{E-mail: eeh55@cam.ac.uk}
Suhail~Dhawan,$^{2}$
Stephen~Thorp,$^{1,3}$
Justin~D.~R.~Pierel,$^{4,5}$ and
Nikki~Arendse$^{3}$
\\
$^{1}$Institute of Astronomy and Kavli Institute for Cosmology, University of Cambridge, Madingley Road, Cambridge CB3 0HA, UK\\
$^{2}$School of Physics and Astronomy, University of Birmingham, Birmingham B15 2TT, UK\\
$^{3}$The Oskar Klein Centre, Department of Physics, Stockholm University, AlbaNova University Centre, SE 106 91 Stockholm, Sweden\\
$^{4}$Space Telescope Science Institute, 3700 San Martin Drive, Baltimore, MD 21218, USA\\
$^{5}$NASA Einstein Fellow\\
}

\date{Accepted 2026 January 13. Received 2026 January 10; in original form 2025 September 28}

\pubyear{2026}

\begin{document}
\label{firstpage}
\pagerange{\pageref{firstpage}--\pageref{lastpage}}
\maketitle

\begin{abstract}
    The delay in arrival time of the multiple images of gravitationally lensed supernovae (glSNe) can be related to the present-day expansion rate of the universe, $H_{0}$. Despite their rarity, Rubin Observatory's Legacy Survey of Space and Time (Rubin-LSST) is expected to discover tens of galaxy-scale glSNe per year, many of which will not be resolved due to their compact nature. Follow-up from ground- and space-based telescopes will be necessary to estimate time delays to sufficient precision for meaningful $H_{0}$ constraints. We present the \textsc{Glimpse} model (\textsc{GausSN} Light curve Inference of Magnifications and Phase Shifts, Extended) that estimates time delays with resolved and unresolved observations together for the first time, while simultaneously accounting for dust and microlensing effects. With this method, we explore best follow-up strategies for glSNe observed by Rubin-LSST. For unresolved systems on the dimmest end of detectability by Rubin-LSST, having peak i-band magnitudes of 22-24 mag, the time delays are measured to as low as 0.7 day uncertainty with 6-8 epochs of resolved space-based observations in each of 4-6 optical and NIR filters. For systems of similar brightness that are resolved by ground-based facilities, time delays are consistently constrained to 0.5-0.8 day precision with 6 epochs in 4 optical and NIR filters of space-based observations or 8 epochs in 4 optical filters of deep ground-based observations. This work improves on previous time-delay estimation methods and demonstrates that glSNe time delays of $\sim10-20$ days can be measured to sufficient precision for competitive $H_{0}$ estimates in the Rubin-LSST era.
\end{abstract}

\begin{keywords}
gravitational lensing: strong -- gravitational lensing: micro -- transients: supernovae -- distance scale -- cosmology: observations -- methods: statistical
\end{keywords}

\section{Introduction}
\label{sec:introduction}
Gravitationally lensed supernovae (glSNe) are observed when the light from a background supernova (SN) is bent by the mass of a foreground galaxy or galaxy cluster such that the SN appears as multiple images. The multiple images will arrive with some delay in time relative to one another owing to geometric differences in path lengths and time dilation as specified by General Relativity (GR). Notably for cosmology, the time delays of the multiple images can be related to the present-day expansion rate of the universe, $H_{0}$, through a time-delay distance, $D_{\rm \Delta t}$\footnote{Along with a time delay, time-delay cosmography requires an accurate source ($z_s$) and lens redshift ($z_l$), an estimate of the external convergence ($\kappa_{\rm ext}$). However, these are not time-critical and can be obtained after the transient has faded \citep[see][for a detailed review]{Treu_2016}.} \citep{Refsdal_1964}. There is currently significant interest in $H_{0}$, motivated by the $5\sigma$ tension between measurements from early-times \citep[e.g., from the cosmic microwave background;][]{Planck_2020} assuming standard cosmology and late-times \citep[e.g., from the local Cepheid-calibrated distance ladder;][]{Riess_2022}. Indeed, this tension may be a sign of new cosmological physics \citep[see e.g.,][for a review]{Mortsell_2018, DiValentino_2021, Verde_2024}, if not due to unknown systematics in one of the measurements. Therefore, as glSNe provide estimates of $H_{0}$ that are independent of early universe physics and of Cepheid-calibrated luminosity distances, they are an ideal probe to understand the $H_{0}$ tension.

The rarity of gravitational lenses and the short-lived nature of SNe, though, has kept the sample of known glSNe small. The wide and deep field of the Vera C. Rubin Observatory's Legacy Survey of Space and Time \citep[Rubin-LSST;][]{Ivezic_2019} is ideally designed for discovering such rare transients. Indeed, Rubin-LSST is expected to discover tens of glSN~Ia per year \citep{Wojtak_2019, Goldstein_2019, SDM_2023, SDM_2024, Arendse_2024, Bronikowski_2025}. Over the course of the survey's ten year lifetime, the projected discovery rates imply a factor of $\sim10-100$ increase in known glSNe from the eight glSNe discovered to date (PS1-10afx, \citealp{Chornock_2013, Quimby_2013, Quimby_2014}; SN~Refsdal, \citealp{Kelly_2015, Kelly_2016_reappearance}; SN~iPTF16geu, \citealp{Goobar_2017}; SN~Requiem, \citealp{Rodney_2021}; SN~C22, \citealp{Chen_2022}; SN~22riv, \citealp{Kelly_2022_riv}; SN~Zwicky, \citealp{Goobar_2023}; SN~H0pe, \citealp{Frye_2024}; SN~Encore, \citealp{Pierel_2024a, Pierel_2025, Suyu_2025}). Despite their rarity, glSNe have already provided two important local measurements of $H_{0}$ at the 7-10$\%$ level \citep{Kelly_2023_Science, Pascale_2024}, though they have yet to reach a level of precision to make definitive claims about the Hubble tension. One major limitation to the precision of these two analyses is the modelling of the lens mass distribution. Both $H_{0}$ estimates from glSNe to date have been from glSNe in cluster-scale lenses, which have very complex mass distributions. Galaxy-scale lenses are easier to model, though the shorter baseline in time delay makes it more difficult to measure time delays precisely. With Rubin-LSST, a significant fraction of glSNe discovered are expected to be lensed by galaxies because of the relative abundance of galaxies compared to clusters of galaxies (see e.g., \citealp{Wojtak_2019, Goldstein_2019, SDM_2023, SDM_2024, Arendse_2024} for galaxy-scale glSNe rate predictions, compared to e.g., \citealp{Bronikowski_2025} for cluster-scale glSNe rate predictions). This expectation makes it imperative to quantify the data required to achieve precise time delays from galaxy-scale glSNe. 

In this work, we focus on how time-delay inference with glSNe discovered by Rubin-LSST can be optimised using follow-up data from additional space- and ground-based facilities. To estimate the time delays of these systems, we present the \textsc{GausSN} Light curve Inference of Magnifications and Phase Shifts, Extended (\textsc{Glimpse}) model with an novel treatment of microlensing to account for chromatic effects and (differential) dust in the lens and host galaxies. Furthermore, as galaxy-scale glSNe have smaller Einstein radii due to the more compact nature of the systems, some galaxy-scale glSNe will have Einstein radii that are smaller than the angular resolution of Rubin-LSST \citep{Arendse_2024}. In this case, the multiple-images will be observed as one unresolved transient. Methods for time-delay estimation with glSNe (e.g., \texttt{sntd}, \citealp{Pierel_2019}; \textsc{GausSN}, \citealp{Hayes_2024}; \textsc{BayeSN-TD}, \citealp{Grayling_2025}; among other statistical models and machine learning implementations, \citealp{Huber_2022, Goncalves_2025, Grupa_2025}) have largely focused on modelling resolved light curves of glSNe. The \texttt{sntd} model has been adapted to fit purely unresolved data from SN~Zwicky with priors from flux ratios \citep{Goobar_2023}; in addition, other statistical and machine learning methods have been developed specifically for unresolved data \citep{Bag_2021, Bag_2024}. Still, resolved and unresolved data have yet to be fit simultaneously with any of these methods. We extend the use of \textsc{GausSN} to model these data together, as will be necessary to achieve accurate time delay for these systems from diverse data sets with different resolutions.

We apply the \textsc{Glimpse} model to simulations of glSNe with a range of lensing configurations, realistic Rubin-LSST observing cadences and space- and ground-based follow-up strategies to determine the quality and quantity of data needed for precise time-delay estimates. Specifically, we consider glSN of Type Ia (glSNe~Ia), as the standardised brightness of the background source can be used to obtain a model independent inference of the lensing absolute magnification \citep{Birrer_2022, Pierel_2024b, Pascale_2024}. This property helps to break modelling degeneracies like the mass-sheet transformation \citep{Falco_1985, Birrer_2022}, which is one of the largest sources of uncertainty in time-delay cosmography with other lensed sources today \citep[e.g., quasars][]{Birrer_2020}. We know that some real-time space-based observations will be necessary, as the lens modelling requires high-resolution imaging of the lensed source to obtain the image positions of the glSN \citep[e.g.][]{Birrer_2019}. However, as space-based follow-up is expensive, competitive, and can take longer to trigger, we expect to complement these observing programs with ground-based follow-up of the glSNe. We explore different exposure times/limiting magnitudes, filter combinations, and follow-up cadences to determine what strategy optimises the time-delay inference with the best use of space- and ground-based follow-up resources.

The paper is structured as follows: in \S\ref{sec:methods}, we describe how the \textsc{Glimpse} model fits unresolved and resolved data simultaneously, with a treatment of microlensing and dust extinction. We describe our simulations of glSNe~Ia systems discovered with Rubin-LSST and followed-up by various space- and ground-based telescopes in \S\ref{sec:simulations}. Using these simulations, we then test the efficacy of different strategies for space-based follow-up of glSNe~Ia on the time-delay and absolute magnification constraints in \S\ref{sec:follow-up}. We discuss the results of this analysis in \S\ref{sec:discussion}. Finally, in \S\ref{sec:conclusions}, we conclude with a summary of the best strategies for follow-up of glSNe~Ia and implications for cosmology. Throughout this work, we assume a fiducial cosmology with $H_{0} = 67.8 \, \text{km} \, \text{s}^{-1} \, \text{Mpc}^{-1}$ and $\Omega_{M} = 0.308$ from \citet{Planck_2020}.

\section{Methodology}
\label{sec:methods}
In \S\ref{sec:methods-gaussn2}, we introduce the \textsc{Glimpse} model, which uses SN light curve templates to model the underlying light curve of the SN and a GP to capture the effects of microlensing affecting each image. In \S\ref{sec:methods-unresolved}, we demonstrate how the model is extended to fit for the time-delays of glSNe using resolved and unresolved data simultaneously. In \S\ref{sec:methods-2dgp}, we extend the \textsc{Glimpse} model to multiple bands, in which the GP treatment of microlensing is configured to model chromatic microlensing. Finally, we describe the treatment of Milky Way, SN host galaxy, and lens galaxy dust extinction in \S\ref{sec:methods-dust}.

In this work, we make use of several vector/matrix operations and probability distributions, which we denote as follows. We denote a dot product of matrix $\bm{A}$ with vector $\bm{b}$ as $\bm{A} \cdot \bm{b}$. For the element-wise multiplication of two matrices with the same dimensions, $\bm{A}$ and $\bm{B}$, we use $\bm{A} \odot \bm{B}$. A uniform distribution is denoted as $\mathcal{U}(a,b)$, with lower bound $a$ and upper bound $b$. We denote an exponential distribution with rate parameter $\lambda$ as Exp($\lambda$).

We use $\mathcal{N}\big(\mu, \sigma^{2}\big)$ to indicate a normal distribution with mean $\mu$ and variance $\sigma^{2}$. In the case of a multivariate normal distribution, we use $\mathcal{N}\big(\bm{\mu}, \bm{\Sigma}\big)$ where $\bm{\mu}$ is the mean and $\bm{\Sigma}$ is the covariance matrix. Thus, we use $\bm{X} \sim \mathcal{N}\big(\bm{\mu}, \bm{\Sigma}\big)$ to denote a multivariate normal vector. For a normal distribution with mean $\mu$ and variance $\sigma^{2}$ that has then been truncated at zero on the left, we use ZLTN$\big(\mu$, $\sigma^{2}\big)$ as in \citet{Uzsoy_2024}. Next, we define a split-normal distribution to be:
\begin{gather}
    P(x| \, \mu, \sigma_{A}^{2}, \sigma_{B}^{2}) = \frac{2}{\sigma_{A}+\sigma_{B}} \begin{cases} \sigma_{A} \, \mathcal{N}(x | \, \mu, \sigma_{A}^{2}) \textrm{ for } x<0 \\ \sigma_{B} \, \mathcal{N}(x | \, \mu, \sigma_{B}^{2}) \textrm{ for } x\geq0 \end{cases}
\end{gather}
where $\mu$ is the mean of the distribution, $\sigma_{A}^{2}$ is the variance of the distribution for $x<0$, and $\sigma_{B}^{2}$ is the variance of the distribution for $x\geq0$. We denote this distribution as SplN$\big(\mu, \sigma_{A}^{2}, \sigma_{B}^{2}\big)$.  

\subsection{The \textsc{Glimpse} Model}
\label{sec:methods-gaussn2}
Consider a glSN system in which the latent light curve in flux of the SN is given by $f(t)$. The light curves of the $M^\text{th}$ image, in flux-space is described by:
\begin{equation}
    f_{M}(t) = \beta_{1,M} \times \epsilon_{M}(t) \times f(t - \Delta_{1,M})
    \label{eq:general-model}
\end{equation}
where $\Delta_{1,M}$ is the time delay of image $M$ relative to image 1; $\beta_{1,M}$ is the time-invariant magnification of image $M$ relative to image 1; and $\epsilon_{M}(t)$ is the time-varying magnification affecting image $M$ arising from microlensing effects. Note that we will define $\Delta_{1,1} \equiv 0$ and $\beta_{1,1} \equiv 1$ for image 1.

For many SNe, we have reliable models of the true underlying light curve. Therefore, we can use a SN light curve template to describe the latent light curve of the SN, $f(t)$. In practice, we can use any template from \texttt{sncosmo} within this framework. For demonstration, we will assume the glSN is a SN~Ia, so we can model the SN light curve with a \texttt{SALT} template \citep{Guy_2007, salt2_2021, salt3, salt3nir}. The \texttt{SALT} template has parameters $t_{0}$ (the time of maximum flux in the $B$-band), $x_{0}$ (the amplitude of the light curve at the time of maximum flux in the $B$-band), $x_{1}$ (light curve stretch parameter), and $c$ (light curve color parameter). These four hyperparameters are shared for all images of the system. We also note that the use of the template has the added benefit of capturing correlations between bands, which were not modelled with the previous \textsc{GausSN} model.

The functional form of the time-varying magnification, $\epsilon_{M}(t)$, on the other hand, is not easily parameterised. We can then use a GP to describe the time-varying magnification such that:
\begin{equation}
    \epsilon_{M}(t) \sim \mathcal{GP}\big(c(t), k(t,t')\big)
\end{equation}
where $c(t)$ is the mean function and $k(t,t')$ is the covariance function, which gives the covariance between $\epsilon_{M}(t)$ and $\epsilon_{M}(t')$. This treatment of microlensing is only possible when a reliable template is available for the SN light curve, as it is not practical to model both the microlensing and the SN light curve as a GP. The mean function of the GP will be $c(t) = 1$ which encourages a return to the template in the absence of a constraint on microlensing, while the covariance captures any time-varying microlensing which affects the images as deviations from the mean.

As we expect the microlensing to vary smoothly in time \citep[see e.g., Figure 4 in][for realistic microlensing curves]{Goldstein_2018}, we will assume a squared exponential kernel for $k(t, t')$ to describe the covariance of $\epsilon_{M}(t)$ and $\epsilon_{M}(t')$ as given by:
\begin{equation}
    k(t, t') = A^{2} \text{exp}\Big[ - (t - t')^{2} / 2\tau^{2}\Big]
    \label{eq:kernel}
\end{equation}
where $A$ is the amplitude and $\tau$ is the length-scale of the kernel fluctuations. Therefore, we can define the covariance matrix, $\bm{K}$, to be the matrix with elements $K_{ij} = k(t_{i}, t_{j})$.

Say that image $M$ is observed at some $N_{M}$ number of times given by the vector $\bm{t}_{M}$. The latent micro-magnification experienced by this image, $\bm{\epsilon}_{M}(\bm{t}_{M})$, is described by:
\begin{equation}
    \bm{\epsilon}_{M}(\bm{t}_{M}) \sim \mathcal{N}(\bm{1}, \bm{K})
\end{equation}
where the vector $\bm{1}$ is a vector of ones.

Under the affine transformation, the vector of latent fluxes of image $M$, $\bm{f}_{M}(\bm{t}_{M})$, can thus be modelled as:
\begin{equation}
    \bm{f}_{M}(\bm{t}_{M}) \sim \mathcal{N}(\bm{T} \cdot \bm{1}, \bm{T}\bm{K}\bm{T}^{\top})
    \label{eq:affine-transform}
\end{equation}
where $\bm{T}$ is a matrix of size $N_{M} \times N_{M}$ with the values of the SN light curve template, $f(t)$, evaluated at times $(\bm{t}_{M} - \Delta_{1,M})$ and magnified by $\beta_{1,M}$ located along the diagonal. Therefore, the new effective mean function consists of the relative macro-magnified template of the image $M$.

Consider a glSN~Ia system with two images. The observed fluxes of each of these images are noisy samples of the latent SN light curve, such that:
\begin{gather}
    \hat{f}_{1}(t_{i}) \sim \mathcal{N}\big(f_{1}(t_{i}), \sigma^{2}_{1,i}\big) \\
    \hat{f}_{2}(t_{i}) \sim \mathcal{N}\big(f_{2}(t_{i}), \sigma^{2}_{2,i}\big)
\end{gather}
where $\sigma^{2}_{M,i}$ is the variance of the flux measurement uncertainty of image $M$ at time $t_{i}$.

Say that image 1 is observed at some times $\bm{\hat{t}}_{1}$, with length $N_{1}$, and image 2 is observed at some times $\bm{\hat{t}}_{2}$, with length $N_{2}$. As in \citet{Hayes_2024}, we concatenate the observation times, flux measurements, and variances of the measurement uncertainties of image 1 and those of image 2 as:
\begin{align}
    \bm{\hat{t}} = 
    &\begin{pmatrix}
        \bm{\hat{t}}_{1} \\[3pt]
        \bm{\hat{t}}_{2}
    \end{pmatrix} \\
    \bm{\hat{f}} =
    &\begin{pmatrix}
        \bm{\hat{f}}_{1} \\[3pt]
        \bm{\hat{f}}_{2}
    \end{pmatrix} \\
    \bm{\hat{\sigma}}^{2} =
    &\begin{pmatrix}
        \bm{\hat{\sigma}}_{1}^{2} \\[3pt]
        \bm{\hat{\sigma}}_{2}^{2}
    \end{pmatrix}
\end{align}
where each of these vectors have length $N_{1} + N_{2}$. We emphasize that the two images do not necessarily have to be observed at the same times nor the same number of times.

The mean of the GP, $\bm{\mu}$, which describes the broad shape of the SN light curve using a template is therefore given by:
\begin{equation}
    \bm{\mu} = 
    \begin{pmatrix}
        \bm{T}_{1} \cdot \bm{1} \\[3pt]
        \bm{T}_{2} \cdot \bm{1}
    \end{pmatrix} \\
    \label{eq:mu}
\end{equation}
where $\bm{T}_{1}$ is a $N_{1} \times N_{1}$ matrix with the latent fluxes of image 1 evaluated at times $\bm{\hat{t}}_{1}$ along the diagonal; $\bm{T}_{2}$ is a $N_{2} \times N_{2}$ matrix with the latent fluxes of image 2 evaluated at times $(\bm{\hat{t}}_{2} - \Delta_{1,2})$ and multiplied by $\beta_{1,2}$ along the diagonal; and $\bm{1}$ is a vector of ones with the same length as $\bm{T}_{1}$ and $\bm{T}_{2}$ respectively in the equation. 

As in Eq. \ref{eq:affine-transform}, the covariance matrix, which we will call $\bm{\Sigma}$, is the element-wise product:
\begin{equation}
    \bm{\Sigma} = \bm{T} \bm{K} \bm{T}^{\top}
\end{equation}
where $\bm{T}$ is:
\begin{equation}
    \bm{T} = 
    \begin{pmatrix}
        \bm{T}_{1} & \bm{0} \\
        \bm{0}^{\top} & \bm{T}_{2} \\
    \end{pmatrix}
    \label{eq:T}
\end{equation}
with $\bm{T}_{1}$ and $\bm{T}_{2}$ as defined above, and $\bm{0}$ is a matrix of zeros of shape $N_{1} \times N_{2}$.

To construct $\bm{K}$, which encodes the microlensing covariance, we must recall that microlensing acts independently on each image, as it is an effect of foreground structure at the positions of the SN images. Therefore, the GP should be independent for each image. If we consider $\bm{K}$ to consist of four quadrants:
\begin{equation}
    \bm{K} =
    \begin{pmatrix}
        \bm{k}(\bm{\hat{t}}_{1}, \bm{\hat{t}}_{1}) & \bm{0} \\
        \bm{0}^{\top} & \bm{k}(\bm{\hat{t}}_{2}, \bm{\hat{t}}_{2})
    \end{pmatrix}
    \label{eq:K}
\end{equation}
where $\bm{k}(\bm{\hat{t}}_{1}, \bm{\hat{t}}_{1})$ is a $N_{1} \times N_{1}$ matrix whose $(i,j)^{\rm th}$ element is $k(t_{1,i}, t_{1,j})$, and $\bm{k}(\bm{\hat{t}}_{2}, \bm{\hat{t}}_{2})$ is a $N_{2} \times N_{2}$ matrix similarly constructed for the second image.

The likelihood of the data is therefore described as:
\begin{equation}
    P(\bm{\hat{f}} | \, \bm{\theta}, \bm{\hat{t}}) = \mathcal{N}(\bm{\mu}, \bm{\Sigma} + \bm{W})
    \label{eq:likelihood}
\end{equation}
where $\bm{\theta} = (A, \tau, t_{0}, x_{0}, x_{1}, c, \Delta_{1,2}, \beta_{1,2})$ refers to the system hyperparameters, assuming a \texttt{SALT} light curve template, and $\bm{W}$ contains the variances of the measurement uncertainties, $\bm{\hat{\sigma}}^{2}$, along the diagonal. 

We can then sample the posterior, given by:
\begin{equation}
    P(\bm{\theta} | \, \bm{\hat{f}}, \bm{\hat{t}}) \propto P(\bm{\hat{f}} | \, \bm{\theta}, \bm{\hat{t}}) P(\bm{\theta})
    \label{eq:posterior}
\end{equation}
where the hyperpriors over the hyperparameters, $P(\bm{\theta})$, will be given for each system in the relevant section. We sample the posterior with a nested sampler as implemented in \texttt{dynesty} for the analyses in this paper \citep{Speagle_2020, Skilling_2004, Skilling_2006}.

Within \texttt{dynesty}, we use slice sampling, developed by \cite{Neal_2003} and first implemented within the nested sampling framework by \cite{Handley_2015a, Handley_2015b}, to propose new live points. We use the default stopping criteria from \texttt{dynesty} \citep{Speagle_2020}. We note that the specifications of the sampling algorithm can be easily adjusted within the \textsc{GausSN}/\textsc{Glimpse} framework, as described in \citet{Hayes_2024}.

\subsection{Extension to Unresolved Data}
\label{sec:methods-unresolved}
In many cases, we expect the multiple images in a system of glSNe to be unresolved by a telescope. The flux of the unresolved object, $f_{1+2}(t)$, is given by:
\begin{equation}
    f_{1+2}(t) = [\epsilon_{1}(t) \times f_{1}(t)] + [\beta_{1,2} \times \epsilon_{2}(t) \times f_{1}(t - \Delta_{1,2})]
\end{equation}
for the case of a system with two images. Again, the observed fluxes are noisy realizations of the latent SN fluxes, such that:
\begin{gather}
    \hat{f}_{1+2, i}(t_{1+2, i}) \sim \mathcal{N}(f_{1+2}(t_{1+2, i}), \sigma^{2}_{1+2, i})
\end{gather}
where $\sigma_{1+2, i}^{2}$ is the variance of the measurement error at time $t_{1+2, i}$ of the unresolved SN.

To account for the fact that the unresolved light curve is a summed version of the resolved light curves in our model, we must make adjustments to the construction of $\bm{\mu}$ and $\bm{\Sigma}$ in Eqs. \ref{eq:mu} - \ref{eq:likelihood}. Suppose that some glSN, which is unresolved, is observed at some $N$ set of times, $\bm{\hat{t}} = (t_{1}, t_{2}, ..., t_{N})$. The matrix $\bm{T}_{1+2}$, as in Eq. \ref{eq:affine-transform}, now becomes:
\begin{equation}
    \bm{T}_{1+2} = \big( \bm{T}_{1}, \bm{T}_{2} \big)
\end{equation}
where:
\begin{equation}
    \bm{T}_{1} = 
    \begin{pmatrix}
        f(t_{1}) & ... & 0 \\
        : & : & : \\
        0 & ... & f(t_{N})
    \end{pmatrix}
\end{equation}
and:
\begin{equation}
    \bm{T}_{2} = 
    \begin{pmatrix}
        \beta_{1,2} f(t_{1} - \Delta_{1,2}) & ... & 0 \\
        : & : & : \\
        0 & ... & \beta_{1,2} f(t_{N} - \Delta_{1,2}) 
    \end{pmatrix}
\end{equation}
Here, $f(t)$ represents the true underlying light curve of the SN, as in Eq. \ref{eq:general-model}. Now, $\bm{T}_{1+2}$ has shape $2N \times N$ for the system with two images. Therefore, the mean, $\bm{\mu}$ is given by the column vector:
\begin{equation}
    \bm{\mu} = \big( \bm{T}_{1+2} \cdot \bm{1} \big)
\end{equation}
where $\bm{1}$ is a vector of ones of length $2N$, giving $\bm{\mu}$ length $N$, as desired. This formulation has the effect of adding together two (or, more generically, $M$) SN light curve templates which are appropriately time delayed and macro-magnified.

To construct $\bm{\Sigma}$, we will use $\bm{T}_{1+2}$, as defined above, with the microlensing covariance, $\bm{K}$. Now, $\bm{K}$ will take the form:
\begin{equation}
    \bm{K} = 
    \begin{pmatrix}
        \bm{k}(\bm{t}, \bm{t}) & \bm{k}(\bm{t}, \bm{t} - \Delta_{1,2}) \\
        \bm{k}(\bm{t} - \Delta_{1,2}, \bm{t}) & \bm{k}(\bm{t}, \bm{t})
    \end{pmatrix}
\end{equation}
where $\bm{k}(\bm{t}, \bm{t}')$ is the matrix whose $(i,j)^{\text{th}}$ element is $k(t_{i}, t_{j}')$. Therefore, $\bm{K}$ is a square matrix of size $2N \times 2N$. Now $\bm{\Sigma}$ becomes:
\begin{equation}
    \bm{\Sigma} = \bm{T}_{1+2} \bm{K} \bm{T}_{1+2}^{\top}
\end{equation}
which has the shape $N \times N$, as desired. This transformation has the effect of summing the appropriately magnified/de-magnified covariance of image 1 with itself, image 1 with image 2 (and vise-versa), and image 2 with itself at times that have been appropriately shifted by some time-delay.

Then, we can proceed to sample the same posterior given in Eq. \ref{eq:posterior} with the likelihood from Eq. \ref{eq:likelihood}, where $\bm{\mu}$ and $\bm{\Sigma}$ are given appropriately for the resolved and/or unresolved data for a glSN system.
 
\subsection{Spectro-temporally Correlated Microlensing}
\label{sec:methods-2dgp}
As of yet, we have not specified how to extend the model to multiple bands, as this extension requires making assumptions about the shape of the microlensing as a function of wavelength. Because of the chromatic expansion of the SN, the microlensing may be unique in each band \citep{Goldstein_2018}. Therefore, we model the microlensing as a GP in time and wavelength to capture smooth spectro-temporal variations in the observed light curve. In addition to allowing for chromatic microlensing, this formulation additionally has the benefit of enabling resolved follow-up in wavelength regimes which overlap with some unresolved optical data from Rubin-LSST to help constrain the microlensing for the unresolved images.

To implement this microlensing treatment, we must define a new covariance matrix $\bm{K}$ to encode both time- and wavelength-correlations. Let us consider an image $M$ which has been observed in two bands -- band $A$ and band $B$. Say that the object was observed in band $A$ at times $\bm{\hat{t}}_{A}$ and in band $B$ at times $\bm{\hat{t}}_{B}$. We can construct $\bm{K}_{M}$ as:
\begin{align}
    \bm{K}_{M} =& 
    \begin{pmatrix}
        \bm{k}_{\rm time}(\bm{\hat{t}}_{A}, \bm{\hat{t}}_{A}) & \bm{k}_{\rm time}(\bm{\hat{t}}_{A}, \bm{\hat{t}}_{B}) \\
        \bm{k}_{\rm time}(\bm{\hat{t}}_{B}, \bm{\hat{t}}_{A}) & \bm{k}_{\rm time}(\bm{\hat{t}}_{B}, \bm{\hat{t}}_{B})
    \end{pmatrix} \nonumber \\
    \odot &
    \begin{pmatrix}
        \bm{k}_{\rm wave}(\bm{\hat{t}}_{A}, \bm{\hat{t}}_{A}) & \bm{k}_{\rm wave}(\bm{\hat{t}}_{A}, \bm{\hat{t}}_{B}) \\
        \bm{k}_{\rm wave}(\bm{\hat{t}}_{B}, \bm{\hat{t}}_{A}) & \bm{k}_{\rm wave}(\bm{\hat{t}}_{B}, \bm{\hat{t}}_{B})
    \end{pmatrix}
\end{align}
where $\bm{k}_{\rm time}(\bm{t}, \bm{t}')$ has an $(i,j)^{\rm th}$ element of $k_{\rm time}(t_{i}, t_{j}')$ and $\bm{k}_{\rm wave}(\bm{t}, \bm{t}')$ has an $(i,j)^{\rm th}$ element of $k_{\rm wave}(t_{i}, t_{j}')$, where a kernel has been defined in each of the time and wavelength dimensions.

In this analysis, we choose to model both the variations in time and wavelength as exponential squared kernels to describe the correlations in time and wavelength. The resulting covariance has one overall amplitude or scaling parameter, and two length-scale parameters -- one describing the time-scale of variations and the other describing the wavelength-scale of variations. The user may specify alternative kernels, including different kernels to describe the correlations in time and those in wavelength.

The matrix $\bm{K}$ for all images is a block diagonal with $\bm{K}_{1}, ..., \bm{K}_{M}$ along the diagonal. There remains zero covariance between the microlensing affecting different images, which makes all off-diagonal blocks zero. To finally give $\bm{\Sigma}$, we need $\bm{T}$, as in Eq. \ref{eq:T}, where now $\bm{T}_{M}$ has firstly the latent fluxes of image $M$ in band $A$, $f_{A}(t)$, at times ($\bm{\hat{t}}_{A} - \Delta_{1,M}$) and then those in band $B$, $f_{B}(t)$, at times ($\bm{\hat{t}}_{B} - \Delta_{1,M}$) along the diagonal, all multiplied by $\beta_{1,M}$, as $\beta_{1,M}$ describes the macro-scale (time- and wavelength-invariant) relative magnification of image $M$. Similarly to in \S\ref{sec:methods-unresolved}, this model can be extended to work for unresolved data.

\begin{figure*}
    \begin{subfigure}{0.375\textwidth}
        \centering
        \vfill
        \includegraphics[width=0.98\textwidth]{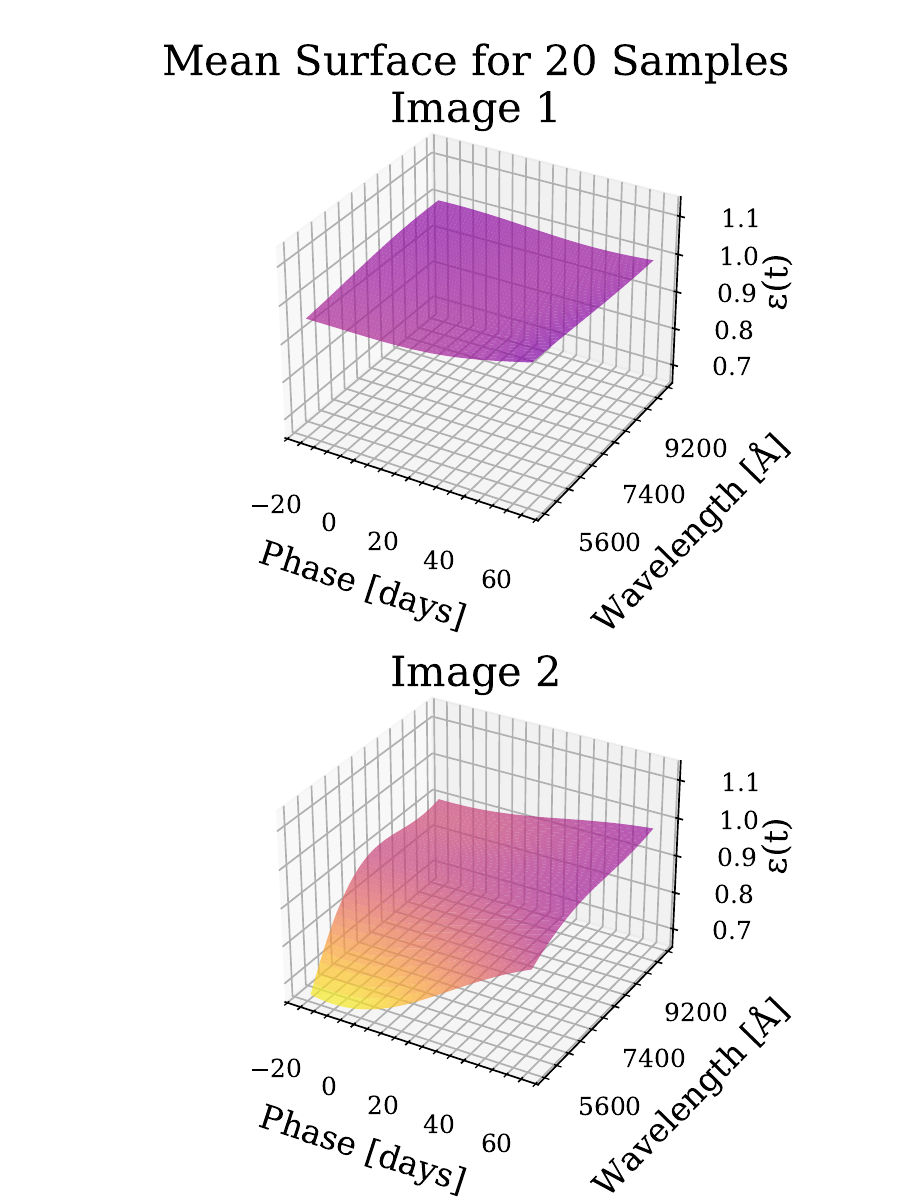}
    \end{subfigure}%
    \begin{subfigure}{0.625\textwidth}
        \centering
        \vfill
        \includegraphics[width=0.98\textwidth]{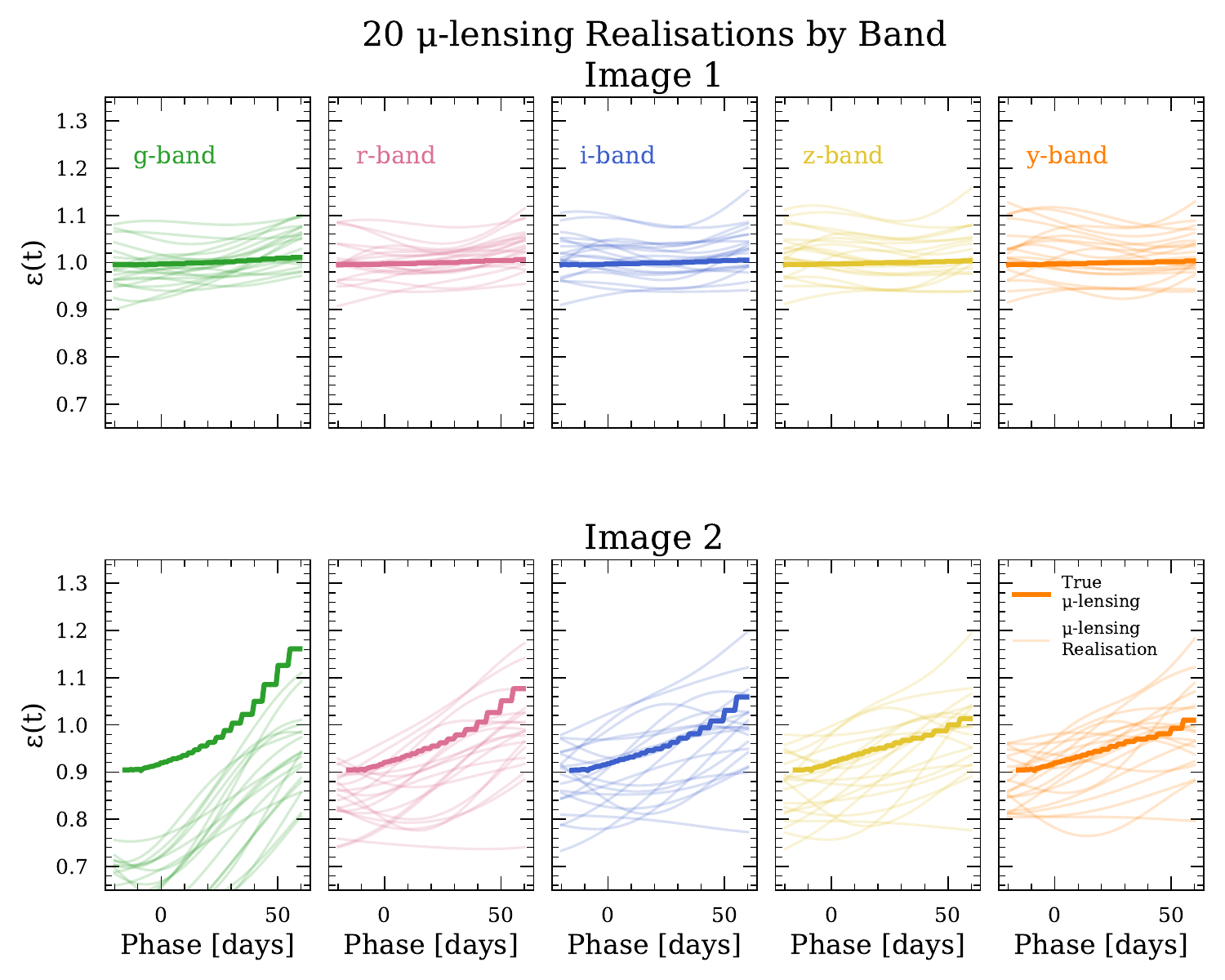}
    \end{subfigure}
    \caption{The fitted microlensing surface described by a 2D GP in time and wavelength for a simulated doubly-imaged glSN~Ia from \citet{Arendse_2024}. The simulated object is observed in the Rubin-LSST $ugrizy$ filters, is assumed to be resolved by Rubin-LSST, and includes a realistic chromatic microlensing treatment. The light curve is very densely sampled and with very good signal-to-noise, which is how a strong constraint on the microlensing surface is obtained. While there is not a significant microlensing contribution affecting image 1, the significant microlensing affecting image 2 is well-captured by the GP. \textbf{Left:} The mean microlensing surface averaged over ten realizations of the fitted GP model. For each realization, we sample the posterior to get the light curve template and GP hyperparameters, divide out the fitted template from the observed data, and condition the GP on the data residuals. \textbf{Right:} Ten samples of the GP microlensing surface sliced at the mean effective wavelength of each filter compared to the true microlensing applied to the light curve in simulations. The \textsc{Glimpse} microlensing realisations generally match the truth very well. We attribute the offset in the recovered microlensing for image 2 in the $g$-band to a poor recovery of dust extinction from the Milky Way, which we fit as a free parameter for this object. The input amount of Milky Way extinction is not available from the simulations, so we cannot correct for this effect before fitting as one would using Milky Way extinction maps for real data. That the shape of the microlensing realisations is broadly correct demonstrates the robustness of this method.}
    \label{fig:2d-gp}
\end{figure*}

In Figure \ref{fig:2d-gp}, we show an example fitted mean 2D GP surface which describes the time-varying chromatic magnification effects for a simulated doubly-imaged glSN~Ia produced by \texttt{lensedSST} for \citet{Arendse_2024}. We fit an object with a very densely sampled and high signal-to-noise light curve to demonstrate that with sufficient data, a meaningful constraint on the shape of the microlensing can be obtained. Being able to back-out the microlensing curves from the light curve fit may additionally provide insights into substructure in the lens, though we emphasize that the fitted microlensing surface may be dependent on the assume light curve template.

\subsection{Dust Treatment}
\label{sec:methods-dust}
Accounting for dust extinction due to dust in the Milky Way, SN host galaxy, and lens galaxy is necessary to reduce possible systematic biases in time-delay estimation. We incorporate a treatment of dust at the light curve template level through functionality provided by \texttt{sncosmo} \citep{Barbary_2016, Barbary_2022}. We assume a \citet{Fitzpatrick_1999} dust law for Milky Way, host, and lens galaxy dust for the analysis in this work, though a different dust law from \texttt{sncosmo} may easily be specified by users. 

To reduce the computational expense to fit each object, we fix $R_{V} = 3.1$ for the Milky Way and $R_{V} = 2.6$ for the host and lens galaxies \citep{Thorp_2021} for all simulations and fits in this work. The line-of-sight $B-V$ dust extinction, $E(B-V)$, for host and lens dust is fit as a hyperparameter of the system. While $E(B-V)_{\text{host}}$ is the same for all images, $E(B-V)_{\text{lens}}$ is fit independently for each image, as it is possible for dust in the lens galaxy to be different along the line-of-sight to each image. The user can easily modify any of these fitting choices (e.g., requiring uniform dust extinction across the lens galaxy) within the \textsc{GausSN}/\textsc{Glimpse} framework.

\section{Simulations of Resolved and Unresolved glSNe Light Curves}
In this section, we describe the methodology for simulating systems of glSNe~Ia; it is based upon the \texttt{lensedSST} tool presented in \citet{Arendse_2024} which we extend to meet the needs of this work. First, we motivate the choice of lens system parameters, including source and lens redshifts, time delays, and magnifications, and SN~Ia light curve parameters, including absolute magnitudes and dust extinction. Then, we describe how the light curves are subsequently produced with a realistic cadence and error model for each telescope\footnote{We make the light curves produced for this analysis publicly available on Github at: \url{https://github.com/erinhay/GausSN}}. Finally, we validate and briefly discuss that the \textsc{Glimpse} model is able to accurately recover the input time delays. 
\label{sec:simulations}
\subsection{System Properties of the glSNe~Ia}
For a full description of the simulation framework, see \citet{Arendse_2024}. As in that work, we begin from the catalog of lenses from \citet{Wojtak_2019} to obtain realistic sets of source and lens redshifts and Einstein radii. In \citet{Arendse_2024}, the lens galaxy mass profile is assumed to be a power law ellipse mass density (PEMD) profile. This profile is implemented in \texttt{lensedSST} using functionality from \texttt{lenstronomy} \citep{Birrer_2018, Birrer_2021}. Besides the Einstein radius, the PEMD model is described by $\gamma_{\textrm{lens}}$, the logarithmic slope of the surface mass density, and $e_{1}$ and $e_{2}$, the ellipticity of the PEMD profile along its two axes. We additionally include the effects of external shear from line-of-sight structures, parameterised in Cartesian coordinates with horizontal ($\gamma_{1}$) and vertical ($\gamma_{2}$) components, which is assumed to be uncorrelated from the lens galaxy orientation. 

\begin{figure}
    \centering
    \includegraphics[width=0.95\linewidth]{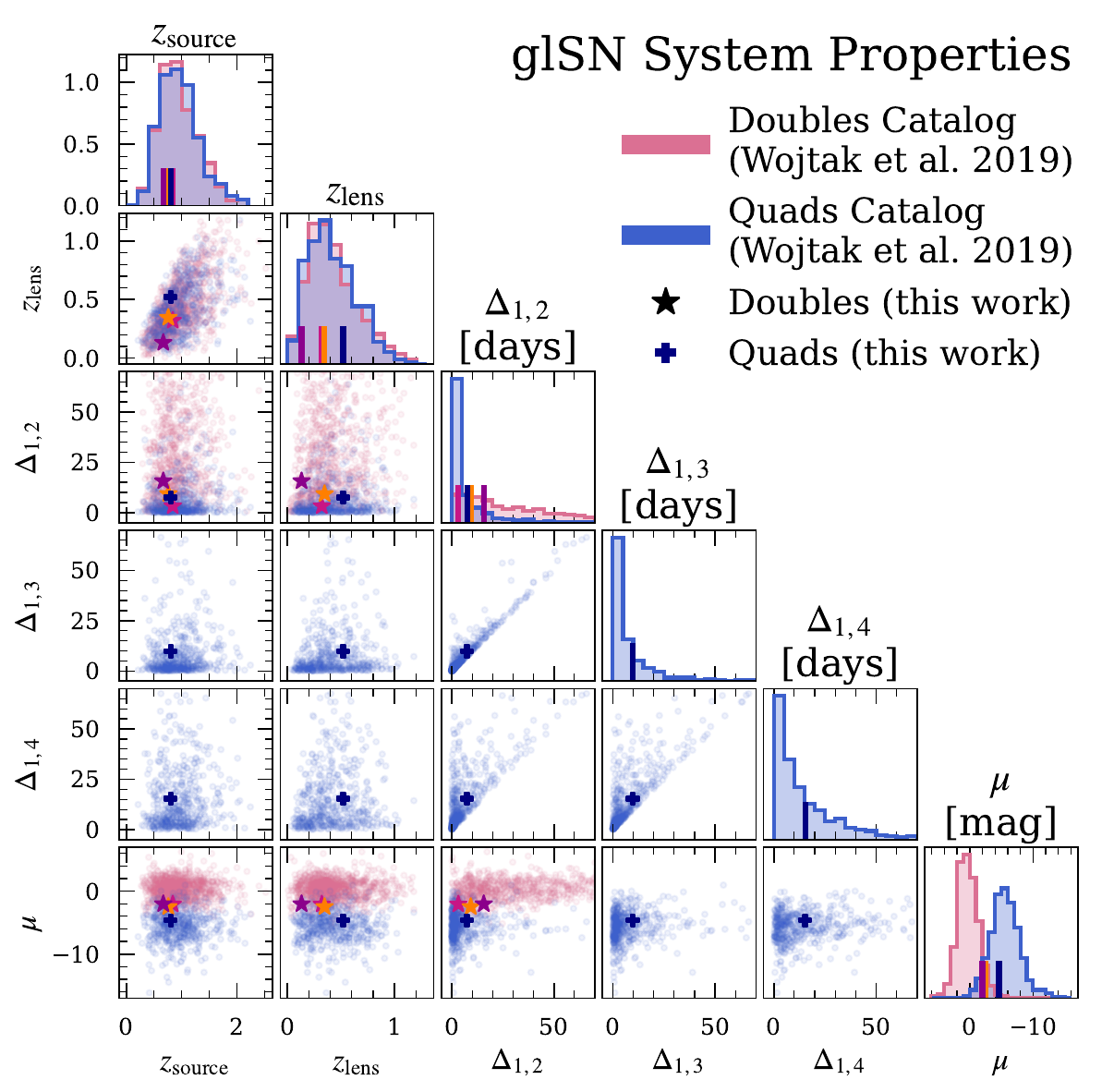}
    \caption{Corner plot for the source redshift, $z_{\text{source}}$, lens redshift, $z_{\text{lens}}$, time delay, $\Delta$, and the total (summed across all images) absolute magnification in magnitude space, $\mu$, for 500 randomly selected objects from the catalogue of simulated glSNe from \citet{Wojtak_2019}. The time delay and absolute magnification are computed once for each draw of source redshift, lens redshift, and Einstein radius with one unique random realisations of the lens mass model and source position parameters. The solid red stars (joint distributions) and solid red lines (marginal distributions) indicate the parameters of the base objects used in this work, as tabulated in Table \ref{tab:base-objects}.}
    \label{fig:sim-corner}
\end{figure}

In Figure \ref{fig:sim-corner}, we show the joint distributions over source redshift, lens redshift, time delay ($\Delta$), and the total (summed across all images) absolute magnification in magnitude space ($\mu$) for 500 randomly-selected double and quad systems from the \citet{Wojtak_2019} catalogue. This catalogue yields a realistic joint distribution of time delays and absolute magnifications for a range of systems. We highlight that the total magnifications, $\mu$, which we define to be the sum of the multiplicative absolute magnification factors affecting each image, are much lower for doubles than they are for quads. Furthermore, the time delays for more highly magnified systems generally tend to be shorter than for low magnification systems.

In Table \ref{tab:base-objects}, we give the system parameters of the four ``base objects'' we analyse in this work. The lens model parameters are all drawn from the parent distributions given in \citet{Arendse_2024}. These objects are indicated as colourful stars/crosses and ticks in Figure \ref{fig:sim-corner}. We select these objects to have representative time delays and magnifications relative to the larger population from \citet{Wojtak_2019} and sample a realistic range expected to be discovered by Rubin-LSST, as given in \citet{Arendse_2024}. We specifically focus on time delays of $<30$ days in this work. The main question we aim to investigate is how follow-up can be optimised to achieve a sub-5\% precision estimate on the time delay. The absolute uncertainty on the time delay needed for a 5\% precision estimate is smaller when the time delay is shorter\footnote{To reach sub-5\% precision on the time delay, $\frac{\sigma_{\Delta t}}{\Delta t} < 0.05$,  when $\Delta t = 30$ requires $\sigma_{\Delta t} < 1.5$ days. On the other hand, when $\Delta t = 10$, reaching sub-5\% precision requires $\sigma_{\Delta t} < 0.5$ days-- a much more significant challenge.}. Therefore, we challenge ourselves in this work to push to smaller uncertainties to determine the lower limits of what a ``long enough'' time delay is to be cosmologically useful (e.g., for a precise $H_{0}$). Furthermore, if this level of uncertainty is achievable for time delays of $<30$ days, then it should be achievable for time delays longer than this threshold. 

\begin{table*}
    \centering
    \caption{The four base objects considered in this analysis.}
    \begin{threeparttable}
        \begin{tabular}{c|c|c|c|c|c|c|c|c|c|c|c|c}
            ID & $z_{\text{source}}$ & $z_{\text{lens}}$ & $\theta_{\rm E}$ & Source Position & $\gamma_{\rm lens}$ & $e_{1}$ & $e_{2}$ & $\gamma_{1}$ & $\gamma_{2}$ & $\Delta$  & Macro-magnification & Peak $i$-band \\
             &  &  & [arcsec] &  &  &  &  &  &  & [days] & [mag] & mag\tnote{a} \\
            \hline
            \hline
            A & 0.839 & 0.321 & 0.270 & (-0.04, 0.08) & 2.13 & 0.21 & 0.13 & -0.00 & -0.00 & -- & -0.82 & 22.8 \\
             &  &  &  &  &  &  &  &  &  & 3.24 & -1.26 & 22.3 \\
            \hline
            B & 0.758 & 0.346 & 0.444 & (0.04, -0.13) & 2.03 & -0.06 & 0.00 & -0.01 & 0.01 & -- & -1.78 & 21.6 \\
             &  &  &  &  &  &  &  &  &  & 9.29 & -0.70 & 22.7 \\
            \hline
            C & 0.667 & 0.132 & 1.064 & (0.12, 0.33) & 1.89 & -0.05 & -0.16 & 0.02 & -0.02 & -- & -1.30 & 21.9 \\
             &  &  &  &  &  &  &  &  &  & 15.67 & -0.73 & 22.5 \\
            \hline
            D & 0.801 & 0.521 & 0.511 & (-0.00, -0.05) & 2.08 & 0.02 & 0.27 & -0.03 & 0.01 & -- & -1.16 & 22.3 \\
             &  &  &  &  &  &  &  &  &  & 7.51 & -1.66 & 21.8 \\
             &  &  &  &  &  &  &  &  &  & 9.82 & -1.24 & 22.2 \\
             &  &  &  &  &  &  &  &  &  & 15.34 & -0.57 & 22.9 \\
            \hline
        \end{tabular}
        \begin{tablenotes}
           \item [a] Unextinguished. These magnitudes are later extinguished by line-of-sight dust for each distinct dust configuration considered in this paper (see \S\ref{sec:simulations} for details).
        \end{tablenotes}
    \end{threeparttable}
    \label{tab:base-objects}
\end{table*}

We add functionality to \texttt{lensedSST} to enable simulation of light curves from the \textsc{BayeSN} spectral energy distribution (SED) model for SNe~Ia \citep{Thorp_2021, Mandel_2022, Grayling_2024}. For simplicity, we assume that the shape parameter $\theta = 0$ and the residual SED variations $\epsilon(t,\lambda) = 0$ for all objects. We also assume that the peak B-band absolute magnitude, $M_{B} = -19.43$ for all SNe~Ia in the simulations. In reality, our models will be an imperfect description of the true SN~Ia light curve, so we will infer the time delays assuming the \texttt{SALT3} template as the \textsc{Glimpse} mean function \citep{Guy_2007, salt3}. Fitting with a template that is not the one from which the data are generated more closely resembles the challenge of fitting for time delays for real objects.

We additionally include treatments of line-of-sight effects, namely dust extinction in the lens and host galaxies. We consider three options for the host and lens dust for each system, assuming $R_{V}=2.6$ for the host and lens in all cases:
\begin{enumerate}[label=(\alph*), leftmargin=*]
    \item $E(B-V)_{\text{host}} = 0.1$, and $E(B-V)_{\text{lens}} = 0.1$

    \item $E(B-V)_{\text{host}} = 0.05$, and $E(B-V)_{\text{lens}} = 0.2$

    \item $E(B-V)_{\text{host}} = 0.05$, $E(B-V)_{\text{lens, im 1}} = 0.1$,\newline $E(B-V)_{\text{lens, im 2}} = 0.3$, $E(B-V)_{\text{lens, im 3}} = 0.2$, and\newline $E(B-V)_{\text{lens, im 4}} = 0.1$
\end{enumerate}
These options span a range of low to high dust extinction cases. In addition, dust configuration "c" tests the ability of the model to constrain differential dust in the lens galaxy \citep[e.g., for iPTF16geu][]{Goobar_2017, Dhawan_2020}.

Notably, we do not include microlensing in these simulations, as the \textit{JWST} filters considered in this work probe redder wavelengths than were modelled for \texttt{lensedSST}. Therefore, we proceed without microlensing to maintain the consistency of the simulations across the entire wavelength regime considered in this work. We reiterate that the \textsc{Glimpse} model that the light curves are fit with does naturally marginalise over chromatic microlensing, meaning the statistical uncertainties include contributions from uncertainty due to microlensing. Furthermore, tests of the \textsc{Glimpse} model on \texttt{lensedSST} simulations that do include microlensing show well-calibrated uncertainties which are not significantly larger than the uncertainties reported for the fits to the glSNe light curves considered in this work (see \S\ref{sec:discussion-microlensing} for a further discussion).

\subsection{Light Curve Cadence and Error Modelling}
\begin{figure}
    \centering
    \includegraphics[width=0.95\linewidth]{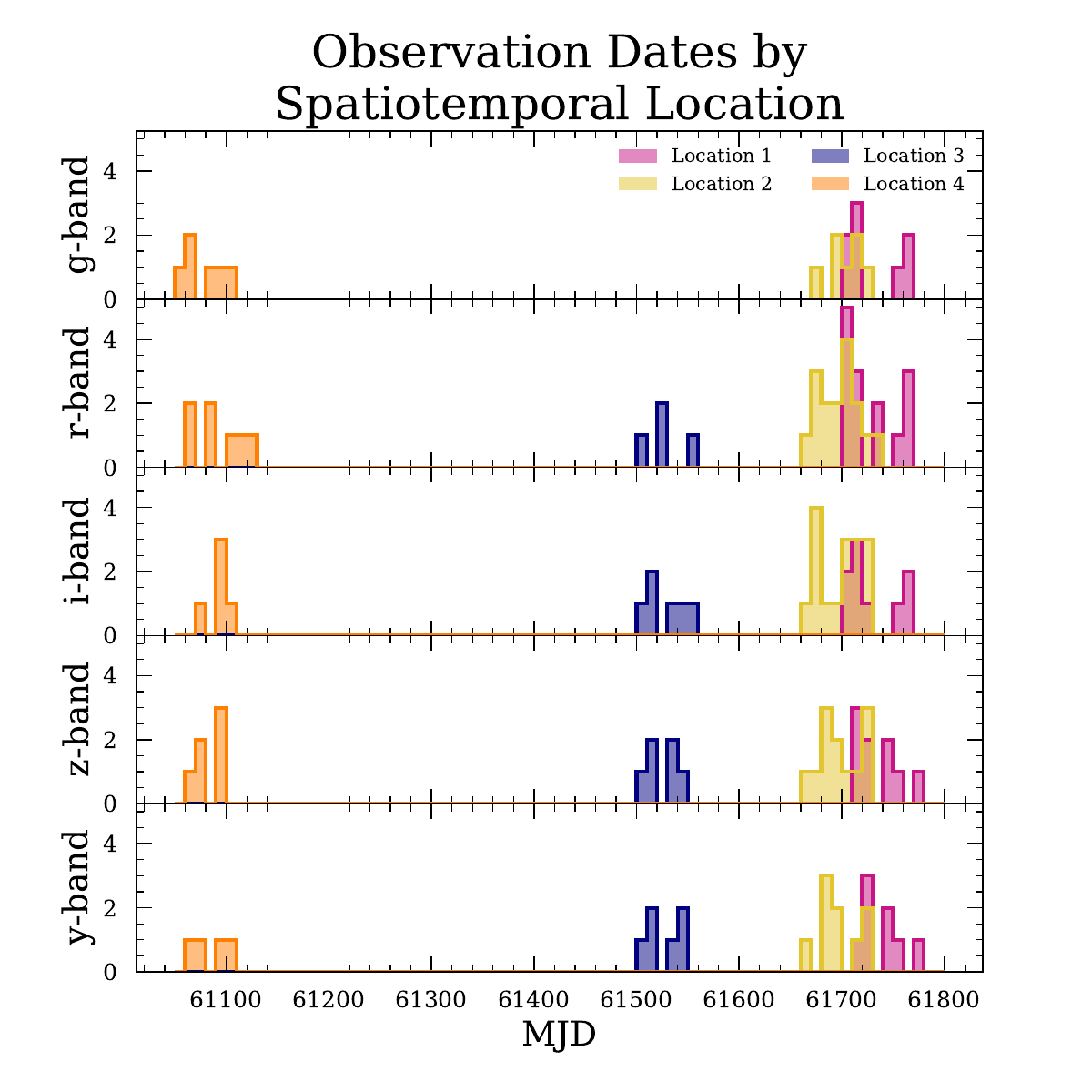}
    \caption{Dates of observation by band for each location in space and time in the Rubin-LSST survey considered in this analysis. For each base object given in Table \ref{tab:base-objects}, we simulate the observed light curve in each of these eight locations.}
    \label{fig:obs-by-location}
\end{figure}

To achieve a realistic cadence for the Rubin-LSST simulations, we assume the survey strategy from baseline v3.4. Information about the simulated observing schedule and nightly conditions is accessed by \texttt{lensedSST} through the Rubin Operations Simulator \citep[\texttt{OpSim};][]{Delgado_2014, Delgado_2016, Naghib_2019} and \texttt{OpSimSummary} \citep{Biswas_2020}. As the times of observation will depend on the location of the glSN in space and time, we simulate each base system at four different spatio-temporal locations which cover a range of possible Rubin-LSST Wide Fast Deep (WFD) observing strategies. These strategies include a rolling observing setting, during which time LSST will observe ``active'' regions at a higher frequency than ``passive'' regions on a rolling basis over the course of the survey, and a non-rolling observing setting, during which time LSST surveys the entire southern sky uniformly \citep[see e.g., \S3.1 in][for more details]{Arendse_2024}. We do not consider any locations in the Deep Drilling Fields (DDFs), as these regions cover only a small portion of the sky and the likelihood of a glSN happening in this region is therefore small \citep[as seen in][]{Arendse_2024}. In Table \ref{tab:locations}, we list the four spatio-temporal locations considered in this analysis. The dates of observation in each band for these locations is visualised in Figure \ref{fig:obs-by-location}. 

\begin{table}
    \centering
    \caption{The four spatio-temporal locations considered in this analysis. Each of the base objected tabulated in Table \ref{tab:base-objects} are simulated four times for each of these four locations to understand the impact of the Rubin-LSST strategy on follow-up needs.}
    \begin{tabular}{c|c|c|c|c}
        ID & RA & DEC & $T_{0}$ [MJD] & Rubin-LSST Survey \\
        \hline
        \hline
        1 & 45 & -22 & 61700 & WFD rolling active \\
        2 & 60 & -18 & 61650 & WFD rolling active \\
        3 & 140 & -35 & 61505 & WFD rolling passive \\
        4 & 140 & -10 & 61050 & WFD non-rolling \\
        \hline
    \end{tabular}
    \label{tab:locations}
\end{table}

There is still some uncertainty about whether the individual glSNe images will be resolved by Rubin-LSST, as a function of the nightly observing conditions and the separation of the images. Therefore, we do not make any assumptions about which base systems will be resolved. Instead, we simulate each base system in each spatio-temporal location twice -- once as two resolved images and once as one unresolved object. These two extremes gives us the range of uncertainty on the time delay we can expect from combining Rubin-LSST data with follow-up. To combine the light from multiple images into one unresolved light curve with realistic measurement uncertainties, we use functionality from \texttt{lensedSST}. We analyse the effect of having resolved vs unresolved Rubin-LSST data on the time-delay inference in \S\ref{sec:follow-up}.

Finally, we consider several options for space-based and ground-based follow-up. For the space-based follow-up from \textit{HST} and \textit{JWST}, we simulate the observations at specified times and in specified bands, depending on the follow-up strategy being considered. More details about the simulation of space-based follow-up are given in Appendix \ref{sec:appendixB}. We describe the follow-up strategies analysed in this work in more detail in \S\ref{sec:follow-up}. All follow-up from \textit{HST} and \textit{JWST} is assumed to be resolved. Given the range of Einstein radii considered in this work, this assumption is reasonable. For more compact systems, like SN~Zwicky ($\theta_{\rm E} = 0.17''$; \citealp{Pierel_2023_Zwicky}), \textit{HST} may not be able to resolve the multiple images in the NIR. In this case, NIR data from \textit{JWST} will be necessary.

When triggering space-based observations, an exposure time in each filter needs to be specified, which will correspond to some signal-to-noise ratio (SNR) for an object of a given magnitude. To simulate measurement uncertainties on space-based observations, we are most interested in the SNR. Using exposure time calculators for \textit{HST}\footnote{UVIS: \url{https://etc.stsci.edu/etc/input/wfc3uvis/imaging/}; IR: \url{https://etc.stsci.edu/etc/input/wfc3ir/imaging/}} and \textit{JWST}\footnote{\url{https://jwst.etc.stsci.edu/}}, we produce one curve of SNR as a function of magnitude per filter for a fixed exposure time. These curves are shown in Appendix \ref{sec:appendixB}. The curve can be re-scaled by a factor of $\sqrt{\textrm{target exposure time} / \, \textrm{input exposure time}}$ to give the SNR as a function of magnitude for different exposure times. These functions enable us to estimate the SNR that would be achieved at some epoch in each image's light curve, given a model for the glSN light curve and some target SNR for the peak magnitude of the brightest image. It is possible to then back-out an estimate of the exposure time from the input exposure time used to produce the function and the specified target SNR. We set the exposure time such that the brighter image would be observed with a SNR of 50 at peak, unless otherwise specified.

We also consider several options for follow-up in the $griz$ filters from ground-based telescopes. We allow the depth and cadence of the ground-based follow-up to vary depending on the size of the telescope, where we consider follow-up from a 2-, 4-, or 8-meter class telescope. The ground-based follow-up is assumed to be ``triggered'' after the peak of the first image. The specifics of the simulated data quality for each telescope configuration are given in \S\ref{sec:ground-followup}. For the ground-based telescopes, we assume the same error model as is used for Rubin-LSST as implemented in \texttt{lensedSST}. 

\begin{figure*}
    \centering
    \includegraphics[width=\linewidth]{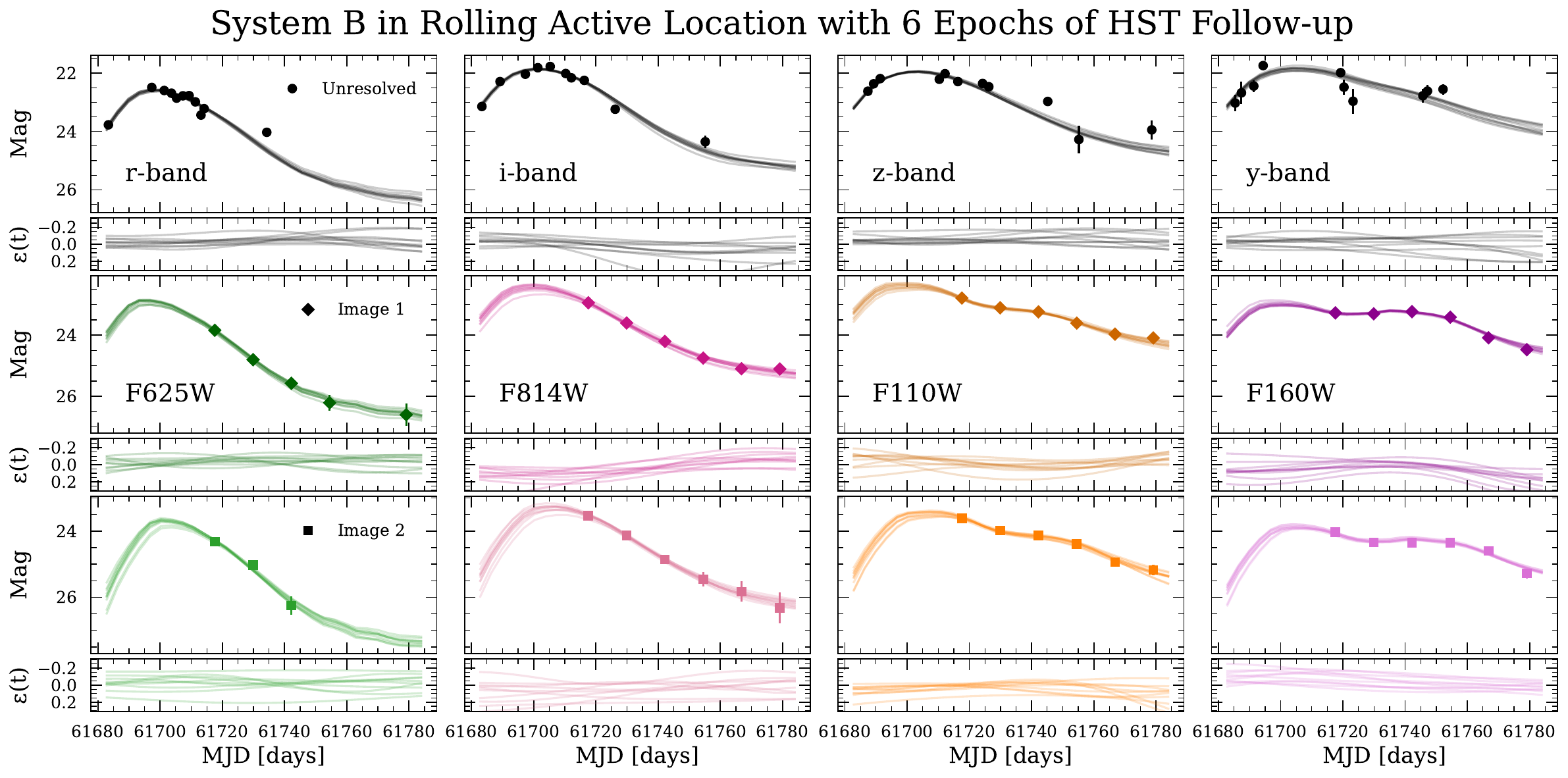}
    \caption{Example light curve for system B at spatiotemporal location 1 under dust configuration b with six epochs of HST follow-up in the \textit{F625W}, \textit{F814W}, \textit{F110W}, and \textit{F160W} filters. Some images show fewer than six observations per filter if the image is too faint to be observed at some point in time (a ``non-detection''). In this realisation of the system, the Rubin-LSST data is assumed to be unresolved. The top row shows the unresolved Rubin-LSST light curves, the middle row shows the resolved image 1 light curves from \textit{HST}, and the bottom row shows the resolved image 2 light curves from \textit{HST}. We show 20 realisations of the \textsc{Glimpse} fit to the data, with the time-vary micro-magnification ($\epsilon(t)$) contributions to the fit shown in the panel below the data. Consistent with the ground truth of no microlensing contributions, the \textsc{Glimpse} model does not find any significant time- or wavelength-dependent deviations from the assumed template.}
    \label{fig:example-lc}
\end{figure*}

We show an example simulated glSN Ia -- system B at location 1 and dust configuration b with six epochs of \textit{HST} follow-up -- in Figure \ref{fig:example-lc}. In summary, we simulate 1,728 unique light curves under these different assumptions about follow-up strategies, which we analyse systematically to determine the best methods for glSN follow-up in \S\ref{sec:follow-up}. The fitting procedure and validation of the model on the simulations is presented in the following \S\ref{sec:sim-analysis}.

\subsection{Validation and Input Parameter Recovery}
\label{sec:sim-analysis}
We fit all objects with the \textsc{Glimpse} model assuming the \texttt{SALT3} template for the true underlying light curve. This model naturally accounts for uncertainty due to chromatic microlensing by the inclusion of the 2D GP microlensing model. The priors over the 11 (for doubly-imaged glSNe) or 17 (for quadruply-imaged glSNe) model parameters are given in Table \ref{tab:sim-priors}. We note that, independent of whether the input has differential dust, the extinction is allowed to be different in the host galaxy and along the different lines sight in the lens galaxy, though we use the same prior for all three extinction parameters. An example \textsc{Glimpse} fit is shown in Figure \ref{fig:example-lc}.

\begin{table}
    \centering
    \caption{The hyperpriors over the GP hyperparameters, \texttt{SALT3} parameters, time delay, and macro-magnification parameters for the \textsc{Glimpse} fit. Extinction is allowed to be different in the host galaxy and along the different lines sight in the lens galaxy, though we use the same prior for all three extinction parameters. For a description of the distribution notation used below, see \S\ref{sec:methods}.}
    \begin{threeparttable}
        \begin{tabular}{c|l}
            Hyperparameter & Hyperprior \\
            \hline
            \hline
            $A$ & ZLTN$(0, 008^{2})$ \\
            $\tau_{\rm time}$ & $\mathcal{U}(30, 100)$ \\
            $\tau_{\rm wave}$ & $\mathcal{U}(300, 4000)$ \\
            $t_{0}$ & $\mathcal{N}(T_{\textrm{brightest}}, 20^{2})$\tnote{a} \\
            $x_{0}$ & $\mathcal{U}(0.1, 1000)$ \\
            $x_{1}$ & $\mathcal{U}(-3, 3)$ \\
            $c$ & $\mathcal{U}(-0.3, 0.3)$ \\
            E(B-V) & Exp$(0.3)$\tnote{b} \\
            $\Delta$ & SplN$(0, 5^{2}, 20^{2})$ \\
            $\beta$ & $\mathcal{U}(0.01, 4)$ \\
            \hline
        \end{tabular}
        \begin{tablenotes}
            \item [a] $T_{\textrm{brightest}}$ refers to the time of maximum flux in any band.
        \end{tablenotes}
    \end{threeparttable}
    \label{tab:sim-priors}
\end{table}

To demonstrate that the \textsc{Glimpse} fits produce accurate time-delay estimates with well-calibrated uncertainties, we estimate the percent of fits where the true time delay falls within the 68\%/95\% confidence interval (CI). We compute the 68\%(95\%) CI from the $16^{\textrm{th}}$($2^{\textrm{nd}}$) and $84^{\textrm{th}}$($97^{\textrm{th}}$) percentiles of the posterior samples. We find that across all 1,728 simulated light curves fit in this work 60.1\% of true time delays fall within the 68\% CI and 85.7\% fall within the 95\% CI. This result shows that the \textsc{Glimpse} uncertainties are broadly well-calibrated. We will not necessarily expect perfect frequentist coverage of the 68\%/95\% CIs, as the sample considered in this work is not constructed from random draws from the prior, but instead is a highly selected subset of the prior distributions on all lensing and light curve parameters.

We also find the median fractional error on the time delay, $(\Delta_{\rm fit} - \Delta_{\rm true})/\Delta_{\rm true}$, to be less than 8\% across nearly all realisations of the systems with varied follow-up. The exception of fits to unresolved Rubin-LSST data alone and unresolved Rubin-LSST data with limited resolved \textit{JWST} follow-up, having a median fractional error of 17\% and 13\% respectively. Given that constraints from unresolved Rubin-LSST data alone or with only a small quantity of resolved follow-up will be weak and sensitive to degeneracies, this result is not concerning. We tabulate these results in Table \ref{tab:pdf-cdf}.

\begin{table*}
    \centering
    \caption{Summary statistics about the \textsc{Glimpse} fits. We compute the 68\%/95\% confidence (CI) from the posterior samples. Note that ``N'' refers to the number of time delays estimated, which will be one per realisation of the system for doubles and three per realisation of the system for quads. Shallow ground-based follow-up is always assumed to be unresolved, regardless of whether the Rubin-LSST data is (un)resolved. If the system is assumed to be (un)resolved by Rubin-LSST data, the medium and deep ground-based follow-up is similarly assumed to be (un)resolved.}
    \begin{tabular}{l|l|c|c|c|c|c}
        &  &  & Median $\Delta_{\rm fit} - \Delta_{\rm true}$ & Median  &  \\
        & Data & $N$ & [days] & $(\Delta_{\rm fit} - \Delta_{\rm true})$/$\Delta_{\rm true}$ & $\Delta_{\rm true}$ in 68\% CI & $\Delta_{\rm true}$ in 95\% CI \\
        \hline
        \hline
         & All light curves & 1728 & -0.10 & -0.01 & 60.1\% & 85.7\% \\
        \hline
        Unresolved & All cases & 864 & 0.09 & 0.01 & 47.7\% & 75.7\% \\
         & Rubin-LSST only & 72 & 2.14 & 0.17 & 59.7\% & 83.3\% \\
         & Rubin-LSST + \textit{HST} & 360 & -0.40 & -0.04 & 49.7\% & 75.8\% \\
         & Rubin-LSST + \textit{JWST} & 72 & 1.61 & 0.13 & 48.6\% & 84.7\% \\
         & Rubin-LSST + \textit{HST} + \textit{JWST} & 72 & 0.25 & 0.03 & 63.9\% & 93.1\% \\
         & Rubin-LSST + shallow ground-based & 72 & 0.41 & 0.05 & 44.4\% & 75.0\% \\
         & Rubin-LSST + mid ground-based & 72 & 0.31 & 0.02 & 37.5\% & 69.4\% \\
         & Rubin-LSST + deep ground-based & 72 & -0.23 & -0.02 & 33.3\% & 56.9\% \\
         & Rubin-LSST + deep ground-based + \textit{HST} & 72 & -0.87 & -0.08 & 36.1\% & 66.7\% \\
        \hline
        Resolved & All cases & 864 & -0.18 & -0.02 & 72.5\% & 95.7\% \\
         & Rubin-LSST only & 72 & -0.43 & -0.04 & 80.6\% & 100.0\% \\
         & Rubin-LSST + \textit{HST} & 360 & -0.24 & -0.02 & 74.4\% & 95.6\% \\
         & Rubin-LSST + \textit{JWST} & 72 & 0.77 & 0.08 & 66.7\% & 97.2\% \\
         & Rubin-LSST + \textit{HST} + \textit{JWST} & 72 & 0.33 & 0.04 & 73.6\% & 98.6\% \\
         & Rubin-LSST + shallow ground-based & 72 & -0.32 & -0.04 & 77.8\% & 95.8\% \\
         & Rubin-LSST + mid ground-based & 72 & -0.36 & -0.04 & 76.4\% & 100.0\% \\
         & Rubin-LSST + deep ground-based & 72 & -0.23 & -0.02 & 69.4\% & 98.6\% \\
         & Rubin-LSST + deep ground-based + \textit{HST} & 72 & -0.51 & -0.04 & 52.8\% & 80.6\% \\
        \hline
    \end{tabular}
    \label{tab:pdf-cdf}
\end{table*}

\section{Comparison of Follow-up Strategies}
\label{sec:follow-up}
We now explore the impact of different follow-up strategies on the time-delay inference. First, we describe the results for time-delay and absolute magnification recovery with Rubin-LSST data alone in \S\ref{sec:lsst-alone}. In \S\ref{sec:hst-followup}, we test the efficacy of several follow-up strategies from \textit{HST} for instances in which resolved light curves can be obtained from Rubin-LSST and those for which they can't. We next explore the impact of follow-up from \textit{JWST}, which can offer a wider wavelength coverage than \textit{HST} but over fewer epochs, in \S\ref{sec:nir-followup}. Finally, we examine the effect of higher-cadence follow-up from ground-based telescopes in \S\ref{sec:ground-followup}. 
\subsection{Rubin-LSST Data Alone}
\label{sec:lsst-alone}
We begin by exploring the precision and accuracy of time-delay and absolute magnification recovery from Rubin-LSST under the assumption of no additional follow-up. We consider both the case of unresolved Rubin-LSST data and resolved Rubin-LSST data for every system.

\subsubsection{Time-Delay Recovery}
Assuming the systems are unresolved, we find that the median uncertainty on the time delay from Rubin-LSST data alone is \LSSTunresolvedmedianTD days. The lowest time-delay uncertainty is from the relative time delay between images 1 and 3 in system D, under the assumption of dust configuration ``a'' and spatiotemporal location 1, at \LSSTunresolvedminTD days. There are differences in the uncertainty on the time delay across the different systems, with system A having a median uncertainty of \LSSTunresolvedAmedianTD days, system B having a median uncertainty of \LSSTunresolvedBmedianTD days, and system C having a median uncertainty of \LSSTunresolvedCmedianTD days. System C, being the brightest overall of the doubles, has the lowest median uncertainty. For the quadruply-imaged system D, the time delay has a median uncertainty of \LSSTunresolvedDDmedianTD days for image 2, of \LSSTunresolvedDDDmedianTD days for image 3, and of \LSSTunresolvedDDDDmedianTD days for image 4. All time delays for system D are measured relative to image 1. With unresolved Rubin-LSST alone, we are unlikely to be able to meaningfully constrain time delays of $\lesssim20$ days (i.e. to $<5\%$ relative error).

For the resolved Rubin-LSST light curves, we find the median uncertainty on the time delay with Rubin-LSST data alone is \LSSTresolvedmedianTD days -- a significant improvement over the fully unresolved Rubin-LSST light curves. When considering each system individually, the median time delay uncertainties are \LSSTresolvedAmedianTD days, \LSSTresolvedBmedianTD, and \LSSTresolvedCmedianTD days for systems A, B, and C, respectively. For system D, the median uncertainties on the time delays of images 2, 3, and 4 relative to image 1 are \LSSTresolvedDDmedianTD days, \LSSTresolvedDDDmedianTD days, and \LSSTresolvedDDDDmedianTD days, respectively.

Therefore, of the twelve realisations of the four base systems, none of the time delays are measured to sub-5\% precision in the unresolved case and only 2.8\% are in the resolved case with Rubin-LSST data alone. While this analysis can be expanded to a larger extent of system parameter values, we have considered a representative range of Rubin-LSST cadences, dust effects, and lensing configurations to give a realistic sense of the accuracy that can be achieved with this data. Such a result motivates the need for follow-up from space- and ground-based facilities.

\subsubsection{Absolute Magnification Recovery}
\label{sec:follow-up-mag-lsst}
A benefit of studying glSNe~Ia is that the absolute magnitudes of SNe~Ia show a remarkably narrow spread, which enables their use as precise distance probes. Having some constraint on the unlensed magnitude of the background source is extremely beneficial to breaking the ``mass-sheet degeneracy'' (MSD) in the lens modelling \citep{Falco_1985}. The mass-sheet degeneracy is the largest source of uncertainty in the lens mass modelling \citep{Birrer_2020}, so additional independent constraints on the distance to the background source are extremely beneficial.

The precision to which the scaling factor of the mass sheet transform (MST), $\lambda$, is measured corresponds to a precision on the observed magnification, $\mu$ of:
\begin{equation}
    \frac{\delta \lambda}{\lambda} = \frac{1}{2} \frac{\delta \mu}{\mu}
\end{equation}
as derived in \citet{Birrer_2022}. To reduce the MSD to a sub-dominant source of uncertainty, $\lambda$ must be estimated to $\sim5\%$ precision. Therefore, the absolute magnification must be estimated to at least $10\%$ precision. For absolute magnifications corresponding to brightening by a few magnitudes, this level of precision corresponds to a constraint to $\sim0.1-0.2$ mag.

A caveat with using the standardisable-candle nature of SNe~Ia to constrain the lens model is that microlensing from stars in the lensing galaxy complicate this by adding stochastic uncertainty to the SNIa luminosities \citep{DoblerKeeton_2006, FoxleyMarrable_2018, Yalahomi_2017, Weisenbach_2024}. However, microlensing can be incorporated in the lens model analysis to still obtain precise measurements of the lens galaxy mass slope \citep{Mortsell_2020, Diego_2022, Arendse_2025}. \textsc{Glimpse} addresses microlensing by modelling it as the time-variable component of the magnification, which allows the macro-magnification to be isolated and used to constrain the lens model. Owing to this caveat, the absolute magnification constraints on the MST parameter should be validated via an independent method. One such alternative uses stellar kinematics, or measurements of the velocity dispersion of stars in the lens galaxy \citep[see e.g.,][]{Shajib_2021}, which requires additional spectroscopy of the lens galaxy after the SN has faded. As these two methods have separate systematic uncertainties, they will provide important cross-checks of one another for the first measurements of $H_{0}$ from galaxy-scale glSNe \citep{Birrer_2022}.

To compute the absolute magnification of each image from the \textsc{Glimpse} fit, we use functionality from \texttt{sncosmo} to retrieve the samples of the observer-frame peak apparent $r$-band magnitude for each image, based on 1,000 draws from the posterior. We then compare each sample to the true unlensed peak apparent $r$-band magnitude to get the posterior on the absolute magnification of each image. We use the $r$-band because we find the differences in the \textsc{BayeSN} SED and \texttt{SALT3} SED to be smallest in this wavelength regime, minimising bias in the absolute magnification recovery. Because the true peak magnitude of the underlying SN~Ia will not be perfectly known, there will be additional uncertainties associated with this process in practice. Especially for glSNe discovered at high redshifts ($z\gtrsim1$), where there may not yet be an appreciable sample of unlensed SNe~Ia, the unlensed peak magnitude for a SN~Ia at a given redshift will need to be interpolated \citep{Pierel_2024b}. Any uncertainties can and would be propagated through in the lens modelling. 

For unresolved Rubin-LSST data, the constraints on absolute magnification are minimal. The median uncertainty from this data alone is \LSSTunresolvedmedianMU mag across all images of all systems. Further to the limitations of unresolved data alone, there is a spread of \LSSTunresolvedstdMU mag in the absolute magnification residuals. There is also a significant bias in the absolute magnification residuals of \LSSTunresolvedbiasMU mag. Unsurprisingly, unresolved data alone does not provide accurate or precise absolute magnification constraints. With resolved Rubin-LSST data, the median absolute magnification uncertainty is a competitive \LSSTresolvedmedianMU mag across all images of all systems and the spread in the residuals is low at \LSSTresolvedstdMU mag. The bias in the residuals is also lower at \LSSTresolvedbiasMU mag.

For this sample, >1\% in the unresolved case and 40\% in the resolved case have  absolute magnifications estimated to sub-10\% precision-- the threshold necessary to constrain $\lambda$ to sub-5\% precision. The absolute magnification constraints from resolved Rubin-LSST data are sufficient in a significant number of cases to reduce the MSD to a sub-dominant source of uncertainty, though this figure can be improved further. There is also still a significant need for additional follow-up in the case of unresolved Rubin-LSST data.

\subsection{Follow-up with \textit{HST}}
\label{sec:hst-followup}
We now consider five different follow-up strategies for following up Rubin-LSST discoveries with \textit{HST}. We choose these five follow-up strategies to balance realism with the goal of understanding how increasing amounts of follow-up data improves time-delay constraints. The five strategies are:
\begin{itemize}
    \item \textbf{HSTopt+nir6}: 6 epochs of data -- every 7 days in the observer-frame from 10 to 45 observer-frame days, relative to the peak of the first image -- in two optical (\textit{F625W} and \textit{F814W}) and two NIR (\textit{F110W} and \textit{F160W}) filters.

    \item \textbf{HSTopt+nir4}: 4 epochs of data -- every 10 days in the observer-frame from 10 to 40 observer-frame days, relative to the peak of the first image -- in two optical (\textit{F625W} and \textit{F814W}) and two NIR (\textit{F110W} and \textit{F160W}) filters.

    \item \textbf{HSTnir4}: 4 epochs of data -- every 10 days in the observer-frame from 10 to 40 observer-frame days, relative to the peak of the first image -- in two NIR (\textit{F110W} and \textit{F160W}) filters.

    \item \textbf{HSTopt4}: 4 epochs of data -- every 10 days in the observer-frame from 10 to 40 observer-frame days, relative to the peak of the first image -- in two optical (\textit{F625W} and \textit{F814W}) filters.

    \item \textbf{HSTopt+nir2}: 2 epochs of data -- at 20 and 40 days in the observer-frame days, relative to the peak of the first image -- in two optical (\textit{F625W} and \textit{F814W}) and two NIR (\textit{F110W} and \textit{F160W}) filters.
\end{itemize}
We do not consider observations from \textit{HST} before 10 observer-frame days, as it is unlikely that the glSN will be identified as strongly lensed much earlier than around the peak of the first image and time to trigger \textit{HST} follow-up must be taken into account.

\subsubsection{Time Delay Recovery}
As expected, the addition of increasing amounts of \textit{HST} follow-up leads to an increasingly precise time delay constraint. The level of improvement, though, depends on whether the system is resolved or unresolved by Rubin-LSST and the brightness of the system being considered.

With just two epochs of \textit{HST} follow-up, under the \textbf{HSTopt+nir2} follow-up configuration, the median time delay uncertainties improve to \LSSTHSTIIunresolvedAmedianTD days, \LSSTHSTIIunresolvedBmedianTD days, and \LSSTHSTIIunresolvedCmedianTD days for each of system A, B, and C. For system D, the median time delay uncertainties are \LSSTHSTIIunresolvedDDmedianTD, \LSSTHSTIIunresolvedDDDmedianTD, and \LSSTHSTIIunresolvedDDDDmedianTD for each of image 2, 3, and 4 relative to image 1. Of course, these constraints are not competitive for time-delay cosmography and there is still significant spread in the range of time delay uncertainties for each system, for example of up to \LSSTHSTIIunresolvedArangeTD days for system A under the \textbf{HSTopt+nir2} follow-up strategy.

We see pronounced improvements again when moving to the \textbf{HSTopt+nir4} strategy, then the \textbf{HSTopt+nir6} strategy. Indeed, the \textbf{HSTopt+nir6} strategy leads to the lowest median uncertainties on the time delays. For system A, B, and C, respectively, the median time delay uncertainty under this strategy is \LSSTHSTVIunresolvedAmedianTD days, \LSSTHSTVIunresolvedBmedianTD days, and \LSSTHSTVIunresolvedCmedianTD days, respectively. For system D, the median time delay uncertainties are \LSSTHSTVIunresolvedDDmedianTD, \LSSTHSTVIunresolvedDDDmedianTD, and \LSSTHSTVIunresolvedDDDDmedianTD days for images 2, 3, and 4. Relative to a fit with Rubin-LSST data alone, the time-delay uncertainties from the fit to the \textbf{HSTopt+nir6} strategy data shows a median improvement of \LSSTHSTVIunresolvedAchangeTD, \LSSTHSTVIunresolvedBchangeTD, and \LSSTHSTVIunresolvedCchangeTD days for systems A, B, and C, and of \LSSTHSTVIunresolvedDDchangeTD, \LSSTHSTVIunresolvedDDDchangeTD, and \LSSTHSTVIunresolvedDDDDchangeTD days for images 2, 3, and 4 of system D.

We tabulate these results in full in Table \ref{tab:hst-results-unres}. Notably, with the \textbf{HSTopt+nir6} strategy, the constraints on the time delays for system C ($\Delta = 15.67$ days) begin to reach the 5\% level of precision. Therefore, a cadenced space-based follow-up strategy can achieve high precision on time delays even for systems that are unresolved by Rubin-LSST. 

When considering the \textbf{HSTopt4} and \textbf{HSTnir4} strategies, we find that the median time delay uncertainties across all systems are \LSSTHSToptunresolvedmedianTD days and \LSSTHSTnirunresolvedmedianTD days, respectively. This result highlights that gaining a wider wavelength coverage with the \textit{HST} NIR filters is preferable over coverage in an overlapping wavelength regime to the Rubin-LSST data. When comparing to the \textbf{HSTopt+nir2} results, which offer a wide wavelength coverage from \textit{HST}, but less time-coverage, the time delay uncertainties from the \textbf{HSTnir4} strategy tends to be lower. Therefore, the better time coverage of this strategy compared to the two epoch strategy indicates that higher cadence data from \textit{HST} is also very important. Still having optical and NIR coverage from \textit{HST}, as in the \textbf{HSTopt+nir4} and \textbf{HSTopt+nir6} follow-up configurations, leads to significant decreases in both the time-delay uncertainties themselves of \LSSTHSTnirunresolvedchangeHSTIVTD-\LSSTHSToptunresolvedchangeHSTVITD days on average relative to when following-up in the optical or NIR alone. Furthermore, with 4-6 epochs of optical and NIR coverage, the time-delay estimates show a smaller range between the minimum and maximum uncertainties across all realisations of a system compared to the optical or NIR only cases.

For the case of resolved Rubin-LSST data, we again find the gains provided by \textit{HST} follow-up to be appreciable, though naturally slightly smaller in absolute value. The results are tabulated by system in Table \ref{tab:hst-results-res}. We find the median time-delay uncertainty decreases from \LSSTHSTIIresolvedmedianTD days for the \textbf{HSTopt+nir2} strategy to \LSSTHSTIVresolvedmedianTD days for the \textbf{HSTopt+nir4}, and further to \LSSTHSTVIresolvedmedianTD days for the \textbf{HSTopt+nir6} strategy. Although gains of a few tenths of a day on the time-delay uncertainty relative to the uncertainty from a fit to resolved Rubin-LSST data alone seem marginal, this difference can be several percentage points improvement in the precision for systems with a time delay of less than 10 days.

We also note a considerable decrease in the spread of the time delay uncertainties when moving from 2 or 4 epochs of \textit{HST} follow-up to the \textbf{HSTopt+nir6} follow-up strategy. The best example of this trend is seen with system C; the spread in the time delay uncertainty decreases from a difference of \LSSTHSTIIresolvedCrangeTD days between the best and worst time-delay constraint for the \textbf{HSTopt+nir2} strategy to one of \LSSTHSTVIresolvedCrangeTD days for the \textbf{HSTopt+nir6} strategy. Similar, though slightly weaker trends are noted for systems A, B, and D. The benefits of \textit{HST} follow-up are not only a decrease in time-delay uncertainty on average, but a more consistent time-delay uncertainty reached for a system of a given brightness. We summarise the results for when systems are assumed to be resolved by Rubin-LSST in Table \ref{tab:hst-results-res}. 

Follow-up in only optical or NIR filters naturally does not lead to as tight a constraint on the time delay as compared to optical and NIR coverage for the same number of epochs. The median time delay uncertainty for the \textbf{HSTopt4} follow-up strategy is \LSSTHSToptresolvedmedianTD days, while the median time delay uncertainty for the \textbf{HSTnir4} follow-up strategy is \LSSTHSTnirresolvedmedianTD days. A follow-up strategy with four epochs of optical or NIR only \textit{HST} follow-up frequently outperforms two epochs of optical and NIR data in terms of time-delay precision. This result again indicates the importance of time coverage of the light curve with \textit{HST}. Interestingly, when the data are resolved from Rubin-LSST, there is a mild preference for the \textbf{HSTopt4} strategy for lower time-delay uncertainties, while the \textbf{HSTnir4} strategy tends to result in lower time-delay uncertainties for the unresolved Rubin-LSST data. While the increased wavelength coverage may be more useful for the unresolved systems, reconciling resolved optical data with NIR data in the case of model misspecification may lead the \textbf{HSTnir4} uncertainties to be larger when combined with resolved Rubin-LSST data.

\begin{figure*}
    \centering
    \includegraphics[width=0.95\linewidth]{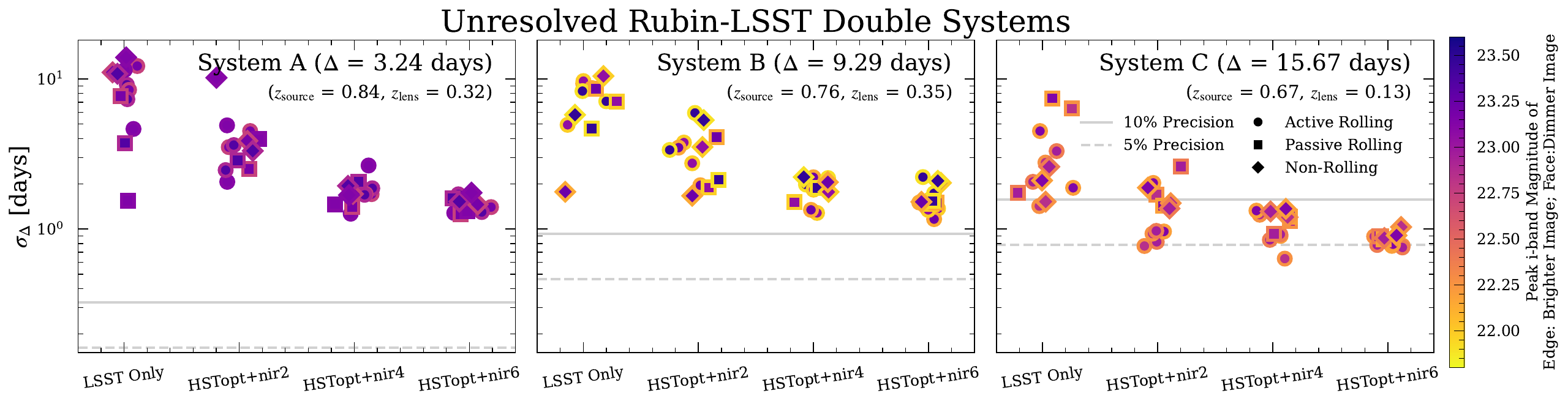}
    \caption{Uncertainty on the time delay as a function of observing strategy, assuming unresolved data from Rubin-LSST. Each sub-panel corresponds to a doubly-imaged lensing system as described in Table \ref{tab:base-objects}. The colours correspond to brightness of the images, with the edge colour representing the peak $i$-band magnitude of the brighter image and the face colour representing the peak $i$-band magnitude of the dimmer image. The circle marker indicates that the glSN is in a WFD rolling active region (Locations 1 and 2 in Table \ref{tab:locations}), the square marker indicates a WFD passive active region (Location 3), and the diamond marker indicates a WFD non-rolling region (Location 4). The solid and dashed gray lines indicate a 10\% and 5\% precision estimate of the time delay, respectively, given the time delay of each system. We focus on time delay uncertainties of less than 3 days in this plot, though there are a significant number of systems which have time delay uncertainties of longer than 3 days in the case of little to no follow-up.}
    \label{fig:td-unc-unresolved}
\end{figure*}

\begin{figure*}
    \centering
    \includegraphics[width=0.95\linewidth]{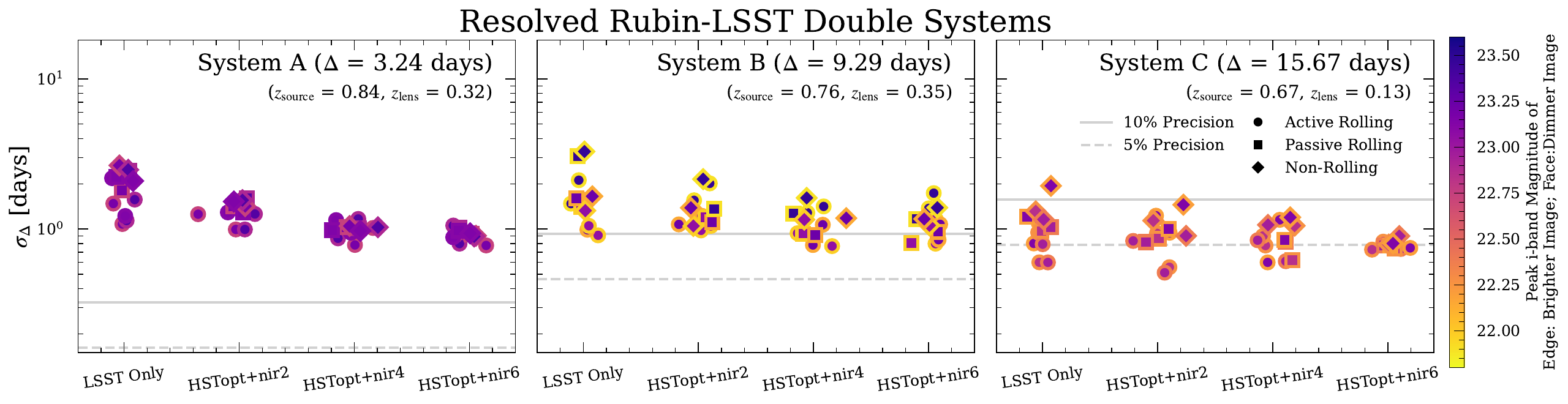}
    \caption{Uncertainty on the time delay as a function of observing strategy, assuming resolved data from Rubin-LSST. Each sub-panel corresponds to a doubly-imaged lensing system as described in Table \ref{tab:base-objects}. The colours correspond to brightness of the images, with the edge colour representing the peak $i$-band magnitude of the brighter image and the face colour representing the peak $i$-band magnitude of the dimmer image. The circle marker indicates that the glSN is in a WFD rolling active region (Locations 1 and 2 in Table \ref{tab:locations}), the square marker indicates a WFD passive active region (Location 3), and the diamond marker indicates a WFD non-rolling region (Location 4). The solid and dashed gray lines indicate a 10\% and 5\% precision estimate of the time delay, respectively, given the time delay of each system.}
    \label{fig:td-unc-resolved}
\end{figure*}

\begin{figure*}
    \centering
    \includegraphics[width=0.95\linewidth]{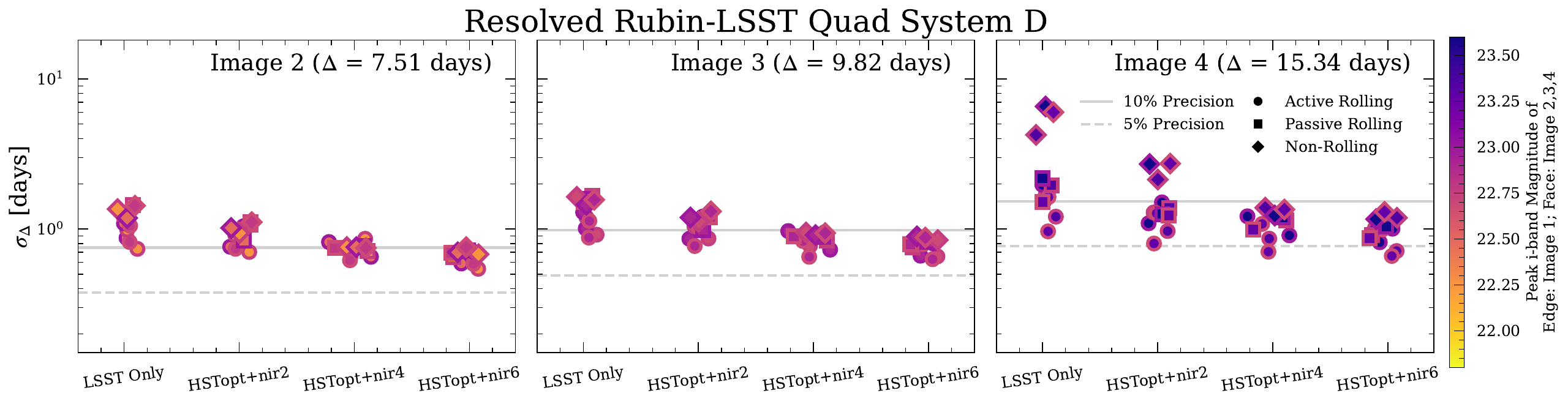}
    \caption{Uncertainty on the time delay as a function of observing strategy, assuming resolved data from Rubin-LSST. Each sub-panel corresponds to the time delay of images 2, 3, and 4, relative to image 1 of the quad system D as described in Table \ref{tab:base-objects}. The colours correspond to brightness of the images, with the edge colour representing the peak $i$-band magnitude of image 1 and the face colour representing the peak $i$-band magnitude of image 2, 3, or 4, as specified in each sub-panel. The circle marker indicates that the glSN is in a WFD rolling active region (Locations 1 and 2 in Table \ref{tab:locations}), the square marker indicates a WFD passive active region (Location 3), and the diamond marker indicates a WFD non-rolling region (Location 4). The solid and dashed gray lines indicate a 10\% and 5\% precision estimate of the time delay, respectively, given the time delay of each image.}
    \label{fig:td-unc-resolved-D}
\end{figure*}

In Figure \ref{fig:td-unc-unresolved} we compare the uncertainty on the time delay estimate across the three optical and NIR follow-up strategies for the unresolved double systems, and in Figure \ref{fig:td-unc-resolved}, we make the same comparison for the resolved double systems. We additionally show in Figure \ref{fig:td-unc-resolved-D} the results for three relative time delays of the resolved quad system. These figures illustrate the decrease in uncertainty in the time-delay estimates with increasing numbers of \textit{HST} epochs, as well as the increased uniformity of the time-delay constraint for increasing numbers of \textit{HST} epochs.

\begin{table*}
    \centering
    \caption{Summary of the results of the \textit{HST} and \textit{JWST} follow-up analysis for systems when they are assumed to be unresolved by Rubin-LSST. The median improvement in $\sigma_{\Delta}$ when considering follow-up is taken to be relative to the time-delay uncertainty from the fit to the Rubin-LSST data alone.}
    \begin{threeparttable}
        \begin{tabular}{c|c|c|l|l|c|c|c}
             & \multicolumn{2}{c}{Rubin-LSST only} &  &  & \multicolumn{3}{c}{with Follow-up} \\
            \hline
            & Median($\sigma_{\Delta}$) & Range($\sigma_{\Delta}$) & \textit{HST} Follow-up & \textit{JWST} Follow-up & Median($\sigma_{\Delta}$) & Range($\sigma_{\Delta}$) & Median Improvement \\
            System & [days] & [days] & Strategy\tnote{a} & Strategy\tnote{b} & [days] & [days] & in $\sigma_{\Delta}$ [days] \\
            \hline
            \hline
            A & 8.79 & 12.49 & \textbf{HSTopt+nir2} & - & 3.58 & 8.14 & 5.00 \\
            ($\Delta=3.24$ days) &  &  & - & \textbf{JWSTopt+nir2} & 2.04 & 1.93 & 6.65 \\
             &  &  & \textbf{HSTopt4} & - & 4.00 & 4.29 & 4.85 \\
             &  &  & \textbf{HSTnir4} & - & 1.80 & 1.02 & 6.72 \\
             &  &  & \textbf{HSTopt+nir4} & - & 1.71 & 1.40 & 7.00 \\
             &  &  & \textbf{HSTopt+nir6} & - & 1.49 & 0.49 & 7.24 \\
             &  &  & \textbf{HSTopt+nir6} & \textbf{JWSTopt+nir2} & 1.20 & 0.49 & 7.54 \\
            \hline
            B & 7.71 & 8.70 & \textbf{HSTopt+nir2} & - & 3.41 & 4.31 & 4.73 \\
            ($\Delta=9.29$ days) &  &  & - & \textbf{JWSTopt+nir2} & 2.20 & 6.25 & 5.55 \\
             &  &  & \textbf{HSTopt4} & - & 4.15 & 4.11 & 3.16 \\
             &  &  & \textbf{HSTnir4} & - & 2.03 & 1.18 & 5.72 \\
             &  &  & \textbf{HSTopt+nir4} & - & 1.89 & 0.94 & 5.87 \\
             &  &  & \textbf{HSTopt+nir6} & - & 1.50 & 1.06 & 5.87 \\
             &  &  & \textbf{HSTopt+nir6} & \textbf{JWSTopt+nir2} & 1.08 & 0.80 & 6.42 \\
            \hline
            C & 2.39 & 6.02 & \textbf{HSTopt+nir2} & - & 1.40 & 1.84 & 1.19 \\
            ($\Delta=15.67$ days) &  &  & - & \textbf{JWSTopt+nir2} & 1.19 & 0.79 & 1.13 \\
             &  &  & \textbf{HSTopt4} & - & 1.33 & 1.49 & 1.03 \\
             &  &  & \textbf{HSTnir4} & - & 1.42 & 1.07 & 0.92 \\
             &  &  & \textbf{HSTopt+nir4} & - & 1.17 & 0.73 & 1.24 \\
             &  &  & \textbf{HSTopt+nir6} & - & 0.87 & 0.28 & 1.55 \\
             &  &  & \textbf{HSTopt+nir6} & \textbf{JWSTopt+nir2} & 0.64 & 0.28 & 1.71 \\
            \hline
            D$_{1,2}$ & 8.00 & 26.54 & \textbf{HSTopt+nir2} & - & 2.68 & 4.34 & 4.32 \\
            ($\Delta=7.51$ days) &  &  & - & \textbf{JWSTopt+nir2} & 1.64 & 2.98 & 4.80 \\
             &  &  & \textbf{HSTopt4} & - & 2.91 & 4.39 & 5.08 \\
             &  &  & \textbf{HSTnir4} & - & 1.75 & 3.21 & 5.40 \\
             &  &  & \textbf{HSTopt+nir4} & - & 1.40 & 1.34 & 6.53 \\
             &  &  & \textbf{HSTopt+nir6} & - & 1.34 & 0.72 & 6.62 \\
             &  &  & \textbf{HSTopt+nir6} & \textbf{JWSTopt+nir2} & 0.97 & 0.47 & 6.87 \\
            D$_{1,3}$ & 6.42 & 21.03 & \textbf{HSTopt+nir2} & - & 3.06 & 4.55 & 3.96 \\
            ($\Delta=9.82$ days) &  &  & - & \textbf{JWSTopt+nir2} & 1.73 & 3.43 & 4.80 \\
             &  &  & \textbf{HSTopt4} & - & 3.06 & 5.55 & 3.00 \\
             &  &  & \textbf{HSTnir4} & - & 1.78 & 2.88 & 4.62 \\
             &  &  & \textbf{HSTopt+nir4} & - & 1.45 & 1.32 & 4.95 \\
             &  &  & \textbf{HSTopt+nir6} & - & 1.40 & 0.91 & 5.22 \\
             &  &  & \textbf{HSTopt+nir6} & \textbf{JWSTopt+nir2} & 1.02 & 0.52 & 5.45 \\
            D$_{1,4}$ & 5.90 & 10.20 & \textbf{HSTopt+nir2} & - & 2.67 & 9.41 & 2.65 \\
            ($\Delta=15.34$ days) &  &  & - & \textbf{JWSTopt+nir2} & 1.95 & 1.13 & 3.96 \\
             &  &  & \textbf{HSTopt4} & - & 2.87 & 3.11 & 2.89 \\
             &  &  & \textbf{HSTnir4} & - & 2.15 & 4.86 & 3.36 \\
             &  &  & \textbf{HSTopt+nir4} & - & 1.43 & 1.19 & 4.49 \\
             &  &  & \textbf{HSTopt+nir6} & - & 1.23 & 1.02 & 4.81 \\
             &  &  & \textbf{HSTopt+nir6} & \textbf{JWSTopt+nir2} & 1.09 & 0.59 & 4.87 \\
            \hline
            \hline
        \end{tabular}
        \begin{tablenotes}
           \item [a] \textbf{HSTopt+nir\textit{N}} refers to \textit{HST} follow-up at \textbf{\textit{N}} epochs in the \textit{F625W}, \textit{F814W}, \textit{F110W}, and \textit{F160W} filters. For the 2 epoch case, the observations are made at (20,40) rest-frame days, for the 4 epoch case, the observations are made at (10, 20, 30, 40) rest-frame days, and for the 6 epoch case, the observations are made at (10, 17, 24, 31, 38, 45) rest-frame days. \textbf{HSTopt4} refers to \textit{HST} follow-up at 4 epochs -- (10, 20, 30, 40) rest-frame days -- in the \textit{F625W} and \textit{F814W} filters, and \textbf{HSTnir4} refers to 4 epochs of follow-up in the \textit{F110W} and \textit{F160W} filters.
           
           \item [b] \textbf{JWSTopt+nir2} refers to \textit{JWST} follow-up at 2 epochs -- (20, 40) rest-frame days -- in the \textit{F070W}, \textit{F090W}, \textit{F115W}, \textit{F150W}, \textit{F200W}, and \textit{F277W} filters. Note, though, that the \textit{F277W} filter drops out for systems B and C as it falls outside the wavelength coverage of the \textsc{BayeSN} model.
        \end{tablenotes}
    \end{threeparttable}
    \label{tab:hst-results-unres}
\end{table*}

\begin{table*}
    \centering
    \caption{Summary of the results of the \textit{HST} and \textit{JWST} follow-up analysis for systems when they are assumed to be resolved by Rubin-LSST. The median improvement in $\sigma_{\Delta}$ when considering follow-up is taken to be relative to the time-delay uncertainty from the fit to the Rubin-LSST data alone.}
    \begin{threeparttable}
        \begin{tabular}{c|c|c|l|l|c|c|c}
             & \multicolumn{2}{c}{Rubin-LSST only} &  &  & \multicolumn{3}{c}{with Follow-up} \\
            \hline
            & Median($\sigma_{\Delta}$) & Range($\sigma_{\Delta}$) & \textit{HST} Follow-up & \textit{JWST} Follow-up & Median($\sigma_{\Delta}$) & Range($\sigma_{\Delta}$) & Median Improvement \\
            System & [days] & [days] & Strategy\tnote{a} & Strategy\tnote{b} & [days] & [days] & in $\sigma_{\Delta}$ [days] \\
            \hline
            \hline
            A & 1.95 & 1.56 & \textbf{HSTopt+nir2} & - & 1.36 & 0.61 & 0.48 \\
            ($\Delta=3.24$ days) &  &  & - & \textbf{JWSTopt+nir2} & 1.41 & 0.74 & 0.50 \\
             &  &  & \textbf{HSTopt4} & - & 1.33 & 0.72 & 0.66 \\
             &  &  & \textbf{HSTnir4} & - & 1.12 & 0.46 & 0.77 \\
             &  &  & \textbf{HSTopt+nir4} & - & 1.01 & 0.38 & 0.91 \\
             &  &  & \textbf{HSTopt+nir6} & - & 0.92 & 0.28 & 1.02 \\
             &  &  & \textbf{HSTopt+nir6} & \textbf{JWSTopt+nir2} & 0.91 & 0.22 & 1.02 \\
            \hline
            B & 1.52 & 2.37 & \textbf{HSTopt+nir2} & - & 1.21 & 1.17 & 0.23 \\
            ($\Delta=9.29$ days) &  &  & - & \textbf{JWSTopt+nir2} & 1.19 & 1.48 & 0.20 \\
             &  &  & \textbf{HSTopt4} & - & 1.21 & 1.00 & 0.23 \\
             &  &  & \textbf{HSTnir4} & - & 1.23 & 0.88 & 0.22 \\
             &  &  & \textbf{HSTopt+nir4} & - & 1.11 & 0.84 & 0.47 \\
             &  &  & \textbf{HSTopt+nir6} & - & 1.12 & 0.93 & 0.48 \\
             &  &  & \textbf{HSTopt+nir6} & \textbf{JWSTopt+nir2} & 0.95 & 0.84 & 0.54 \\
            \hline
            C & 1.00 & 1.34 & \textbf{HSTopt+nir2} & - & 0.92 & 0.94 & 0.13 \\
            ($\Delta=15.67$ days) &  &  & - & \textbf{JWSTopt+nir2} & 0.94 & 0.74 & -0.01 \\
             &  &  & \textbf{HSTopt4} & - & 0.79 & 0.66 & 0.27 \\
             &  &  & \textbf{HSTnir4} & - & 1.00 & 0.80 & -0.03 \\
             &  &  & \textbf{HSTopt+nir4} & - & 0.85 & 0.59 & 0.16 \\
             &  &  & \textbf{HSTopt+nir6} & - & 0.78 & 0.17 & 0.24 \\
             &  &  & \textbf{HSTopt+nir6} & \textbf{JWSTopt+nir2} & 0.60 & 0.23 & 0.37 \\
            \hline
            D$_{1,2}$ & 1.14 & 0.71 & \textbf{HSTopt+nir2} & - & 0.97 & 0.41 & 0.14 \\
            ($\Delta=7.51$ days) &  &  & - & \textbf{JWSTopt+nir2} & 0.91 & 0.39 & 0.19 \\
             &  &  & \textbf{HSTopt4} & - & 0.89 & 0.35 & 0.20 \\
             &  &  & \textbf{HSTnir4} & - & 0.85 & 0.29 & 0.31 \\
             &  &  & \textbf{HSTopt+nir4} & - & 0.75 & 0.24 & 0.36 \\
             &  &  & \textbf{HSTopt+nir6} & - & 0.70 & 0.21 & 0.43 \\
             &  &  & \textbf{HSTopt+nir6} & \textbf{JWSTopt+nir2} & 0.67 & 0.21 & 0.44 \\
            D$_{1,3}$ & 1.37 & 0.78 & \textbf{HSTopt+nir2} & - & 1.08 & 0.54 & 0.20 \\
            ($\Delta=9.82$ days) &  &  & - & \textbf{JWSTopt+nir2} & 1.00 & 0.51 & 0.31 \\
             &  &  & \textbf{HSTopt4} & - & 1.05 & 0.46 & 0.28 \\
             &  &  & \textbf{HSTnir4} & - & 1.06 & 0.48 & 0.23 \\
             &  &  & \textbf{HSTopt+nir4} & - & 0.89 & 0.31 & 0.43 \\
             &  &  & \textbf{HSTopt+nir6} & - & 0.79 & 0.26 & 0.53 \\
             &  &  & \textbf{HSTopt+nir6} & \textbf{JWSTopt+nir2} & 0.73 & 0.25 & 0.58 \\
            D$_{1,4}$ & 1.81 & 5.60 & \textbf{HSTopt+nir2} & - & 1.28 & 1.92 & 0.47 \\
            ($\Delta=15.34$ days) &  &  & - & \textbf{JWSTopt+nir2} & 1.29 & 3.84 & 0.46 \\
             &  &  & \textbf{HSTopt4} & - & 1.28 & 1.44 & 0.60 \\
             &  &  & \textbf{HSTnir4} & - & 1.41 & 1.81 & 0.42 \\
             &  &  & \textbf{HSTopt+nir4} & - & 1.17 & 0.68 & 0.70 \\
             &  &  & \textbf{HSTopt+nir6} & - & 1.00 & 0.64 & 0.86 \\
             &  &  & \textbf{HSTopt+nir6} & \textbf{JWSTopt+nir2} & 0.95 & 0.52 & 0.76 \\
            \hline
            \hline
        \end{tabular}
        \begin{tablenotes}
           \item [a] \textbf{HSTopt+nir\textit{N}} refers to \textit{HST} follow-up at \textbf{\textit{N}} epochs in the \textit{F625W}, \textit{F814W}, \textit{F110W}, and \textit{F160W} filters. For the 2 epoch case, the observations are made at (20,40) rest-frame days, for the 4 epoch case, the observations are made at (10, 20, 30, 40) rest-frame days, and for the 6 epoch case, the observations are made at (10, 17, 24, 31, 38, 45) rest-frame days. \textbf{HSTopt4} refers to \textit{HST} follow-up at 4 epochs -- (10, 20, 30, 40) rest-frame days -- in the \textit{F625W} and \textit{F814W} filters, and \textbf{HSTnir4} refers to 4 epochs of follow-up in the \textit{F110W} and \textit{F160W} filters.
           
           \item [b] \textbf{JWSTopt+nir2} refers to \textit{JWST} follow-up at 2 epochs -- (20, 40) rest-frame days -- in the \textit{F070W}, \textit{F090W}, \textit{F115W}, \textit{F150W}, \textit{F200W}, and \textit{F277W} filters. Note, though, that the \textit{F277W} filter drops out for systems B and C as it falls outside the wavelength coverage of the \textsc{BayeSN} model.
        \end{tablenotes}
    \end{threeparttable}
    \label{tab:hst-results-res}
\end{table*}

Therefore, we conclude that the \textbf{HSTopt+nir6} strategy unsurprisingly provides the best time-delay constraints. While this follow-up strategy is the most costly, it achieves the lowest median time-delay uncertainty and with a high level of uniformity in the time-delay constraint across variations in amounts of Rubin-LSST data and line-of-sight effects. We can be more confident in our ability to achieve an optimal time-delay estimate because the difference between best and worst case-scenario is much smaller. In the case of unresolved data, 5.6\% of the 72 time delays estimated in this case represent a sub-5\% precision estimate and in the case of resolved data, 11.1\% represent a sub-5\% precision estimate. In the case of unresolved data, only system C is measured to sub-5\% precision (33.3\% of the 12 realisations of this system). When the systems are resolved, 50\% of the time delays in realisations of system C and 16.7\% of the time delays of image 4 relative to image 1 in realisations of system D are measured to sub-5\% precision. Again, none of the realisations of the other systems are measured to this level of precision. As this represents a 300\% and 60\% increase to the number of glSNe~Ia with sub-5\% precision for the \textbf{HSTopt+nir6} strategy compared to the \textbf{HSTopt+nir4} strategy, in the unresolved and resolved cases respectively. Given this significant gain compared to even the \textbf{HSTopt+nir4} strategy, investing in the extra epochs is worthwhile. 

We additionally note that fitting with more \textit{HST} data occasionally leads to an increase in the time delay uncertainty, particularly when at least one of the images is brighter than 22.5 mag in the $i$-band. We attribute this unexpected behaviour to the mismatch between the model which we simulate from, \textsc{BayeSN}, and the model which we fit with, \texttt{SALT3}. With higher quality data, the results become sensitive to an imperfect model which can lead to these anomalies. We explore this result in greater detail in \S\ref{sec:discussion-mismatch}.

Finally, we highlight that the error floor on the time-delay depends heavily on the brightness of the glSN~Ia system. The trend is very clear with both the doubles and the quad considered, though the quad tends to have lower median uncertainties than the doubles for comparable image brightnesses, likely because of the larger quantity of data with which to constrain shared system parameters (e.g., stretch, colour). This result is unsurprising, as brighter systems will have higher SNR data from Rubin-LSST, but the effect is appreciable enough to make a difference of up to a few percentage points in precision on a $\sim$ 10 day time delay. We test the limits of the time delay precision with a brighter system and discuss these results in \S\ref{sec:discussion-highmag}.

\subsubsection{Absolute Magnification Recovery}
The absolute magnification constraints from Rubin-LSST data alone are weak, but the addition of \textit{HST} follow-up significantly improves the recovery precision. Under the \textbf{HSTopt+nir2} strategy, the median absolute magnification uncertainty decreases to \LSSTHSTIIunresolvedmedianMU mag. Increasing the time coverage also improves the absolute magnification constraints, likely by better constraining on the colour evolution of the SNe~Ia and allowing that effect to be disentangled from time-invariant dust extinction. The \textbf{HSTopt+nir4} and \textbf{HSTopt+nir6} strategies both reach a median absolute magnification uncertainty of \LSSTHSTVIunresolvedmedianMU mag. Furthermore, in both the cases of 4 and 6 epochs of \textit{HST} data, the standard deviation of the residuals is reduced to \LSSTHSTIVunresolvedstdMU mag. The bias in the absolute magnification residuals is also reduced to below \LSSTHSTVIunresolvedbiasMU mag, which is within the variation expected for a sample size of 120 estimated absolute magnifications. 

The wide wavelength coverage from \textit{HST} is very important in the case of unresolved Rubin-LSST data. With \textbf{HSTopt4}, the median absolute magnification uncertainty is \LSSTHSToptunresolvedmedianMU mag, and with \textbf{HSTnir4}, it is \LSSTHSTnirunresolvedmedianMU mag. Unlike for the time delays, the \textbf{HSTopt4} strategy gives mildly better results than the \textbf{HSTnir4}. In this case, we attribute the difference to the need for the optical \textit{HST} to break degeneracies in the Rubin-LSST unresolved optical data in order to take advantage of this data. These results also show higher dispersion in the absolute magnification residuals compared to fits which include both optical and NIR coverage from \textit{HST} in general. There is some trade-off with time coverage, though, as the \textbf{HSTopt4} does outperform the \textbf{HSTopt+nir2} strategy according to these two metrics. 

When the Rubin-LSST data is assumed to be resolved, the absolute magnification constraints again show less dramatic improvements when adding follow-up observations. The \textbf{HSTopt+nir2} strategy does not result in any significant change to the median absolute magnification uncertainty compared to Rubin-LSST data alone, though the dispersion of the residuals is lowered to \LSSTHSTIIresolvedstdMU mag. Under the \textbf{HSTopt+nir4} and \textbf{HSTopt+nir6} observing strategies, the absolute magnification constraints improve to a median of \LSSTHSTVIresolvedmedianMU mag in both cases. For both of these observing strategies, the dispersion in the residuals remains low at \LSSTHSTVIresolvedstdMU mag. All biases in the residuals are found to be small when Rubin-LSST data is assumed to be resolved, irrespective of amount of follow-up.

\begin{figure*}
    \centering
    \includegraphics[width=0.95\linewidth]{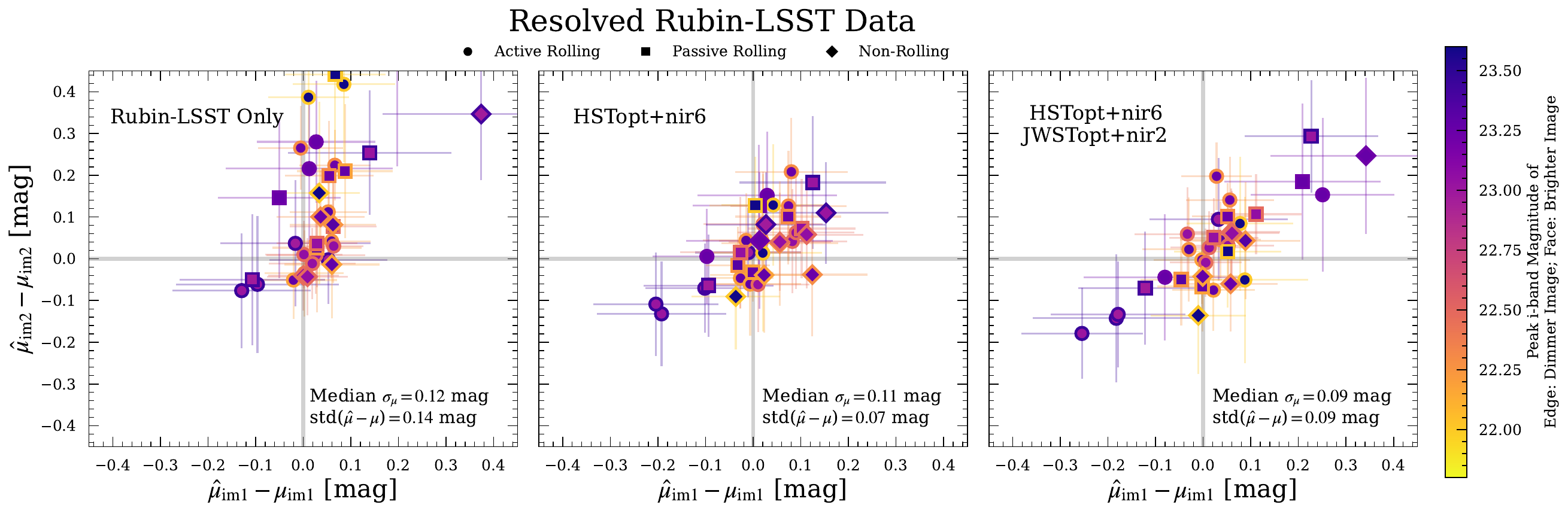}
    \caption{The recovery of the absolute magnification of image 1 (x-axis) and image 2 (y-axis) for three follow-up configurations. We use $\hat{\mu}_{\rm im \, M}$ to denote the absolute magnification of image M from the \textsc{Glimpse} fit and $\mu_{\rm im \, M}$ to denote the true absolute magnification of image M. \textbf{Left:} Resolved Rubin-LSST data alone, \textbf{Center:} with 6 epochs of HST data in the \textit{F625W}, \textit{F814W}, \textit{F110W}, and \textit{F160W} filters, and \textbf{Right:} additionally with 2 epochs of \textit{JWST} data in the \textit{F070W}, \textit{F090W}, \textit{F115W}, \textit{F150W}, \textit{F200W}, and \textit{F277W} filters. The edge colour of the points indicate the $i$-band peak magnitude of the brighter image and the face colour of the points indicate the $i$-band peak magnitude of the dimmer image. The markers indicate the amount of Rubin-LSST data as a function of survey setting. The uncertainties on the individual systems do not significantly decrease with more data, but the overall dispersion around the true absolute magnifications is much lessened when \textit{HST} data is included. Comparing the results with and without \textit{JWST} data is difficult because a different template is used to fit the light curves with \textit{JWST} data (\texttt{SALT3-NIR}), than without it (\texttt{SALT3}). Still, the additional \textit{JWST} data does not appreciably improve the results, which we attribute model misspecification as it will particularly affect the results when redder data is included in the fit.}
    \label{fig:mag-recovery-res}
\end{figure*}

In Figure \ref{fig:mag-recovery-res}, we show the absolute magnification constraints on the absolute magnifications of each of the two images of the doubles (systems A, B, and C) considered in this work. We show the results when the absolute magnifications are computed from the fit to the resolved Rubin-LSST data alone, and when they are computed from the fit to the resolved Rubin-LSST with \textbf{HSTopt+nir6} follow-up. This plot particularly illustrates the improved accuracy of the absolute magnification constraints when including \textit{HST} follow-up. 

As in the case of unresolved Rubin-LSST data, the \textbf{HSTopt4} strategy gives mildly better constraints on the absolute magnification than the \textbf{HSTnir4}. For \textbf{HSTopt4}, the median absolute magnification uncertainty is \LSSTHSToptresolvedmedianMU with a dispersion in the residuals of \LSSTHSToptresolvedstdMU mag. The \textbf{HSTnir4} strategy shows a marginally increased median absolute magnification uncertainty of \LSSTHSTnirresolvedmedianMU mag, but does not have as competitive a dispersion in the residuals, at \LSSTHSTnirresolvedstdMU mag, though the uncertainties are well-calibrated. Interestingly, the \textbf{HSTopt4} result is an improvement on the results from the \textbf{HSTopt+nir2} strategy, emphasising the need for 1) time coverage to constrain colour evolution, and 2) \textit{HST} wavelength coverage that overlaps with the Rubin-LSST wavelength coverage. Still the \textbf{HSTopt+nir4} strategy shows lower residual dispersion, so the additional NIR coverage is useful to absolute magnification constraints. Therefore, at least four epochs of \textit{HST} data in preferably four \textit{HST} filters across the optical and NIR are necessary to achieve the most precise possible absolute magnification constraints while maintaining a high level of accuracy.

These results generally indicate that \textsc{Glimpse} can provide some constraint on the absolute magnifications, with it kept in mind that the recovered absolute magnification necessarily depends on the assumed underlying template used with \textsc{Glimpse}. We find that the results of this section are strongly dependent on the choice of filter we use to compare the true unlensed magnitude from a \textsc{BayeSN} template to the measured lensed magnitude from a \texttt{SALT3} fit. Using a different filter than the $r$-band resulted in greater biases on the absolute magnification. The differences in the way colour is handled by \textsc{BayeSN} and \texttt{SALT} may be one contributing factor to this effect, as the dust effects begin to take on different meanings in these two contexts. As a result, we occasionally find that increasing amount of NIR data in particular leads to less precise absolute magnification constraints, as was seen for the time-delay constraints. Models of SN~Ia SEDs are known to differ more in the NIR than in the optical \citep{salt3nir, Jones_2022, Peterson_2023}, which aligns with this finding. Further to this point, the challenge of accurate recovery of dust extinction is known to be a significant source of systematic uncertainty in SNe~Ia modelling \citep{Freedman_2009, Popovic_2023, Vincenzi_2024, Thorp_2024, Popovic_2024, Popovic_2025, Dhawan_2025}, as well.

For these reasons, it is unsurprising that there be significant uncertainties regarding dust and absolute magnification recovery in the \textsc{Glimpse} fit, especially when there are other compounding effects with wavelength dependence (e.g., microlensing). In practice, we will estimate the true unlensed apparent magnitude of a SNe~Ia at some redshift from e.g., Pantheon+ \citep{Brout_2022}, which has its own set of uncertainties to consider that may make any such biases from absolute magnification recovery small in comparison \citep{Pierel_2024b}. Still, much care must be taken when considering absolute magnification and dust constraints from light curve fitting, and these estimates should be used with appropriate uncertainties.

We conclude that the general trends seen in this work affirm the need for a wide wavelength coverage of resolved observations to get sufficiently precise absolute magnification constraints. These results also interestingly highlight the importance of time coverage to absolute magification constraints. The added time coverage may be helping to disentangle the colour evolution of the SNe~Ia light curves from dust and chromatic microlensing effects in order to best estimate absolute magnifications. Furthermore, better time coverage gives more precise time-delay constraints, which will lead to improvements on the absolute magnification recovery. We leave to future work a full investigation of how these three wavelength-dependent components are constrained by the \textsc{Glimpse} model.  

\textit{HST} follow-up strategies with 4-6 epochs in the optical and NIR provide good enough constraints to reduce the MSD to a sub-dominant source of uncertainty for the majority of systems considered, in both the case of unresolved and resolved data, in the absence of other uncertainties. We refer the reader to \citet{Pierel_2024b} for a reference regarding the contribution of additional sources of uncertainty to the absolute magnification constraints. We finally reiterate that the absolute magnification for individual objects must be carefully considered, especially in the context of light curve template choice, which affects the assumed true underlying colours of the transient and therefore absolute magnification inference. 

\subsection{Follow-up with \textit{JWST}}
\label{sec:nir-followup} 
The depth and resolution offered by \textit{JWST} in the NIR is a unique advantage of the instrument; \textit{JWST} probes a redder rest-frame wavelength regime than is possible with either Rubin-LSST or \textit{HST}, which may improve constraints on line-of-sight dust extinction effects and, therefore, magnifications. Furthermore, high-resolution spectroscopy from \textit{JWST} is useful for spectroscopic time-delays \citep{Chen_2024} and studies of the redshift evolution of SNe \citep{Dhawan_2024}. Because of the significant overheads and limited visibility windows, though, it will only be possible to obtain a few epochs of data from \textit{JWST}. Therefore, we next explore the impact of two epochs of \textit{JWST} follow-up in up to six filters -- \textit{F070W}, \textit{F090W}, \textit{F115W}, \textit{F150W}, \textit{F200W}, and \textit{F277W}. We assume that two epochs of \textit{JWST} in each filter are obtained at 20 and 40 rest-frame days. We refer to this strategy as \textbf{JWSTopt+nir2}. Note that these are the same times of observation as the \textbf{HSTopt+nir2} strategy, but offset from the times of observation for the \textbf{HSTopt+nir6} strategy.

We use the \texttt{SALT3-NIR} extension of the \texttt{SALT3} model from \citet{salt3nir} to fit the light curves with \textit{JWST} observations. All other fitting settings are kept consistent with those in the previous section. Owing to the lower redshifts of systems B and C, the \textit{F277W} filter is not able to be modelled by \texttt{SALT3-NIR}. Firstly, we fit the Rubin-LSST data with the \textbf{JWSTopt+nir2} follow-up, and secondly, we fit the Rubin-LSST data with the \textbf{HSTopt+nir6} and \textbf{JWSTopt+nir2} follow-up. 

\textit{JWST} is also well-suited for the discovery and/or follow-up of high-redshift glSNe~Ia ($z \gtrsim 1$) owing to its redder wavelength coverage, but Rubin-LSST is unlikely to discover glSNe~Ia at those distances \citep{Huber_2022}. Therefore, we maintain the focus of this work on an intermediate redshift range.

\subsubsection{Time Delay Recovery}
The \textbf{JWSTopt+nir2} follow-up strategy leads to significant improvements in the time-delay recovery compared to when Rubin-LSST data is fit alone. This strategy demonstrates an improvement on the results from the \textbf{HSTopt+nir2} strategy and is, in some cases, competitive in with the results from the \textbf{HSTopt4} and \textbf{HSTnir4} strategies. When Rubin-LSST data is assumed to be unresolved, the two epochs of resolved \textit{JWST} follow-up in five to six filters results in a median time-delay uncertainty of \LSSTJWSTunresolvedmedianTD days. Compared to the \textbf{HSTopt+nir2}, which has a median time delay uncertainty of \LSSTHSTIIunresolvedmedianTD days across all systems, the \textbf{JWSTopt+nir2} strategy is appreciably better with its additional redder wavelength coverage and higher sensitivity for faint objects (see Appendix \ref{sec:appendixB}). This result demonstrates that the additional wavelength coverage does add leverage to the time-delay constraints.

We also highlight that the biggest gains relative to Rubin-LSST data alone are seen for the dimmest and highest redshift system, system A. With two epochs of \textit{JWST} optical and NIR data, system A has a median time-delay uncertainty of \LSSTJWSTunresolvedAmedianTD days, reduced by \LSSTJWSTunresolvedAchangeTD days on average compared to Rubin-LSST data alone. 

We summarise the results from this test for all systems in Table \ref{tab:hst-results-unres}. This table makes clear that the \textbf{JWSTopt+nir2} strategy, despite being only two epochs of data, produces similarly precise results to the \textbf{HSTopt4} and \textbf{HSTnir4} follow-up strategies. We conclude that a wide wavelength coverage may, in some cases, be more beneficial for unresolved systems than better time coverage of data. Having resolved data that overlaps with the optical, allowing the unresolved data to be better utilised, and that covers the NIR is found to be very beneficial for time-delay recovery.

When the Rubin-LSST data is resolved, the gains are less significant in absolute value, but still generally represent an improvement on the results from Rubin-LSST data alone. The median time-delay uncertainty under the \textbf{JWSTopt+nir2} follow-up strategy is \LSSTJWSTresolvedmedianTD days, which is a median improvement of \LSSTJWSTresolvedchangeTD days relative to the results from Rubin-LSST data alone across all systems. This result is very comparable to the \textbf{HSTopt+nir2} follow-up results for resolved Rubin-LSST data. We give these results by system in Table \ref{tab:hst-results-res}. We re-iterate that fractions of a day can correspond to several percentage points in precision on time delays of less than e.g., 10 days, so this improvement can be very impactful. 

Indeed, we occasionally find that the addition of the \textit{JWST} data leads to an increase in the uncertainty on the time delay. We attribute the occasional increase in time-delay uncertainty to model misspecification (see \S\ref{sec:discussion-mismatch}), which is known to be more pronounced in the NIR than in the optical \citep{salt3nir, Jones_2022, Peterson_2023}. Therefore, adding redder data poses a bigger challenge to the fitting in the case of model misspecification, which may lead to an increase in time-delay uncertainty. The issue of discrepancies between templates will be exacerbated for brighter systems, which have higher SNR data, and we see this reflected in our results in Table \ref{tab:hst-results-res}. The variation of the template with redshift may explain further differences in outcomes seen for the different systems, as \texttt{SALT3-NIR} and \textsc{BayeSN} will show different evolutions and may vary in similarity across redshift. Finally, low number of fits being averaged over for each system may result in statistical fluctuations in each bin. 

As the benefits of the \textit{JWST} follow-up are seen most significantly for the faintest objects, and these systems happen to have shorter time delays in our sample, we do not see an increase to the fraction of systems with time delays estimated to sub-5\% precision relative to with Rubin-LSST data alone. As the sample considered in this work was chosen to test a range of properties, rather than be representative of all systems expected to be discovered by Rubin-LSST, this result should not be taken to conclude that \textit{JWST} data does not help to give precise time delays in conjunction with Rubin-LSST. Instead, it suggests that the properties of the glSNe~Ia should be considered when deciding what type of follow-up is most appropriate.

A more aggressive follow-up approach may include observations from both \textit{HST} and \textit{JWST}. We finally test the case of six epochs of \textit{HST} follow-up in the four filters covering the optical and NIR with two epochs of \textit{JWST} follow-up in five to six filters, depending on the redshift of the system, covering the optical and NIR. It may be more realistic to expect two epochs of \textit{HST} follow-up with two epochs of \textit{JWST} follow-up. Seeing as the \textbf{JWSTopt+nir2} follow-up strategy gives similar results to the \textbf{HSTopt+nir2} follow-up strategy, we might expect that two epochs each from \textit{HST} and \textit{JWST} in the optical and NIR spread out over 10-40 rest-frame days would give similar, if not slightly improved, results to the \textbf{HSTopt+nir4} strategy that has already been considered. Therefore, we only test the more unique case of the combined \textbf{HSTopt+nir6} and \textbf{JWSTopt+nir2} strategies.

The combination of these follow-up approaches leads to appreciable additional gains in time-delay precision, regardless of whether the systems are resolved or unresolved by Rubin-LSST. For systems that are unresolved from the ground, we find that the median time-delay uncertainty is \LSSTHSTJWSTunresolvedmedianTD days overall, or \LSSTHSTJWSTunresolvedAmedianTD days for system A, \LSSTHSTJWSTunresolvedBmedianTD days for system B, \LSSTHSTJWSTunresolvedCmedianTD days for system C, \LSSTHSTJWSTunresolvedDDmedianTD days for image 2 of system D, \LSSTHSTJWSTunresolvedDDDmedianTD days for image 3 of system D, and \LSSTHSTJWSTunresolvedDDDDmedianTD days for image 4 of system D. The median improvement in the time-delay uncertainty is \LSSTHSTJWSTunresolvedchangeTD days compared to the uncertainty on the fits of Rubin-LSST data alone. Comparing to the \textbf{HSTopt+nir6} strategy, this result represents a 0.25 day median further reduction across all realisations of all systems. 

For a individual system that is unresolved by Rubin-LSST, we compare the results for the case of no follow-up, two epochs of \textit{JWST} follow-up, six epochs of \textit{HST} follow-up, and this \textit{JWST} and \textit{HST} follow-up together in Figure \ref{fig:space-corner}. We show the full posterior over the light curve, time delay, and magnification parameters for system A in a rolling active location under dust configuration ``c,'' which assumes differing amount of dust extinction along the different lines of sight. With two epochs of \textit{JWST} optical and NIR observations, the time-delay uncertainty is reduced to 2.55 days from an essentially unconstrained time delay using unresolved Rubin-LSST data alone. The uncertainty is further reduced to 1.48 days when \textit{JWST} follow-up is combined with \textit{HST} follow-up. Although this result is not competitive for cosmology, largely owing to the short time delay of only $3.24$ days, the combination of unresolved data with relatively few resolved observations and only weak priors for this level of time-delay is a significant step forward for time-delay estimation.

If the system is resolved by Rubin-LSST, we find that the median time delay uncertainty is \LSSTHSTJWSTresolvedmedianTD days overall. System C, being the brightest, see the lowest median time-delay uncertainty at \LSSTHSTJWSTresolvedCmedianTD days when averaged across spatiotemporal locations and dust configurations. System C also sees the greatest gains from the addition of two epochs of \textit{JWST} data in five filters compared to the \textbf{HSTopt+nir6} strategy alone, as it shows a median further reduction of 0.18 days. We also note a decrease in the spread of the time-delay uncertainties when these strategies are combined, which holds across almost all system configurations. The additional data therefore results in greater consistency in the uncertainties being achieved on the time delays for these systems, as well as a reduced time-delay uncertainty in general.

The combined \textit{HST} and \textit{JWST} follow-up strategy results in 16.7\% of all realisations of all systems are measured to sub-5\% precision in the case of unresolved Rubin-LSST data. This fraction represents a 200\% increase compared to the case of \textbf{HSTopt+nir6} follow-up alone. When the systems are resolved, 18.1\% of all time delays estimated for this test reach sub-5\% precision, which is a 550\% increase relative to Rubin-LSST data alone and a 62.5\% increase relative to the \textbf{HSTopt+nir6} strategy. Thus, the complementary wavelength and time coverage provided by \textit{JWST} are a highly valuable investment for improving the precision of time delay constraints. Indeed, now nearly all realisations of systems C are measured to sub-5\% precision, regardless of whether the Rubin-LSST data is resolved or unresolved, though it is still the case that no image pairs with time delays of shorter than 15 days reach this precision threshold.

\subsubsection{Absolute Magnification Recovery}
With an increased wavelength range covered by observations, the absolute magnifications are theoretically easier to constrain because the particular time-invariant chromaticity of dust extinction can be disentangled from the colour evolution of SNe~Ia \citep{Thorp_2021}. However, the limitations of SNe~Ia SEDs may limit the benefits of redder data, as seen with the time-delay estimation. When the systems are assumed to be unresolved by Rubin-LSST, we find the median uncertainty on the magnification to be \LSSTJWSTunresolvedmedianMU mag across all images of all systems. This level of precision falls between the that from the \textbf{HSTopt+nir2} follow-up strategy and from the \textbf{HSTopt+nir4} and \textbf{HSTopt+nir6} follow-up strategies. The same trend is found for the dispersion of the absolute magnification residuals, which is \LSSTJWSTunresolvedstdMU mag for the \textbf{JWSTopt+nir2} strategy. The bias on the constraints is consistent with zero for this same size.

If the systems are resolved by Rubin-LSST, we find that the median uncertainty on the absolute magnifications is \LSSTJWSTresolvedmedianMU mag and a dispersion in the residuals of \LSSTJWSTresolvedstdMU mag under the \textbf{JWSTopt+nir2} follow-up strategy. Again, the bias in the residuals is negligible. As with the case of unresolved Rubin-LSST data, these results fall at an intermediate point between the 2 and 4-6 epoch follow-up strategies from \textit{HST}. In all, these results suggest that the wider wavelength coverage that \textit{JWST} offers is very beneficial, though an extra 2-4 epochs of time coverage is still more valuable to the absolute magnification constraints.

The absolute magnification constraints provided under the \textbf{JWSTopt+nir2} follow-up strategy achieve sub-10\% precision for 33\% of all images considered in this work when systems are taken to be unresolved by Rubin-LSST and 50\% of all images when systems are taken to be resolved by Rubin-LSST. Strategies with more epochs have higher rates of sub-10\% precision, but these percentages still represent a significant portion of the sample. Furthermore, the increase in the sample of unresolved systems with well-measured absolute magnifications -- from less than a percent without follow-up to 33\% with \textit{JWST} follow-up -- is notable. Thus follow-up from \textit{JWST} is very useful and sufficient in many cases to reduce the MSD to subdominant uncertainty levels.

The combination of the \textbf{JWSTopt+nir2} and \textbf{HSTopt+nir6} strategies does not necessarily yield significant gains for the magnification constraints compared to the \textbf{HSTopt+nir6} strategy alone. We note, though, that because a different model is used to fit the combination of Rubin-LSST, \textit{HST}, and \textit{JWST} data (\texttt{SALT3-NIR}) compared to the Rubin-LSST and \textit{HST} data (\texttt{SALT3}), it is not a completely fair comparison. When systems are unresolved from Rubin-LSST, the median magnification uncertainty is \LSSTHSTJWSTunresolvedmedianMU mag overall, which is a mild improvement on the results from \textbf{HSTopt+nir6} follow-up alone. We do see an increased spread in the magnification residuals of \LSSTHSTJWSTunresolvedstdMU mag. The bias in the magnification residuals remains insignificant. These constraints produce sub-10\% precision estimates on the magnifications for 55.8\% of images fit in this work, though these constraints should be considered with care given additional potential systematic uncertainties from model misspecification as previously discussed.

\begin{figure*}
    \centering
    \includegraphics[width=0.95\linewidth]{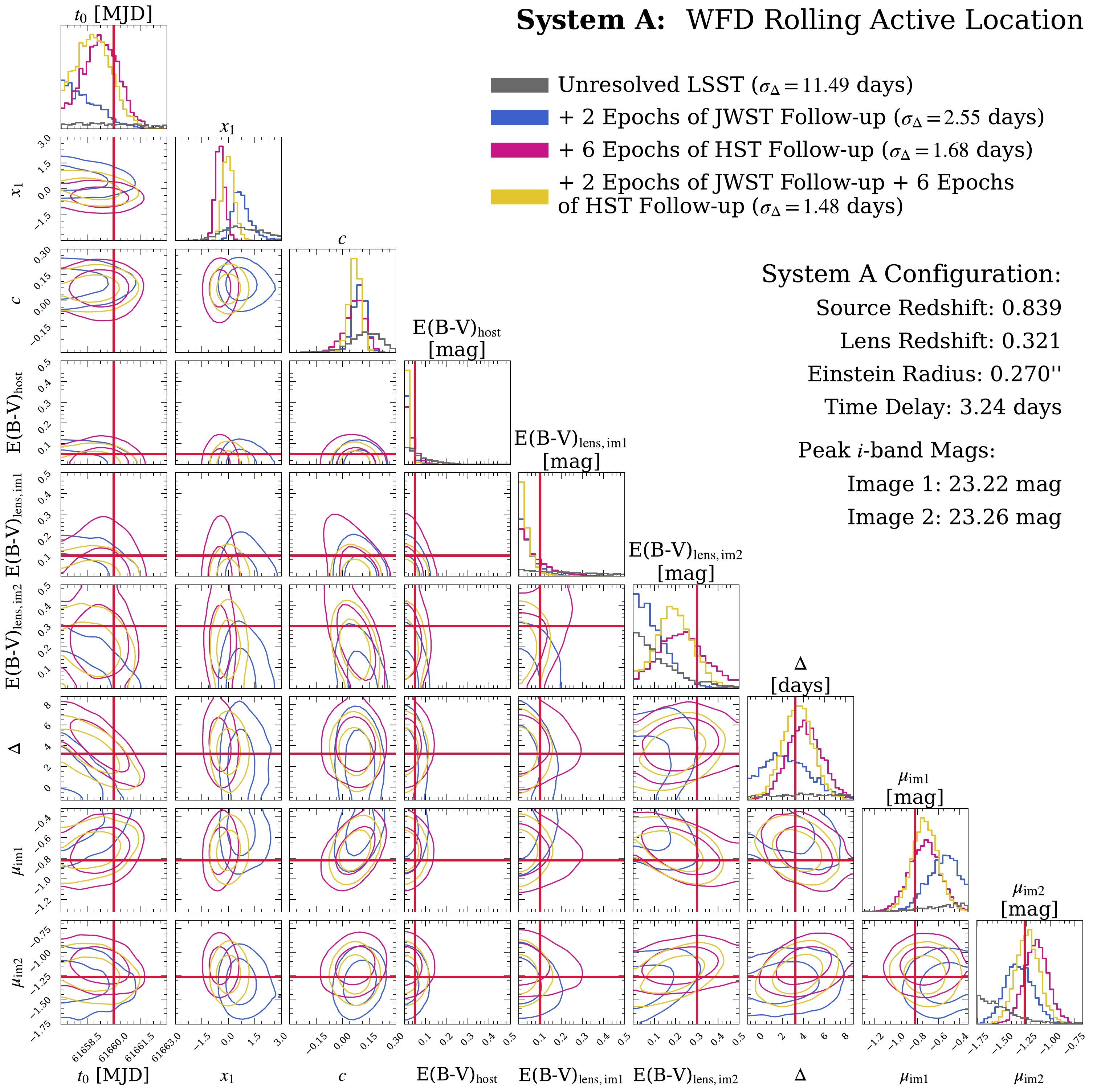}
    \caption{The posterior over the SN light curve parameters ($t_{0}$, $x_{1}$, $c$, E(B-V)$_{\rm host}$, E(B-V)$_{\rm lens, im1}$, and E(B-V)$_{\rm lens, im 2}$), time delay ($\Delta$), and absolute magnification of each image ($\mu_{\rm im 1}$, $\mu_{\rm im 2}$) for system A at location 1 with dust configuration c (differential lens dust). The blue posterior shows the fit with two epochs of \textit{JWST} follow-up in the \textit{F070W}, \textit{F090W}, \textit{F115W}, and \textit{F150W}, \textit{F200W}, and \textit{F277W} filters, the pink posterior shows the fit with six epochs of \textit{JWST} follow-up in the \textit{F625W}, \textit{F814W}, \textit{F110W}, and \textit{F160W} filters, and the yellow posterior shows fit from the combination of these two follow-up strategies. We additionally show in gray in the 1D histograms the results from the fit to the Rubin-LSST only light curve. Because the parameters are only poorly constrained in the case of no follow-up, we do not show the gray joint distribution contours.}
    \label{fig:space-corner}
\end{figure*}

Again, the posterior is shown for a realisation of system A under four different assumptions about the amount of space-based follow-up in Figure \ref{fig:space-corner}. Compared to without any space-based follow-up, the magnification constraints are much improved with the addition of two epochs of \textit{JWST} data. In this case, when the \textit{JWST} follow-up is combined with \textit{HST}, we also find that the recovered magnifications are much more accurate than with \textit{JWST} follow-up alone, and with a slightly better precision. Furthermore, the addition of the \textit{HST} enables the model to recover differing amounts of dust in the lens at the two image positions, as the dust along the line of sight to image 1 is found to be consistent with zero and along the line of sight to image 2 is found to be non-zero. Still, accurate recovery of magnifications remains one of the significant challenges for light curve modelling in the context of model misspecification in general.

In the case of resolved Rubin-LSST data, the median magnification uncertainty is \LSSTHSTJWSTresolvedmedianMU mag across all realisations of all systems. As with unresolved Rubin-LSST data, we see an increase in the dispersion of the magnification residuals to \LSSTHSTJWSTresolvedstdMU mag, compared to 0.07 mag from the results without the additional \textit{JWST} data. We also find that 52.5\% of all images have precise enough magnification constraints to reduce the MSD to a sub-dominant source of uncertainty, which is a mild decrease compared to the results from the \textbf{HSTopt+nir6} strategy alone. Still, this fraction is an improvement on the constraints from Rubin-LSST data alone.

This result is demonstrated in Figure \ref{fig:mag-recovery-res}, which shows the constraints on the absolute magnifications of the two images in the double systems considered in this work. In addition to the fit from the resolved Rubin-LSST data alone and that with \textbf{HSTopt+nir6} follow-up (discussed in \S\ref{sec:hst-followup}), we show the results from the fit to the Rubin-LSST data with \textbf{JWSTopt+nir2} and \textbf{HSTopt+nir6} follow-up. The addition of the \textit{JWST} data does not lead to more precise magnification constraints. In some cases, it produces more dispersion in the magnification residuals likely owing to greater uncertainty in NIR models and discrepancy among NIR models.

In summary, the combination of the \textbf{JWSTopt+nir2} and \textbf{HSTopt+nir6} strategies do not improve the magnification constraints as cleanly as it does the time-delay constraints. Some improvement is seen for the fraction of systems with magnifications measured to sub-10\% precision, sufficient to break the MSD, though the accuracy can be a challenge. While no significant biases are noted, we see an increased dispersion in the magnification residuals, as illustrated in Figure \ref{fig:mag-recovery-res}. As the \textit{JWST} data extends further into the NIR, where SN~Ia SED models are more uncertain, we attribute the worsening of the results in some senses when \textit{JWST} data is added to the fits to model misspecification.

The main limitation to improving the precision and accuracy of these constraints is uncertainty in the SN~Ia light curve template underlying the \textsc{Glimpse} model in the NIR wavelength regime. Not only are SN~Ia models worse constrained in the NIR, making it more difficult to leverage this important data for dust constraints, but they vary across templates more than in the optical. Given the difficulty of estimating the absolute magnification and dust parameters based on an assumed light curve template, we again emphasise that these constraints should be used carefully in any further analysis with consideration for the template being used with \textsc{Glimpse}, or any template-based time-delay estimation approach. With improved models, the additional data in a redder wavelength regime, as provided by \textit{JWST} may prove even more useful in the future.
\subsection{Follow-up from Ground-based Facilities}
\label{sec:ground-followup}
With current and upcoming ground-based telescopes, it will be easier to obtain follow-up on glSNe candidates at a higher cadence and more quickly after discovery than is possible with space-based resources. Therefore, we set out to determine what quality and quantity of ground-based follow-up will be most useful to improving time-delay and absolute magnification estimates.

We consider three types of possible ground-based follow-up from 2-, 4-, and 8-meter class telescopes which observe to different depths and with difference cadences. For the 2-meter class telescope follow-up, we assume a daily cadence with a depth of 21.5, 22.0, 22.0, 21.0, 20.5 mag respectively in each of the $grizy$ bands\footnote{Based on estimates assuming feasible exposure times and seeing conditions using the Liverpool Telescope (LT) exposure time calculator \url{https://telescope.livjm.ac.uk/TelInst/calc/}.}. We consider this case to be the ``shallow'' ground-based follow-up case. For the 2-meter class telescopes, we take all follow-up data to be unresolved, regardless of whether a system is resolved by Rubin-LSST, as the resolution of Rubin-LSST images is expected to be much higher than current such ground-based facilities.

For follow-up from a 4-meter class telescope, we assume a 3-day cadence with a depth of 22.0, 23.0, 23.0, 22.0, 21.5 mag respectively in each of the $grizy$ bands\footnote{Based on estimates assuming feasible exposure times and seeing conditions using the ESO Faint Object Spectrograph and Camera version 2 (EFOSC2) exposure time calculator: \url{https://www.eso.org/observing/etc/bin/gen/form?INS.NAME=EFOSC2+INS.MODE=imaging}.}. We take this case to be the ``mid'' or ``medium'' depth ground-based follow-up. Finally, we take a 5-day cadence for follow-up from an 8-meter class telescope with a depth of 24.0, 25.0, 24.0, 23.5, 23.0 mag respectively in the $grizy$ bands\footnote{Based on estimates assuming feasible exposure times and seeing conditions using the Gemini Multi-Object Spectrograph (GMOS) exposure time calculator \url{https://www.gemini.edu/instrumentation/gmos/exposure-time-estimation}.}. This case is referred to as the ``deep'' ground-based follow-up. We assume that the system is (un)resolved by the 4- and 8-meter class telescope if it is (un)resolved by Rubin-LSST. For the $grizy$ filters, we use the Dark Energy Camera (DECam) bandpasses from \citet{DESDR1_2018} as implemented in \texttt{sncosmo} for the transmission functions.

\subsubsection{Time Delay Recovery}
When combined with the unresolved Rubin-LSST light-curves, the shallow ground-based follow-up results in a median time-delay uncertainty of \LSSTshalunresolvedmedianTD days, the medium ground-based follow-up in a median time-delay uncertainty of \LSSTmediunresolvedmedianTD days, and the deep ground-based follow-up in a median of \LSSTdeepunresolvedmedianTD days. In general, the deep ground-based follow-up leads to the lowest time-delay uncertainties, indicating that the higher SNR with slightly sparser sampling is preferred over shallowers, but higher cadence observations. Although the resulting time-delay uncertainties are unlikely be precise enough to provide useful cosmological constraints with this data alone, the combination of unresolved data from Rubin-LSST and an 8m-class telescope can give time-delay constraints that are competitive with resolved Rubin-LSST data. Indeed, for system D, the median time-delay uncertainty when unresolved Rubin-LSST data is combined with unresolved deep ground-based follow-up is \LSSTdeepunresolvedDDmedianTD days for image 2, \LSSTdeepunresolvedDDDmedianTD days for image 3, and \LSSTdeepunresolvedDDDDmedianTD days for image 4, which is comparable to results from resolved Rubin-LSST data alone. This result demonstrates that unresolved data can be very impactful for constraining time delays with the \textsc{Glimpse} model. 

For the systems which are resolved by Rubin-LSST, the biggest improvements to the time-delay constraints are also seen for the deep ground-based follow-up strategy. Across all systems, the median time delay uncertainty for this follow-up strategy is \LSSTdeepresolvedmedianTD days. This level of precision is comparable to the \textit{HST} follow-up strategy with 6 epochs of data in the optical and NIR (median $\sigma_{\Delta} = 0.82$ days). For system C, resolved Rubin-LSST data with deep ground-based follow-up yields a median time-delay uncertainty of \LSSTdeepresolvedCmedianTD days, with a range between the best and worst uncertainties of just \LSSTdeepresolvedCrangeTD days. This result demonstrates that for the brightest systems, the higher cadence possible from ground-based facilities is more beneficial to the time-delay constraints than fewer observations with increased SNR from \textit{HST}. On the other hand, the greater depth that can be achieved by \textit{HST} relative to an 8-meter telescope is necessary for precise time-delay constraints for fainter systems, like system A. We emphasise that at least one epoch of high-resolution imaging e.g., from space while the SN is active will still be necessary to get precise image positions for the lens modelling aspect of $H_{0}$ estimation.

The shallow ground-based follow-up does not generally improve upon the time-delay constraints from resolved Rubin-LSST data alone, with the median uncertainty on the time delay remaining at \LSSTshalresolvedmedianTD days. As the shallow follow-up is always assumed to be unresolved, and we have seen in this work that unresolved data does not have the same constraining power as resolved data, the contribution to the time-delay constraint was expected to be minimal. Even so, it is notable that the extra time coverage provided by the daily cadence of this strategy, with the resolved data for degeneracy-breaking, does not generally improve constraints. The medium ground-based follow-up shows an intermediate level of precision between the shallow and deep ground-based follow-up, with a median time-delay uncertainty of \LSSTmediresolvedmedianTD days. We summarise the results from each of the shallow, medium, and deep ground-based follow-up strategies in Table \ref{tab:gb-results-unres} for the unresolved systems and in Table \ref{tab:gb-results-res} for the resolved systems.

\begin{table}
    \centering
    \caption{Summary of the results of the ground-based follow-up analysis for systems when they are assumed to be unresolved by all ground-based facilities. The median improvement in $\sigma_{\Delta}$ when considering follow-up from ground-based facilities is taken to be relative to the time-delay uncertainty from the fit to the Rubin-LSST data alone.}
    \begin{tabular}{c|c|c|c|c}
         &  & Median $\sigma_{\Delta}$ & Range($\sigma_{\Delta}$) & Median Improv. \\
        System & Strategy & [days] & [days] & in $\sigma_{\Delta}$ [days] \\
        \hline
        \hline
        A & Shallow & 9.26 & 11.83 & 0.12 \\
         & Medium & 7.06 & 12.32 & 0.28 \\
         & Deep & 6.92 & 11.10 & 0.59 \\
        \hline
        B & Shallow & 7.05 & 11.09 & -0.37 \\
         & Medium & 6.16 & 8.79 & 0.45 \\
         & Deep & 6.05 & 7.13 & 0.21 \\
        \hline
        C & Shallow & 2.52 & 8.41 & 0.29 \\
         & Medium & 2.82 & 9.32 & -0.48 \\
         & Deep & 2.48 & 11.14 & 0.14 \\
        \hline
        D$_{1,2}$ & Shallow & 5.35 & 14.52 & 0.41 \\
         & Medium & 2.30 & 19.16 & 4.78 \\
         & Deep & 2.58 & 9.02 & 3.18 \\
        D$_{1,3}$ & Shallow & 5.40 & 13.61 & 0.66 \\
         & Medium & 3.27 & 9.96 & 2.56 \\
         & Deep & 1.28 & 5.50 & 5.19 \\
        D$_{1,4}$ & Shallow & 3.45 & 12.27 & 1.70 \\
         & Medium & 2.08 & 6.18 & 2.39 \\
         & Deep & 1.28 & 13.79 & 3.22 \\
        \hline
        \hline
    \end{tabular}
    \label{tab:gb-results-unres}
\end{table}
\begin{table}
    \centering
    \caption{Summary of the results of the ground-based follow-up analysis for systems when they are assumed to be resolved by Rubin-LSST, the 4m-class (medium-depth) telescope follow-up, and the 8m-class telescope (deep) follow-up. The shallow ground-based follow-up is always taken to be unresolved. The median improvement in $\sigma_{\Delta}$ when considering follow-up from ground-based facilities is taken to be relative to the time-delay uncertainty from the fit to the Rubin-LSST data alone.}
    \begin{tabular}{c|c|c|c|c}
         &  & Median $\sigma_{\Delta}$ & Range($\sigma_{\Delta}$) & Median Improvement \\
        System & Strategy & [days] & [days] & in $\sigma_{\Delta}$ [days] \\
        \hline
        \hline
        A & Shallow & 1.62 & 1.36 & 0.06 \\
         & Medium & 1.58 & 1.04 & 0.22 \\
         & Deep & 1.23 & 0.60 & 0.50 \\
        \hline
        B & Shallow & 1.57 & 2.09 & 0.04 \\
         & Medium & 1.34 & 1.25 & 0.20 \\
         & Deep & 0.81 & 0.41 & 0.67 \\
        \hline
        C & Shallow & 0.98 & 0.87 & -0.01 \\
         & Medium & 0.87 & 0.71 & 0.06 \\
         & Deep & 0.46 & 0.34 & 0.39 \\
        \hline
        D$_{1,2}$ & Shallow & 1.28 & 0.85 & -0.02 \\
         & Medium & 1.06 & 0.50 & 0.08 \\
         & Deep & 0.79 & 0.30 & 0.30 \\
        D$_{1,3}$ & Shallow & 1.56 & 1.08 & -0.06 \\
         & Medium & 1.25 & 0.63 & 0.11 \\
         & Deep & 0.79 & 0.30 & 0.49 \\
        D$_{1,4}$ & Shallow & 1.91 & 7.48 & 0.00 \\
         & Medium & 1.53 & 1.52 & 0.23 \\
         & Deep & 0.92 & 0.37 & 0.87 \\
        \hline
        \hline
    \end{tabular}
    \label{tab:gb-results-res}
\end{table}

This test demonstrates that greater depth is more valuable than a higher cadence in ground-based follow-up for the relatively faint glSN systems considered in this work. This trend is further contextualised by considering the number of $5\sigma$ observations added by each ground-based follow-up strategy. For shallow follow-up, an average of 10 $5\sigma$ observations are added to each light curve across all five bands. When considering medium-depth follow-up, this number increases to 21 $5\sigma$ observations when averaging across all glSN systems. The deep ground-based follow-up sees the largest number of $5\sigma$ observations added to each light curve at 41 observations on average, which is $\sim8.2$ observations per filter. Although the shallow follow-up has a higher cadence than the medium and deep follow-up, the greater SNR of data from deeper observations means that the deep follow-up strategy yields more better quality observations than the other two strategies. The reasons for our findings when comparing ground-based strategies are easily interpreted with this context, and provides an additional metric to be used when comparing ground-based follow-up to space-based follow-up strategies. 

Unfortunately, the addition of ground-based follow-up for systems that are unresolved from the ground does not improve the time-delay constraints enough to produce any sub-5\% precision estimates in this sample. We reiterate that resolved space-based imaging will be essential to precise time-delay estimates for systems that are unresolved by Rubin-LSST. In the case of systems that are resolved from the ground, though, we see an increase of 550\% to the number of time delays in our sample that are estimated to sub-5\% precision compared to with Rubin-LSST data alone, for a total of 18.1\% of the sample. Notably, all of these cases are for system C and D, which have the longest time delays ($15-16$ days). 

Because the deep ground-based follow-up case can improve time delay uncertainties by up to several days, which results in constraints comparable to \textit{HST} follow-up in some cases, we proceed to test whether combining these ground- and space-based follow-up strategies offers any additional benefits. When considering all systems, we find that the median time delay uncertainty is \LSSTHSTGBunresolvedmedianTD days for systems that are assumed to be unresolved from the ground and \LSSTHSTGBresolvedmedianTD days for systems that are assumed to be resolved from the ground when there is both deep ground-based follow-up and 6 epochs of optical and NIR \textit{HST} follow-up. These uncertainties represent median decreases in the time-delay uncertainty of \LSSTHSTGBunresolvedchangeTD days and \LSSTHSTGBresolvedchangeTD days, respectively, relative to the case of Rubin-LSST data alone. Therefore, this combination of follow-up provides the most precise time-delay constraints of all strategies considered.

\begin{figure}
    \centering
    \includegraphics[width=\linewidth]{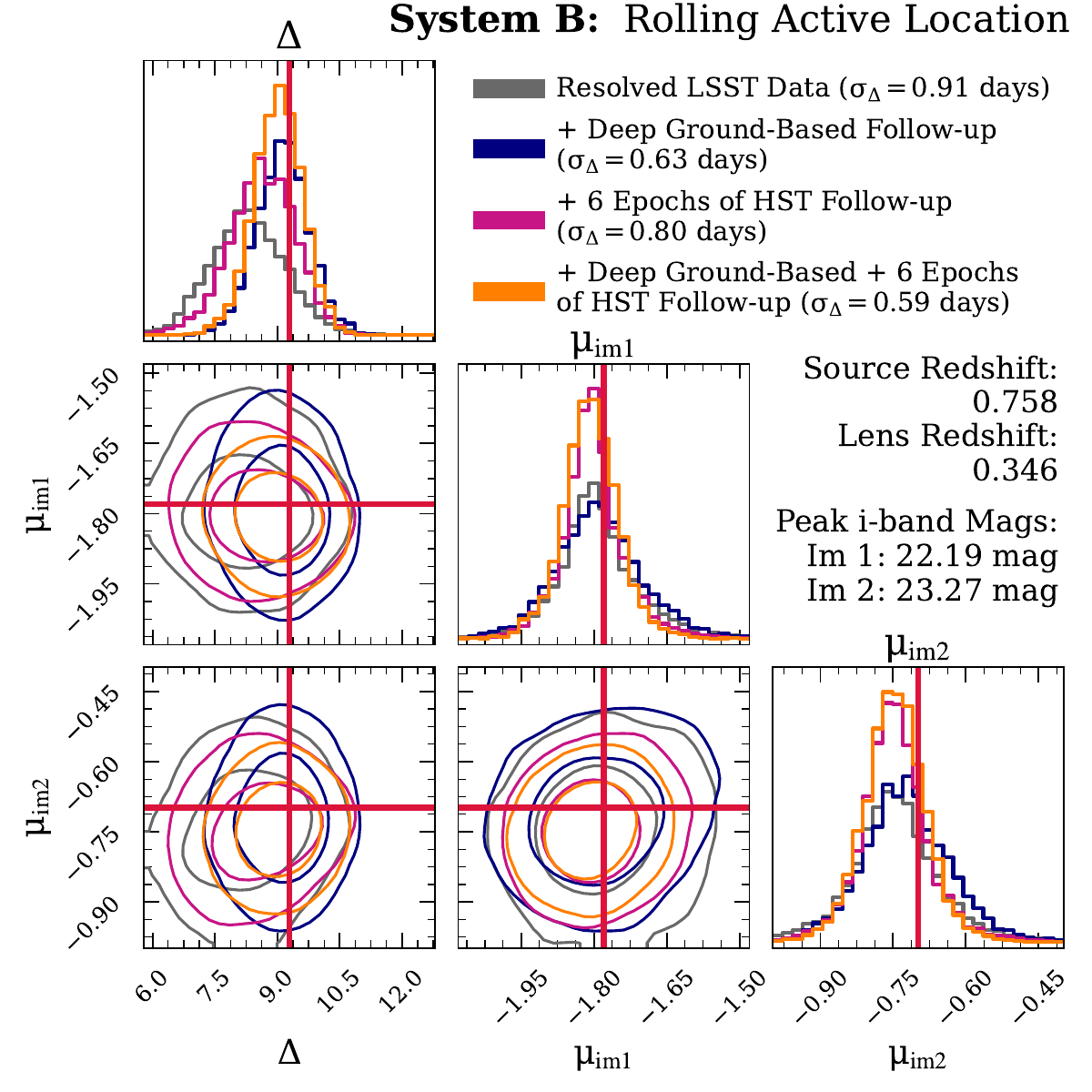}
    \caption{The posterior on the time delay and absolute magnifications of both images from the \textsc{Glimpse} fits to system B under three follow-up strategies: 1) with deep resolved imaging from an 8-meter class telescope assuming a 5-day cadence (blue), 2) with 6 epochs of optical and NIR \textit{HST} data (pink), and 3) with both deep resolved imaging from an 8-meter class telescope assuming a 5-day cadence and 6 epochs of optical and NIR \textit{HST} data (orange). In all cases, the ground- and space-based data is assumed to be resolved. This figure demonstrates that combining 8-meter class ground-based follow-up with \textit{HST} follow-up yields the most accurate and precise constraints on the time-delay and absolute magnifications. This realisation of system B is at spatiotemporal location 1 and dust configuration ``a.''}
    \label{fig:ground-based-test-corner}
\end{figure}

The benefits of this combined space- and ground-based follow-up strategy is demonstrated for an individual system in Figure \ref{fig:ground-based-test-corner}. We show a comparison of the posteriors on the time delay and absolute magnifications of each image assuming three different follow-up strategies for system B, which is assumed to be resolved by Rubin-LSST. We consider the following follow-up strategies: 1) resolved deep ground-based follow-up with a 5-day cadence, 2) 6 epochs of resolved optical and NIR \textit{HST} data, and 3) both deep resolved imaging from an 8-meter class telescope assuming a 5-day cadence and 6 epochs of optical and NIR \textit{HST} data. The uncertainty on the time delay is lower for case 1 than case 2, but the \textit{HST} optical and NIR data is important to accurate constraints on the image absolute magnifications. Indeed, combining follow-up from both deep ground-based imaging and \textit{HST} imaging, as in case 3, leads to the most accurate and precise time delay and absolute magnification estimates in general. 

With the combination of both deep ground-based imaging and 6 epochs of \textit{HST} follow-up in the optical and NIR, we find that 6.9\% of unresolved systems from the ground and 30.6\% of resolved systems from the ground have time delays estimated to sub-5\% precision. Again, all of these precise estimates are for systems C and D, which have the longest time delays. In the unresolved case, the addition of the deep ground-based follow-up does not represent a significant improvement from the results with only the \textbf{HSTopt+nir6} follow-up. However, when the system can be resolved from the ground, the addition of the higher-cadence deep ground-based imaging results in an increase of 175\% in the number of systems with precise time delays relative to with just \textbf{HSTopt+nir6} follow-up. For any system that is resolved from the ground, this combination of follow-up is the most likely to produce a precise time-delay estimate.

\subsubsection{Absolute Magnification Recovery}
For systems which are unresolved by ground-based facilities, the addition of ground-based follow-up data does result in mild improvements to the precision of the absolute magnification constraints compared to Rubin-LSST data alone. The median uncertainty on the absolute magnification with shallow ground-based follow-up is \LSSTshalunresolvedmedianMU mag, with medium-depth ground-based follow-up is \LSSTmediunresolvedmedianMU mag, and with deep ground-based follow-up is \LSSTdeepunresolvedmedianMU mag. Again, because the deep ground-based follow-up yields the largest number of $5\sigma$ observations, it is unsurprising that this strategy gives the tightest absolute magnifications constraints of the three. Still, there remains a large dispersion and bias in the absolute magnification residuals, of similar size as with Rubin-LSST data alone. Furthermore, this level of precision is unlikely to be sufficient for breaking the MSD.

If a system is resolved by Rubin-LSST, additional unresolved ground-based follow-up from a 2 meter class telescope does not result in any improvements to the absolute magnification constraints compared to absolute magnifications estimated from Rubin-LSST data alone. That said, the medium and deep ground-based follow-up strategies offer more constraining power. We find that the median absolute magnification uncertainty is \LSSTmediresolvedmedianMU mag for medium ground-based follow-up and \LSSTdeepresolvedmedianMU mag for deep ground-based follow-up, compared to 0.12 mag from resolved Rubin-LSST data alone. Furthermore, we see a lower dispersion of \LSSTmediresolvedstdMU mag in the absolute magnification residuals for medium ground-based follow-up and \LSSTdeepresolvedstdMU mag for deep ground-based follow-up, compared to 0.14 mag from Rubin-LSST data alone. These benefits again indicate that additional time-coverage leads to improved absolute magnification constraints in addition to better time-delay constraints.

Naturally, combining both deep ground-based follow-up with optical and NIR follow-up from \textit{HST} leads to the tightest absolute magnification constraints in general. When we consider the combined deep ground-based and \textbf{HSTopt+nir6} follow-up strategy, we find that the absolute magnification estimates have a median uncertainty of \LSSTdeepHSTunresolvedmedianMU mag for systems that are unresolved from the ground and \LSSTdeepHSTresolvedmedianMU mag for systems that are resolved from the ground. Both of these figures represent an improvement in the constraints from either deep ground-based follow-up or \textbf{HSTopt+nir6} follow-up on their own. For the systems which are unresolved from the ground, we find that the \textit{HST} follow-up is also particularly powerful for reducing the bias in the absolute magnification recovery from -0.3 to -0.5 mag from ground-based follow-up alone to consistent with zero with resolved \textit{HST} follow-up.

Therefore, we conclude that in the case of systems which are unresolved from the ground, additional ground-based follow-up offers benefits to the absolute magnification constraints compared to Rubin-LSST data alone. This follow-up strategy, without any resolved space-based imaging, is unlikely to be sufficient for a sub-5\% estimate of $\lambda$, though. When resolved from the ground, additional ground-based follow-up across the $grizy$ filters leads to much improved absolute magnification constraints compared to Rubin-LSST data alone. Rubin-LSST data with deep ground-based follow-up data is likely to be sufficient for a 10\% precision estimate on the absolute magnification, as necessary to break the MSD. Still, the addition of NIR wavelength coverage from space-based follow-up to Rubin-LSST data with ground-based follow-up leads to constraints with the smallest uncertainties on average and the lowest levels of bias in the absolute magnification residuals. In the future, further studies of follow-up strategies in the NIR may be explored to determine the relative benefits of higher SNR but lower cadence space-based NIR data compared to lower SNR but higher cadence ground-based NIR data. Again, though, we highlight the need for improved models of SNe~Ia in the NIR to best leverage follow-up data in this wavelength regime.

\section{Discussion}
\label{sec:discussion}
\subsection{What systems should we follow-up? What quantity and quality of follow-up data is needed?}
In this work, we consider four glSNe systems on the faint end of detectability by Rubin-LSST (e.g., with peak i-band magnitudes of between 22-24 mag) to determine the upper limits of precision on the time-delays using the \textsc{Glimpse} model for time-delay estimation. For these systems, we find that time delays can be consistently estimated to a precision of 0.5-0.8 days when Rubin-LSST data is combined with sufficient \textit{HST}, \textit{JWST}, and/or deep ground-based facility follow-up. As the \textsc{Glimpse} model simultaneously marginalises over line-of-sight effects, these uncertainties include a contribution from dust extinction in the host and lens galaxies and microlensing (discussed further in \S\ref{sec:discussion-microlensing}), as well as a contribution from model misspecification (discussed further in \S\ref{sec:discussion-mismatch}). Notably, we find that there is a strong dependence of the time-delay constraints on the overall brightness of the glSN system considered, indicating that time-delay uncertainties may be smaller for systems which are brighter than those considered in this work (discussed further in \S\ref{sec:discussion-highmag}).

\begin{figure*}
    \centering
    \includegraphics[width=0.95\linewidth]{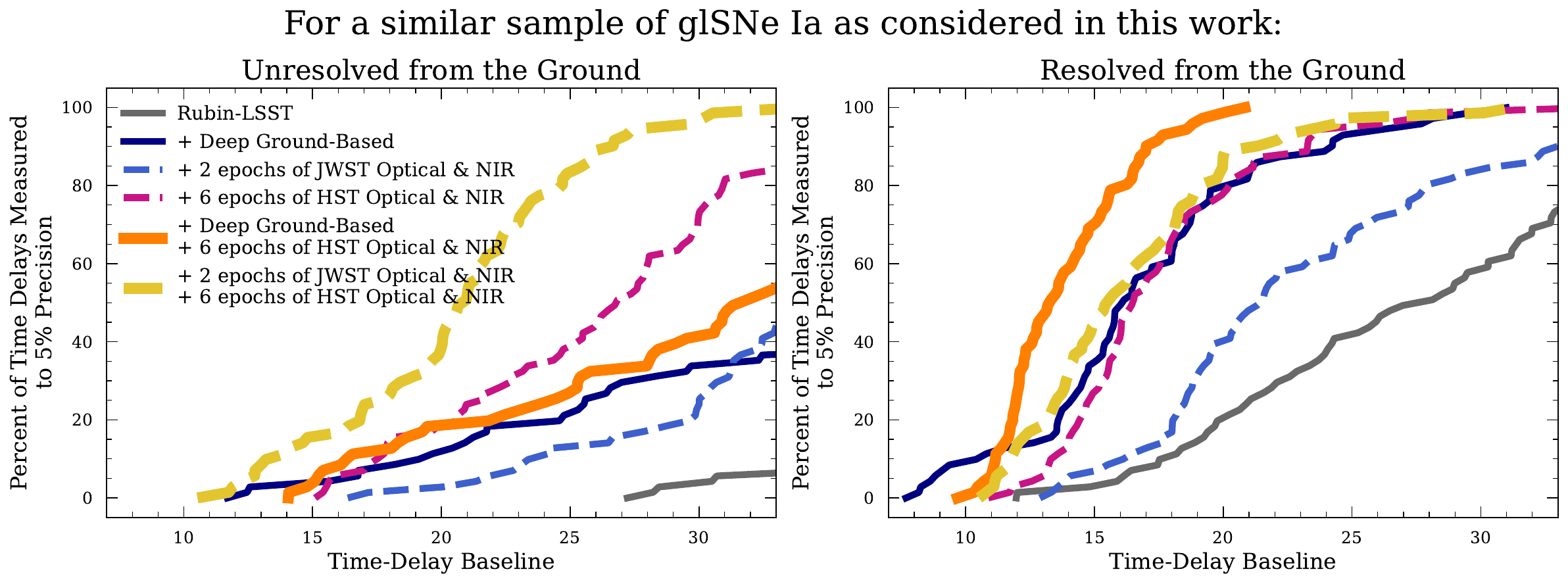}
    \caption{The percent of time delays that can be expected to be measured to 5\% precision for a sample of glSNe~Ia with similar data quality to those considered in this work (e.g., of similar brightnesses or data quality) as a function of the time-delay baseline by which precision is estimated. (Left) The systems are assumed to be unresolved by ground-based facilities. (Right) The systems are assumed to be resolved by ground-based facilities that are of 4 meter class or greater. Different line colours indicate different amounts of follow-up data. This figure demonstrates that e.g., 70\% of time delays can be estimated to 5\% precision under the assumption of a 15 day time delay for the combination of deep ground-based and \textbf{HSTopt+nir6} follow-up for systems that are resolved by all telescopes.}
    \label{fig:percentiles}
\end{figure*}

The best strategies for follow-up to turn glSNe~Ia as discovered by Rubin-LSST into precise cosmological probes vary significantly depending on whether the system is resolved or unresolved from the ground. These results are visually represented in Figure \ref{fig:percentiles}, which shows the percent of time delays estimated to 5\% precision under different follow-up strategies if the underlying system had data of similar quality to the sample considered in this work under the assumption of different time-delay baselines (between $\sim7$ and $37$ days). As there will be some covariance between the peak apparent brightness of the images, the redshift of the source, the absolute magnification of the images, and the length of the time delay, we opt to frame this result in the context of data quality of the systems rather than specific lensing parameters. In general, this figure demonstrates that increasing amounts of follow-up data leads to larger fractions of systems with time delays measured to sub-5\% precision. 

For unresolved ground-based data with resolved space-based observations, the combination of the \textbf{JWSTopt+nir2} and \textbf{HSTopt+nir6} strategies clearly gives the best outcomes in terms of time-delay precision. If we assume a similar quality to the data of the sample considered in this work, it is unlikely that Rubin-LSST data alone will be sufficient to constrain the time delays to the necessary precision for these faint systems. With the combined \textit{JWST} and \textit{HST} follow-up, 16\% of these systems would be measured to sub-5\% precision if we assume a time-delay baseline of 15 days. This result assumes a distribution of time-delay uncertainties that is identical to that produced from the fits done in this work. For comparison, the combined follow-up represents an increase from >1\% with 6 epochs of \textit{HST} or 2 epochs of \textit{JWST} follow-up alone. Additional unresolved ground-based imaging is not as effective as the higher resolution, but lower cadence space-based imaging in this case.

As an aside, we note that the fits to the light curves with a combination of deep ground-based follow-up and six epochs of \textit{HST} follow-up results in a distribution of time-delay uncertainties in this sample which has a tail of more uncertain estimates compared to the fits to light curves which only have \textit{HST} follow-up. Thus, the addition of the deep ground-based unresolved follow-up appears to make the fits worse. In these fits, the unresolved data is going to have a relatively larger influence on the time-delay posterior because of the data from Rubin-LSST and another follow-up facility. As unresolved data may be more sensitive to degeneracies, the effective up-weighting of this data in the likelihood may lead to increases in time-delay uncertainties. We plan to investigate this behaviour in a more complete sample of glSNe as expected to be discovered by Rubin-LSST in future work.

For systems that are always resolved from the ground, we advocate for a combination of deep higher-cadence follow-up from the ground ($\sim8$ epochs of $5\sigma$ observations in a few filters) with 6 epochs of high SNR optical and NIR imaging from space. If we assume a time-delay baseline of 15 days, the uncertainty on the time-delay estimates from the combined deep ground-based follow-up with \textbf{HSTopt+nir6} strategy leads to a 5\% precision estimate on more than 70\% of systems in a sample with resolved images which are similar in quality of data to the sample considered in this work. For comparison, 27\% of such systems are expected to be estimated to 5\% precision with the \textbf{HSTopt+nir6} strategy on its own and 35\% for the deep ground-based strategy on its own. The combination of the \textbf{HSTopt+nir6} and \textbf{JWSTopt+nir2} strategies yields 44\% of systems with a 15 day time delay having a sub-5\% precision constraint. Thus, the combination of deep ground-based and \textit{HST} follow-up provides the most precise and most consistent time-delay estimates when the systems are resolved by Rubin-LSST and all follow-up telescope facilities. 

We thus conclude that less highly magnified glSNe~Ia systems, which are more likely to have longer time delays, larger image separations, and are expected to be more populous in the Rubin-LSST data stream, can be powerful tools of cosmology with the follow-up advocated for in this analysis. This work demonstrates that systems with images that peak in apparent magnitudes in the i-band between 22-24 mag must have time-delays of $>15$ days to be useful for time-delay cosmography. For systems which are similar in brightness and therefore data quality to those considered in this work, we find that less than 1\% can be measured to 5\% precision under the assumption of a 10 day time-delay baseline even under the most aggressive follow-up strategies. 

Finally, we re-iterate that the deep, high-resolution space-based follow-up in real-time will be necessary for time-delay cosmography regardless of time-delay estimation to get accurate image positions from the lens modelling. As such, we can expect to have some amount of space-based follow-up, though a higher cadence of \textit{HST} or \textit{JWST} data does offer more benefits to both time-delay and absolute magnification estimation, particularly for fainter systems. This data has the potential to turn systems which are unresolved by Rubin-LSST and other ground based facilities into useful cosmological probes. Having resolved data from Rubin-LSST and other ground-based facilities, though, yields a significant improvement in the constraints on time delays. If light curves for individual images can be recovered from unresolved imaging via de-blending methods (discussed further in \S\ref{sec:discussion-blending}), the fraction of systems with precise time-delay estimates will increase.

\subsection{Simulations with Microlensing}
\label{sec:discussion-microlensing}
To confirm that the results from the \textsc{Glimpse} model are not affected when there is microlensing affecting the light curves, we test the model on a set of 19 simulated double-imaged resolved glSNe~Ia from \citet{Arendse_2024}. These simulations cover only the $g,r,i,z,y$ filter wavelength range, for which a model of the microlensing can be computed from the \texttt{lensedSST} code. Each of these objects has a realisation of the light curve that includes a contribution from microlensing and one which does not include any microlensing effects. We can therefore compare the performance of \textsc{Glimpse} across these two versions of the same underlying system to determine the effect of microlensing on the inference.

We find that the uncertainties on the fits are broadly well-calibrated regardless of whether or not there is microlensing included in the simulations. Without microlensing in the simulated data, 63.2\% of the true time delays fall within the 68\% CI and 94.7\% fall within the 95\% CI. When the simulated data includes microlensing, 78.95\% of the true time delays fall within the 68\% CI and 100.0\% fall within the 95\% CI. Furthermore the median time-delay uncertainty for the simulations without microlensing is 5.05 days, while that for the simulations with microlensing is 5.08 days. Overall, there is a median increase in time-delay uncertainty of 0.17 days when comparing the uncertainty from the fit to the light curve without microlensing to that from the fit to the light curve with microlensing. This level of increase is insignificant given the small sample size and may be attributed to a statistical fluctuation. We have also shown in this work that the brightness of the system is strongly correlated with the resulting time-delay uncertainty. As microlensing affects the brightness of the system, the difference in uncertainty across the two fits will also reflect the relative (de)magnification. Indeed, we find that the objects which are demagnified by microlensing are more likely to have larger time-delay uncertainties from the fit to the light curve with microlensing than the fit to the light curve without microlensing.

Therefore, we conclude that the inference from \textsc{Glimpse} is not significantly affected by the presence of microlensing in the simulations. The results of this work are likely to remain unchanged if the analysis was re-run with simulations that include microlensing. We re-iterate though that it is not possible to re-run the analysis in full with microlensing applied to the simulated light curves because the wavelength range probed by the filters considered in this work is not fully covered by the microlensing simulations from \texttt{lensedSST}. 

\subsection{Impact of Model Misspecification}
\label{sec:discussion-mismatch}
In this work, we present results assuming a mismatch between the model from which we simulate the data and the model with which we fit the data. Realistically, we will not have a perfect template for the true underlying SN light curve, as the majority of light curve templates are empirical models. Because of the choice to present the results from \textsc{Glimpse} with model mismatch, the projected constraints are a more realistic representation of real data.

One consequence of model mismatch is that as the quality of some data improves, the inference will become more sensitive to model misspecification. In this section, we explore the impact of increasing the SNR of the \textit{HST} data to test the limit at which the sensitivity to template mismatch becomes a limiting factor in time-delay precision. For the fiducial analysis in this work, we set the \textit{HST} exposure time such that an observation in each filter of an object, which has the same magnitude as the peak $i$-band magnitude of the brighter image, would have an SNR of 50. Now, we additionally test the impact of a higher target SNR of 100 for the resolved systems A, B, and C. 

\subsubsection{Time Delay Recovery}
On average, the median time delay uncertainty remains largely unchanged from \LSSTHSTresolvedABCmedianTD days in the fiducial SNR 50 case to \LSSTHSTresolvedABCmedianCTD days in the SNR 100 case. For system A -- the dimmest overall system -- the improvements are more pronounced, with the median uncertainty on the time delay decreasing from \LSSTHSTIVresolvedAmedianTD days to \LSSTHSTIVresolvedAmedianCTD days when moving from the SNR 50 to SNR 100 case for the \textbf{HSTopt+nir4} strategy. Similar reductions in the time delay uncertainty are seen for the \textbf{HSTopt+nir2} and \textbf{HSTopt+nir6} strategies for this system. For the brighter two systems B and C, the median time-delay uncertainty actually increases from the SNR 50 to the SNR 100 case for the \textbf{HSTopt+nir4} -- from \LSSTHSTIVresolvedBCmedianTD days to \LSSTHSTIVresolvedBCmedianCTD days. Although the \textbf{HSTopt+nir2} and \textbf{HSTopt+nir6} strategies do not show such extreme differences, this behaviour for the brighter systems suggests a sensitivity to model mismatch.

To determine whether these unexpected trends can be attributed to the model misspecification, we compare the SNR 50 case to the SNR 100 case for a set of simulations which assume a \texttt{SALT3} template. For these simulations, we take $x_{1}=0$, $c=0$, and $x_{0}$ scaled such that the peak absolute magnitude is -19.43 in the $B$-band. The median uncertainty on the time delay from the SNR 50 case to the SNR 100 case now sees a median decrease of \saltLSSTHSTresolvedABCchangeCTD days overall -- from a median of \saltLSSTHSTresolvedABCmedianTD days to \saltLSSTHSTresolvedABCmedianCTD days. With the correct template, the SNR 100 more generally represents an much more significant improvement in time-delay uncertainty, and again the largest gains are seen for the dimmest systems. Finally, we find that the minimum time delay uncertainty is decreased from \saltLSSTHSTresolvedABCminTD days with model mismatch to \saltLSSTHSTresolvedABCminCTD days when assuming the correct template.

This result suggests that an error floor of up to 0.25 days from model misspecification contributes to the uncertainty on the time delay from \textsc{Glimpse}. The brighter systems, which tend to have more precise time-delay estimates, are more limited by the error floor arising from model mismatch than dimmer systems. Given that dimmer objects will have worse quality Rubin-LSST data, the challenge of reconciling the Rubin-LSST data in optical filters with the \textit{HST} data in optical and NIR filters will not be as difficult. For bright systems, though, matching behaviour in high-quality data across a wide wavelength regime becomes a limiting factor in the presence of model misspecification.

We also conclude that higher SNR observations from \textit{HST} can be beneficial, particularly for dim systems. Of course, obtaining higher SNR observations of dim systems will likely require more than one orbit per epoch, which becomes expensive. If the projected gains of up to $\sim$0.50 days would represent a significant improvement in the precision of the time-delay estimate (e.g., for systems with time days of 8-10 days, where 0.50 days corresponds to up to 7 percentage points in precision), higher SNR \textit{HST} should be considered. 

\subsubsection{Absolute Magnification Recovery}
For the case of model misspecification, when the target SNR of the \textit{HST} data at peak is increased to 100 from the fiducial case of 50, the median absolute magnification uncertainty for systems A, B, and C remains broadly consistent at \LSSTHSTresolvedABCmedianCMU mag, compared to \LSSTHSTresolvedABCmedianMU mag. The dispersion of the absolute magnification residuals remains unchanged, having a standard deviation of \LSSTHSTresolvedABCstdCMU regardless of the SNR of the \textit{HST} follow-up. 

The absolute magnification estimates are much improved when the assumed underlying template matches the template from which the light curves were simulated. When considering the set of simulations from the \texttt{SALT3} template with \textit{HST} SNR of 50 at peak, we find that the median uncertainty on the absolute magnification is \saltLSSTHSTIIresolvedABCmedianMU mag under the \textbf{HSTopt+nir2} strategy and \saltLSSTHSTVIresolvedABCmedianMU mag under both the \textbf{HSTopt+nir4} and \textbf{HSTopt+nir6} strategies. This level of precision is a marked improvement from the fiducial analysis. Furthermore, there is a reduced spread in the absolute magnification residuals of $\sim$\saltLSSTHSTVIresolvedABCstdMU mag compared to \LSSTHSTVIresolvedABCstdMU-\LSSTHSTIIresolvedABCstdMU mag in the model mismatched case. This result already highlights how imperfect modelling is a major limitation to the precision of the absolute magnification constraints.

When the SNR is increased to 100 at peak for the \textit{HST} observations, the constraints are improved further, though the gains are mild. The median uncertainty on the absolute magnifications across all \textit{HST} follow-up strategies remains at \saltLSSTHSTresolvedABCmedianCMU mag, comparable to the results with 4 or 6 epochs of \textit{HST} observations with target SNR of 50. The spread in the residuals remains at \saltLSSTHSTresolvedABCstdCMU mag across all \textit{HST} follow-up strategies with target SNR of 100, though is seen to be slightly reduced to \saltLSSTHSTVIresolvedABCstdCMU mag for the \textbf{HSTopt+nir6} strategy. Again, the uncertainties remain well-calibrated. While increasing the SNR of the \textit{HST} data does not lead to any significant gains in the absolute magnification recovery, these results continue to represent much more precise constraints than in the case of model misspecification.

Therefore, we conclude that model misspecification contributes to a error floor on the absolute magnification recovery of up to 0.06 mag, which sets a lower limit on the precision of $\lambda$ of $\sim 3$\%, assuming a absolute magnification of $\sim 1$ mag. As discussed in \S\ref{sec:hst-followup} an estimate of the MST scaling factor, $\lambda$, to sub-5\% precision would make the MSD a sub-dominant source of uncertainty in the final $H_{0}$ estimate. We re-iterate, though, that there will be additional sources of uncertainty in this constraint from estimates of the true unlensed peak magnitude of a SN~Ia. In the case of sources at $z \lesssim 1$, the uncertainty will be small owing to the well-understood distribution of SNe~Ia peak magnitudes in this regime, but it will increase for higher-redshift sources \citep{Pierel_2024b}. Therefore, model misspecification contributes a significant amount to the uncertainty on the absolute magnification, but the absolute magnifications may still be well-enough constrained to reduce the MSD to a sub-dominant source of uncertainty for $H_{0}$ from glSNe~Ia with \textsc{Glimpse}.

\subsubsection{Future Work}
\label{sec:discussion-mismatch-future}
Light curve templates, such as from \texttt{SALT} or \texttt{BayeSN}, are very useful models for SNe~Ia in cosmology. However, they are limited by the amount of training data available to produce reliable models in the time- and wavelength-regimes of interest, as they are not built from physics first principles (e.g., radiative transfer). In the future, physical models may be preferred to describe the time- and wavelength dependent evolution of the underlying source. The \textsc{Glimpse} framework naturally allows any deterministic model to be used as the model ``template'' or mean function, providing another advantage of this light-curve modelling approach for time-delay estimation. Physical parameters describing the source progenitor could therefore be constrained alongside the time delays and magnifications. Advances in fast forward modelling of observations based on full radiative transfer codes from tools such as \textsc{redback} \citep{Sarin_2024} or \textsc{riddler} \citep{Magee_2024} would be central to the computational feasibility of this work. In a similar vein, this methodology would allow for the combination of spectra and photometry for time-delay fitting, though the effects of microlensing on spectra will need to be carefully considered, particularly at late times \citep{Suyu_2020}.

Furthermore, the recent publication of the Carnegie Supernova Project II (CSP-II) data set \citep{Lu_2023} provides an extensive sample of spectra in the optical and NIR which can be used to re-train an empirical SNe~Ia SED model. The CSP-II data set would increase the sample of NIR spectra upon which \texttt{SALT3-NIR}, for example, is trained on by a factor of ten. Thus, it would better capture the true observed diversity of SNe~Ia in the NIR, leading to a lower systematic uncertainty relating to template selection in time-delay estimation.

\subsection{Higher Magnification System}
\label{sec:discussion-highmag}
We chose the base objects presented in Table \ref{tab:base-objects} based on the median redshifts and absolute magnifications of underlying glSNe~Ia population based on the lens catalogue from \citet{Wojtak_2019}. Our analysis suggests that one of the key determinants of the expected time-delay uncertainty is the brightness of the systems, which particularly affects the quality of the Rubin-LSST data. Therefore, we test the limit on the time-delay precision through an analysis of a higher absolute magnification (overall brighter) system. While higher absolute magnification systems are rarer, we may be more likely to find such an object because of photometric selection effects.

In this section, we analyse an double glSN~Ia with $z_{\rm source} = 0.604$, $z_{\rm lens} = 0.237$, and $\theta_{E} = 0.483''$. The object has a time delay of 1.56 days and absolute magnifications of -3.1 mag and -2.6 mag for image 1 and 2, respectively. As we are interested in the brightest case, we only consider dust configuration ``a,'' which has the lowest line-of-sight extinction effects. Therefore, the peak $i$-band magnitude of image 1 is 20.49 mag and of image 2 is 20.93 mag. When unresolved, the system reaches 19.93 mag at its brightest point. As in the fiducial analysis, we consider both the resolved and unresolved cases of observations from Rubin-LSST. We will analyse the same five strategies of \textit{HST} follow-up as described in \S\ref{sec:hst-followup}.

When the system is unresolved by Rubin-LSST, the time delays are measured to 3.71 day precision without any additional follow-up. This uncertainty level is consistent with the results seen in the analysis of systems A-D when unresolved. The accuracy in the time-delay recovery is very poor for system H, with the \textsc{Glimpse} estimate of the time delay being more than 2$\sigma$ away from the truth for three of four spatiotemporal locations considered. The low number of realisations of the system considered make this result sensitive to statistical fluctuations. We also find that the posteriors on the time delays are often multi-modal, suggestive of the degeneracies present when fitting with unresolved data alone, so a simple mean and standard deviation do not capture well the model's time-delay estimate and associated uncertainty. Still, it is likely that model misspecification plays some role in the inaccuracies seen, as the Rubin-LSST data alone is high-quality for such a bright object making it more challenging to reconcile data simulated from one model and fit with another.

The results with \textit{HST} follow-up in the optical and NIR for the unresolved system see a reduction in time-delay uncertainty, though the estimates remain inaccurate. The median time-delay uncertainty is 2.40 days under the \textbf{HSTopt+nir2} strategy, 0.75 days under the \textbf{HSTopt+nir4} strategy, and 0.64 days under the \textbf{HSTopt+nir6} strategy. Upon visual inspection of the fitted light curves, \textsc{Glimpse} has to flex to match the \textit{HST} data, as evidenced by systematic deviations of the $\epsilon(t)$ realisations away from zero in the NIR filters, indicating challenges in matching the colour evolution observed with that of the \texttt{SALT3} SED. Although the time-delay estimates from unresolved Rubin-LSST data and resolved \textit{HST} data are not completely unbiased across the set of four realisations of the system, the biases in the time-delay estimates are reduced by over 1.2 days when including any amount of resolved \textit{HST} follow-up in the fit compared to fits with unresolved Rubin-LSST data alone. We refrain from drawing any further conclusions on whether the results from \textsc{Glimpse} are biased owing to low statistics for this test.

If the system is resolved by Rubin-LSST, the time-delay constraints for this highly magnified system are extremely precise, even in the case of Rubin-LSST data alone. Using only Rubin-LSST data in the fit, \textsc{Glimpse} reaches a median time-delay uncertainty of 0.24 days. This level of uncertainty corresponds to up to a sub-5\% precision estimate on time delays of less than 5 days. We also find that the median of the time-delay residuals is consistent with zero given the sample size, though again statistics are limited for this test.

Naturally, model misspecification becomes a significant challenge to time-delay inference when including \textit{HST} observations in the fits with resolved Rubin-LSST data. Indeed, we do not see any systematic improvement in the time-delay constraints when including \textit{HST} follow-up for system H. The fitted $\epsilon(t)$ realisations show even stronger deviations from the underlying template, particularly in the \textit{F160W} filter, indicating differences in the colour evolution of the \texttt{SALT3} template compared to the observed data. We do note a slight reduction in the range of time-delay uncertainties across the four spatiotemporal realisations of the system from 0.30 days without \textit{HST} follow-up to 0.17 days under the \textbf{HSTopt+nir6} observing strategy. Therefore, there is still a benefit to \textit{HST} follow-up for these bright, resolved systems.

When considered as a whole, these results demonstrate the power of Rubin-LSST data to reach remarkable precision on time-delay estimates for glSNe that are much brighter than the single-exposure limiting magnitude of the survey. The depth, higher cadence, and early-time light curve coverage that Rubin-LSST offers helps to reduce the time-delay uncertainty of resolved systems to a median of 0.24 days, even in the absence of follow-up from \textit{HST}. The \textit{HST} data, though, is still important to 1) improve the consistency of the time-delay constraints, and 2) reduce systematic biases and break degeneracies in the time-delay inference, particularly for when the Rubin-LSST data is unresolved. Indeed, the time-delay uncertainties with \textit{HST} follow-up remain consistently low at a quarter of a day for resolved Rubin-LSST data and reach uncertainties of just half a day for unresolved Rubin-LSST data. This result demonstrates a realistic lower limit of precision on time delays from systems discovered by Rubin-LSST, resolved or unresolved. Therefore, we conclude that sub-5\% precision on the time delay is possible for systems with as low as a 5 day time delay if the data quality for such a system from Rubin-LSST and follow-up is similar to that of the higher-magnification system considered in this section. Therefore, just one of these galaxy-scale glSNe may yield precise cosmological results.

\begin{figure}
    \centering
    \includegraphics[width=0.98\linewidth]{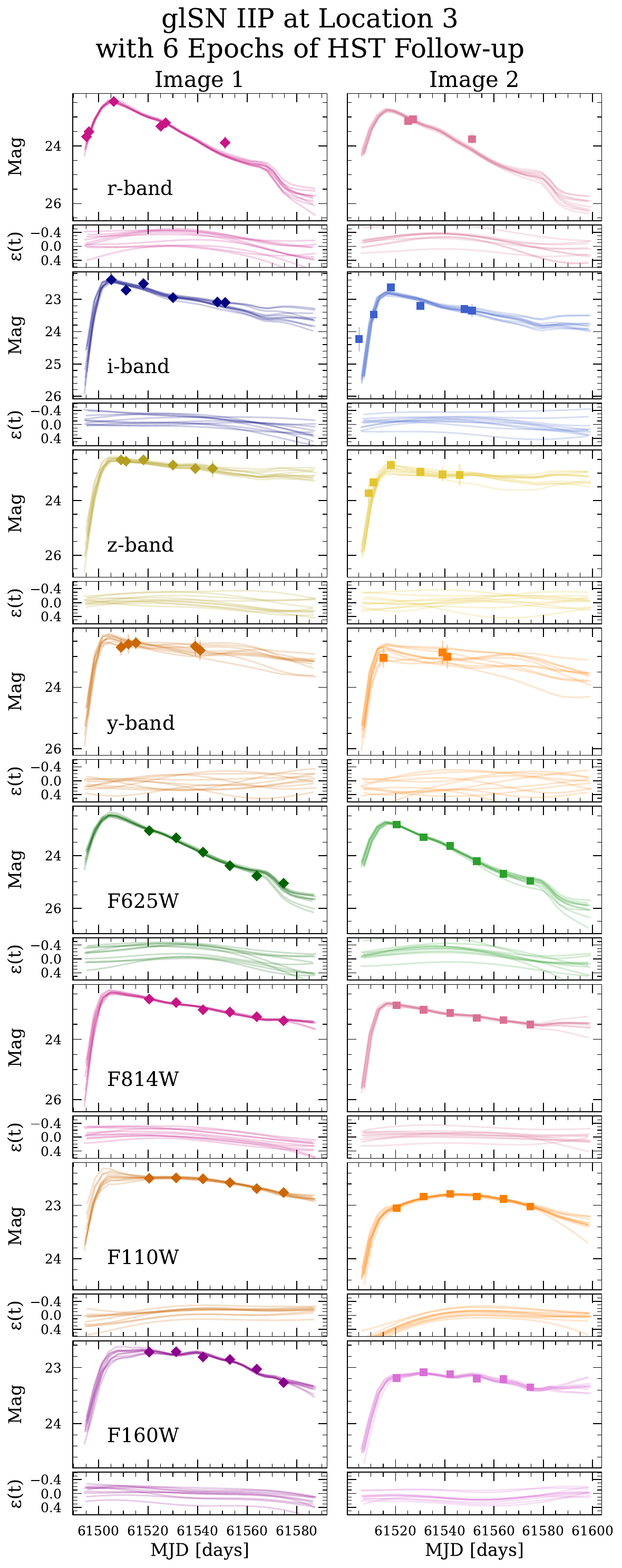}
    \caption{The fitted light curves of a glSN~IIP system as described in \S\ref{sec:cc-sn} performed under the assumption of model misspecification. The left column shows the fits to image 1 and the right column shows the fits to image 2. The subpanel below each light curve fit shows the microlensing contribution to the fitted light curve. }
    \label{fig:glsn-cc}
\end{figure}

\subsection{Deblended Resolved Data}
\label{sec:discussion-blending}
In this work we have presented results under the assumption that the systems will be either entirely resolved or entirely unresolved by Rubin-LSST. For those systems that are unresolved by Rubin-LSST, tools such as \texttt{STARRED} \citep{Millon_2024} and \texttt{scarlet2} (\citealp{Ward_2025}; Krishnaraj et al. \textit{in prep.}) will enable resolved light curves to be extracted from the unresolved images given a resolved observation of the system (from e.g., \textit{HST}, \textit{JWST}, Keck) from which the glSN image positions can be precisely determined. In future work, we intend to apply \textsc{Glimpse} to light curves extracted from images via these methods to determine the precision to which time delays can be estimated from more realistic light curve data. We anticipate the results will fall somewhere between the best and worst case-scenarios considered in this work, being fully resolved and fully unresolved data, respectively. Given that resolved Rubin-LSST data resulted in much more precise time-delay estimates than unresolved Rubin-LSST data, the potential to obtain resolved light curves via deblending is exciting for precision cosmology. This additional benefit of resolved space-based imaging is further motivation for timely follow-up of glSNe discovered by Rubin-LSST.

\subsection{Application to Core Collapse SNe}
\label{sec:cc-sn}
While we have focused on glSNe~Ia in this work, the \textsc{Glimpse} model can be applied to any type of lensed transient for which we have a template, including core collapse (CC)~SNe. We adapt the \texttt{lensedSST} code to enable the simulation of glSNe~Ib/c, glSNe~IIn, and glSNe~IIP using templates from \texttt{sncosmo}. For this limited test, we focus on SNe~IIP, which we model according to the \texttt{snana-2007kw} template from \citet{Kessler_2010}. We assume the SN~IIP is at $z_{\rm source} = 0.604$ with a peak absolute B-band magnitude of -17.5 mag. We place the lens at $z_{\rm lens} = 0.332$ and with $\theta_{E} = 0.778''$. The resulting doubly imaged glSN~IIP has a time-delay of 12.39 days and absolute magnifications of -2.8 mag and -2.4 mag for images 1 and 2, respectively. We only estimate the relative magnification in this case, as we do not expect to be able to back out the individual image absolute magnifications for these systems. Without dust extinction, the system has peak i-band magnitudes of 22.0 mag for image 1 and 22.3 mag for image 2.

As with the fiducial analysis of this work, we do not simulate microlensing effects for the CC~SN light curves. To produce realistic microlensing, it is necessary to understand the chromaticity of the expansion of the SN, as microlensing in this context is caused when the expanding SN photosphere passes behind compact objects in the lens galaxy. The diversity of CC~SNe makes it difficult to estimate the chromatic expansion of these objects. We leave to future work a full analysis of time delay estimation for non~Ia glSNe including microlensing modelling with \textsc{Glimpse}.

We fit both with the correct template and under the assumption of model misspecification. For the case of model misspecification, we assume the \texttt{snana-2007ld} template in the \textsc{Glimpse} fit. This template roughly matches the colour and time evolution in the optical filters, but does diverge more significantly in the NIR. Template selection will be a source of systematic uncertainty in time-delay estimation with any template-based method, including the \textsc{Glimpse} model, owing to the great diversity of CC~SNe light curves. For a time-delay estimation method which is agnostic to SN type, we refer the reader to the \textsc{Glimpse} model from \citet{Hayes_2024}.

First, we present the results under the assumption of the correct underlying template. With Rubin-LSST data alone, the median time-delay uncertainty is \IIPmatchLSSTunresolvedmedianTD days if unresolved by Rubin-LSST and \IIPmatchLSSTresolvedmedianTD days if resolved by Rubin-LSST. In both cases, the $1\sigma$ uncertainties capture well the 68\% CI. The uncertainties are further reduced to 0.53 days, 0.44 days, and 0.41 days for 2, 4, and 6 epochs of optical and NIR \textit{HST} observations, respectively, in the case of unresolved Rubin-LSST data. When the system is resolved by Rubin-LSST, the median time-delay uncertainties are 0.39 days, 0.38 days, and 0.35 days, respectively for the three \textit{HST} follow-up strategies. These results show similar trends to those found in the fiducial analysis, though owing to the correctly specified model and the brighter nature of this system, the uncertainties are a bit tighter in general. This study demonstrates the \textsc{Glimpse} model works well to estimate the time-delays of multiply-imaged sources regardless of the shape of the underlying light curve of the source, if a suitable model for its light curve is available.

When we assume an incorrect template, the results are significantly degraded compared to the above results. When unresolved, the \textsc{Glimpse} model achieves a median time-delay uncertainty of \IIPLSSTunresolvedmedianTD days with Rubin-LSST data alone. Naturally, the results when fitting with an incorrect model are more uncertain than when the model is correctly specified. Furthermore, the $1\sigma$ uncertainties do not capture well the 68\% CI, though this behaviour is not unexpected given the discussion in \S\ref{sec:sim-analysis}. The results improve in uncertainty when the systems are assumed to be resolved, with a median time-delay uncertainty of \IIPLSSTresolvedmedianTD days, though the error bars remain unrepresentative of the 68\% and 95\% CIs (with 50.0\% of time delays in the 68\% CI and 66.7\% in the 95\% CI). Including \textit{HST} data in the fit does not result in a consistent improvement to time-delay estimates, as seen in \S\ref{sec:discussion-mismatch}, owing to the greater sensitivity of higher SNR data from \textit{HST} to model misspecification. When the systems are unresolved by Rubin-LSST, the median time-delay uncertainty ranges from 0.67-0.96 days depending on the amount of \textit{HST} follow-up and when the systems are resolved by Rubin-LSST, the median time-delay uncertainty ranges from 0.57-0.69 days depending on the amount of \textit{HST} follow-up.

While the accuracy of time-delay estimates may be an issue, uncertainties of less than a day approach 5\% precision time-delay measurements in the case of a $\sim12$ day time delay. In Figure \ref{fig:glsn-cc}, we show the fitted light curves of one realisation of the glSN~IIP with 6 epochs of \textit{HST} follow-up assuming an incorrect template. For this fit, the time-delay estimate is accurate and precise: $12.03 \pm 0.59$ days (4.7\% precision). There is evidence of model misspecification in the microlensing residuals, which show systematic deviations from the template across wavelength and time. The quick rise-time and sharp turn over at peak captured with Rubin-LSST data, in addition to the relatively bright nature of the system, are the likely explanations for the impressive time-delay precision.

The results from the analysis indicate that it is possible to accurately and precisely estimate the time-delays of glSNe~CC with \textsc{Glimpse}, though the results are highly sensitive to model misspecification. \textit{HST} follow-up can improve the precision on the time delays, though its benefits are more limited when the incorrect model is assumed. The issue of model misspecification is exacerbated in the case of unresolved Rubin-LSST data, owing to the challenge of degeneracy breaking in unresolved light curves made more difficult by an ill-chosen model. Realistically, alternative approaches to time-delay estimation would need to be explored for unresolved glSNe~CC systems, as a template-based method is not found to be suitable for such systems based on this work.

This brief analysis demonstrates the challenge of estimating time-delays with template-based methods for glSNe~CC. Custom-made templates, such as was constructed for SN Refsdal \citep{Kelly_2023_ApJ}, may provide better constraints on the time delays of glSNe~CC, though this approach may not be feasible for large samples. We leave a more thorough investigation of the systematic uncertainties introduced by template assumptions for glSNe~CC to future work. Further tests should also be carried out for other types of CC~SNe, some of which do exhibit more uniformity than SNe~IIP.

\section{Conclusions}
\label{sec:conclusions}
In this work, we present the \textsc{Glimpse} model,\footnote{The source code, with jupyter notebook tutorials, can be found at \url{https://github.com/erinhay/GausSN}.} a Bayesian Gaussian Process approach to time-delay estimation which simultaneously models resolved and unresolved data with treatments of chromatic microlensing and differential dust extinction across the lens galaxy. This model uses SN light curve templates \texttt{sncosmo} as the mean function, allowing a 2D covariance in time and wavelength to capture the effect of microlensing as a perturbation of the SN template. This formulation allows data from all wavelength regimes to inform the constraint on the microlensing effects, while still allowing for chromaticity.

With this model, we examine the limits of the precision on time-delays from Rubin-LSST data with varying amounts of space- and ground-based follow-up. For this test, we examine a sample of glSNe~Ia simulated from the \textsc{BayeSN} SED, using the \texttt{lensedSST} tool to achieve realistic observing conditions. To further enhance the realism of the analysis, we fit assuming a \texttt{SALT3} template as the mean function within the \textsc{Glimpse} model. As the assumed model is not going to perfectly describe the true observed light curve, simulating and fitting from different underlying templates improves the realism of this analysis, and enables a quantification of systematic uncertainty from model misspecification.

We examine four glSNe~Ia systems – three doubles and one quad – which vary in redshift and lensing parameters (e.g., time delays and absolute magnifications). These systems are at the faint end of detectability by Rubin-LSST (see Appendix \ref{sec:appendixA}), with each system having images that peak between 21.5-23.5 mag in the $i$-band. Furthermore, we take realisations of each system across spatiotemporal locations in the Rubin-LSST survey, as being both resolved and unresolved by Rubin-LSST, and with varied dust extinction in the host and lens galaxies. For a limited number of test cases, we also fit a highly magnified doubly imaged glSN~Ia, for which both images peak around 20-20.5 mag in the $i$-band. Finally, we simulate and fit a doubly imaged glSN~IIP both under the assumption of a correct template and an incorrect template to demonstrate the efficacy of the method for glSNe~CC. 

We find that time delays can consistently be estimated to 0.2-0.8 day precision for systems that are resolved by Rubin-LSST with 6 epochs of \textit{HST} follow-up and/or deep ground-based imaging from an 8 meter class telescope with a $\sim$5 day cadence. These constraints are strongly dependent on the brightness of the system, and thus the SNR of the Rubin-LSST data. This result highlights that the time coverage provided by the Rubin-LSST observations, particularly at early times before it will be possible to realistically obtain follow-up, can have a significant impact on the time-delay precision. We find that the consistency of the constraints by system, though, is achieved with increasing amounts of follow-up data, whether that be from \textit{HST}, \textit{JWST}, or ground-based facilities, or a combination of both.

In the case of glSNe that are unresolved by Rubin-LSST, time-delays are estimated to a 1-2 day level of precision with 6 epochs of \textit{HST} follow-up in the optical and NIR. Although these results are not as competitive with those from resolved glSNe systems, they may be improved if resolved light curves can be extracted from unresolved ground-based imaging via de-blending methods (\citealp{Millon_2024, Ward_2025}; Krishnaraj et al. \textit{in prep.}). We leave to future work an analysis of the time-delay estimates for more highly realistic light curves derived from de-blended images and therefore the cosmological utility of unresolved glSNe systems.

Based on these results, we advocate for the following strategies for follow-up observations of glSNe~Ia: If an unresolved system is determined to have at least two images with a time delay of $\gtrsim$15 days that are each brighter than 22 mag in the $i$-band at peak, six to eight epochs of space-based follow-up in at least four optical and NIR filters is likely to achieve a 5\% precision time delay estimate. If a similarly bright system is resolved by Rubin-LSST, a precise time-delay estimate can be achieved for time delays of as short as 5 days with only two epochs of space-based follow-up in the optical and NIR. This follow-up need not be collected pre-peak; we assume the system is first observed from space between 10-20 rest-frame days post-peak. However, we assume good early-time coverage of these systems from Rubin-LSST, which is likely given the conditions necessary for discovery. If the system is able to be resolved by other ground-based facilities with sufficient depth, a 5-day cadence post-peak in at least four optical to NIR filters is also likely to provide a precise enough time delay in combination with Rubin-LSST data. In this case, only one epoch of high-resolution imaging from space will be necessary for time-delay cosmography, with a second epoch to be collected after the SN~Ia has faded.

For resolved systems which have peak $i$-band magnitudes of between 22-24 mag, we find that a combination of six epochs of space-based imaging in at least four optical and NIR filters and eight epochs of deep ground-based imaging in at least four optical filters make a precise time-delay estimate possible for time delays of $\gtrsim$10 days. We note that it is necessary for follow-up observations to be deep enough to achieve a sufficient number of $5\sigma$ observations of follow-up across a wide wavelength regime. An unresolved system of similar brightness would require a time delay of $\gtrsim$20 days and at least eight space-based observations in at least six optical and NIR filters to reliably achieve 5-10\% precision on the time-delay estimate. Of course, the follow-up for a specific glSN~Ia system should carefully consider in context the redshift of the system, the brightnesses and separations of the images, the time delays predicted from preliminary lens modelling, and the visibility windows and depth of relevant follow-up facilities to decide what follow-up should be obtained for a precise time-delay estimate.

Therefore, we conclude that appropriately allocated follow-up makes 5\% precision time-delay estimates achievable for a broad range of systems that will be discoverable by Rubin-LSST, including systems which are dimmer than 22 mag at peak in the $i$-band, which make up the bulk of systems in the projected galaxy-scale lens populations, and those that are unresolved from the ground. This work takes into account key systematic uncertainties from microlensing, differential dust extinction, and model misspecification to make these projections as realistic as possible. With sufficiently precise lens modelling (at the 5\% level), a 7\% precision $H_{0}$ estimate is possible from a single one of these glSNe~Ia systems. For comparison, the time-delay uncertainty dominates the $H_{0}$ error budget from time-delay cosmography for $\gtrsim$15\% precision level (e.g., $\sigma_{\Delta t} = 1.5$ days compared to $\sigma_{\Delta t} = 0.5$ days on a 10-day time delay).

The assumed 5\% lens modelling uncertainties despite the MSD is largely considered to be possible for galaxy-scale lenses \citep{Suyu_2020, Birrer_2022}. As shown in this work, absolute magnification constraints from \textsc{Glimpse} are precise enough to break the MSD using light-curve data. We find that the follow-up requirements for precise time delays are sufficient to achieve a precise absolute magnification constraint. For the faint unresolved systems, only four to six epochs of \textit{HST} follow-up in four optical and NIR filters, and indeed fewer epochs for brighter and/or resolved systems, is sufficient to reach 0.1 mag precision on the absolute magnifications. As in \S\ref{sec:follow-up-mag-lsst}, $\lesssim$0.1 mag precision on factor of $\gtrsim$1 absolute magnifications represents a constraint on the MST parameter, $\lambda$, at the sub-5\% level.

Alternatively, stellar kinematics provide an avenue for breaking the MSD in glSNe~Ia using spectroscopic data of the lens galaxy after the SN~Ia has faded. These two methods have independent systematics and utilise independent data, so they will provide important cross-checks of one another. Especially as absolute magnification constraints from light-curve data are potentially subject to systematics from microlensing, it will be important to carry out both measurements for the first $H_{0}$ estimates from galaxy-scale glSNe~Ia. Between these two methods, lens modelling uncertainties including uncertainties from the MSD can reliably be reduced to below the 5\% level in total.

We finally consider this conclusion in the context of the precision necessary to achieve a $>5\sigma$ test of the Hubble tension, which requires an $H_{0}$ estimate at the sub-5\% level. \citet{Suyu_2020} estimated the precision on $H_{0}$ from the sample of glSNe~Ia expected in ten years of Rubin-LSST data with ideal system parameters for time-delay cosmography (resolved, $\Delta t>20$ days, $z<0.7$), coming to 20 glSNe~Ia. They assume each system has a 7\% total error budget, from time-delay estimation and lens modelling. In this work, we demonstrate that 7\% precision is achievable for a much larger sample of glSNe~Ia from the \textsc{Glimpse} model with appropriately allocated follow-up. Thus, the 1.3\% precision $H_{0}$ estimate that \citet{Suyu_2020} predicts for ten years of Rubin-LSST observations is achievable within just three years of Rubin-LSST, considering projected glSN~Ia rates from \citet{Arendse_2024}. Thus, time-delay cosmography with galaxy-scale glSNe~Ia has the potential to provide an unprecedented independent local test of the Hubble tension in the next three years.

Rubin-LSST will produce tremendous amounts of data among which will be tens to hundreds of glSNe -- potentially increasing our sample of glSNe by a factor of ten. This work provides guidance for how best to follow-up glSNe which are expected to be discovered by Rubin-LSST. The challenge then becomes discovering glSNe in Rubin-LSST data in real-time. A few works have explored methods for identifying glSNe \citep[e.g.,][]{Arendse_2024, Townsend_2024, SC_2024}, though further work is needed to ensure these methods work in real-time and scale-up to Rubin-LSST with a sufficiently low false positive rate. The biggest barrier to developing more sophisticated search methods is the availability of high-quality, uniform simulated samples of glSNe and potential contaminants (e.g., unlensed SNe, quasars, tidal disruption events). There is ongoing effort from JOLTEON\footnote{\url{https://portal.nersc.gov/cfs/lsst/jolteon/}} to produce such a training set for the development of identification methods at the light-curve level. Furthermore, new tools such as the Strong Lensing Simulator (\texttt{SLSim})\footnote{\url{https://github.com/LSST-strong-lensing/slsim}} will enable the development of image-based datasets for training and testing of search methods in the near future.

This data-rich era of astronomy provides the opportunity to answer some of the most pressing questions in the field. In particular, glSNe may be able to reveal the true value of $H_{0}$ in the local universe, among other mysteries in cosmology, if a sufficient number of systems are discovered and appropriately followed-up. This work, with future progress in glSN discovery efforts, will enable the field to take advantage of the full power of Rubin-LSST.

\section*{Acknowledgements}
The authors thank Charlotte Ward and Matthew Grayling for helpful discussion.

E.E.H.\ is supported by a Gates Cambridge Scholarship (\#OPP1144) and acknowledges travel support provided by STFC for UK participation in LSST through grant ST/S006206/1.
S.D.\ is supported by UK Research and Innovation (UKRI) under the UK government’s Horizon Europe funding Guarantee EP/Z000475/1. 
S.T.\ was supported by funding from the European Research Council (ERC) under the European Union's Horizon 2020 research and innovation programmes (grant agreement no. 101018897 CosmicExplorer).  
S.T.\ also thanks the Isaac Newton Institute for Mathematical Sciences, Cambridge, for support and hospitality during the programme `Accelerating statistical inference and experimental design with machine learning', where some work on this paper was undertaken. This was supported by EPSRC grant EP/Z000580/1.
S.T.\ and N.A.\ were supported by the research project grant `Understanding the Dynamic Universe' funded by the Knut and Alice Wallenberg Foundation under Dnr KAW 2018.0067.
J.D.R.P.\ is supported by NASA through a Einstein Fellowship grant No. HF2-51541.001 awarded by the Space Telescope Science Institute (STScI), which is operated by the Association of Universities for Research in Astronomy, Inc., for NASA, under contract NAS5-26555.

This research utilized the Sunrise HPC facility supported by the Technical Division at the Department of Physics, Stockholm University. This work was performed using resources provided by the Cambridge Service for Data Driven Discovery (CSD3) operated by the University of Cambridge Research Computing Service (\url{www.csd3.cam.ac.uk}), provided by Dell EMC and Intel using Tier-2 funding from the Engineering and Physical Sciences Research Council (capital grant EP/T022159/1), and DiRAC funding from the Science and Technology Facilities Council (\url{www.dirac.ac.uk}).

\section*{Data Availability}
The simulated data are publicly available on Github at \url{https://github.com/erinhay/GausSN}.

\bibliographystyle{mnras}
\bibliography{bib} 

\appendix

\section{Detectability}
\begin{table}
    \centering
    \caption{Detectability of each of the base systems, depending on its position in space and time during Rubin-LSST survey operations. The \magnification \ symbol indicates that the system is detectable by the magnification method and the \multiplicity \ symbol indicates that the system is detectable by the multiplicity method, as described in \citet{Wojtak_2019}.}
    \begin{tabular}{c|cccc}
         & System A & System B & System C & System D \\
        \hline
        \hline
        Location 1 & \multiplicity & \multiplicity & \magnification \, /  \multiplicity & \multiplicity \\[4pt]
        Location 2 & \multiplicity & \multiplicity & \magnification \, /  \multiplicity & \multiplicity \\[4pt]
        Location 3 & \multiplicity & \multiplicity & \magnification \, /  \multiplicity & \multiplicity \\[4pt]
        Location 4 & \multiplicity & \multiplicity & \magnification \, /  \multiplicity & \multiplicity \\[4pt]
    \end{tabular}
    \label{tab:detectability}
\end{table}

In Table \ref{tab:detectability}, we predict whether each of the three base objects would be detected as a glSN at each of the four spatiotemporal location considered in this analysis. We use functionality from \texttt{lensedSST} to determine whether the systems are detectable by the ``magnification'' method and/or by the ``multiplicity'' method. The magnification method requires that the unresolved object is at least 0.7 mag brighter than an unlensed SNe~Ia at the lens redshift in at least one filter. The multiplicity method requires that two or more resolved images are brighter at peak than 0.2 mag above a filter's limiting magnitude in at least one filter. Furthermore, there must be a separation of at least 0.5 arcsec between at least two of the images. While these definitions of detectability are only rough estimates, this grid demonstrates how the observing strategy will affect different systems of glSNe. Based on these criteria, the systems considered in this work are within the detection limits of Rubin-LSST and, therefore, reasonably discoverable by the survey.

\label{sec:appendixA}

\section{Simulated Space-Based Data}
\begin{figure}
    \centering
    \includegraphics[width=\linewidth]{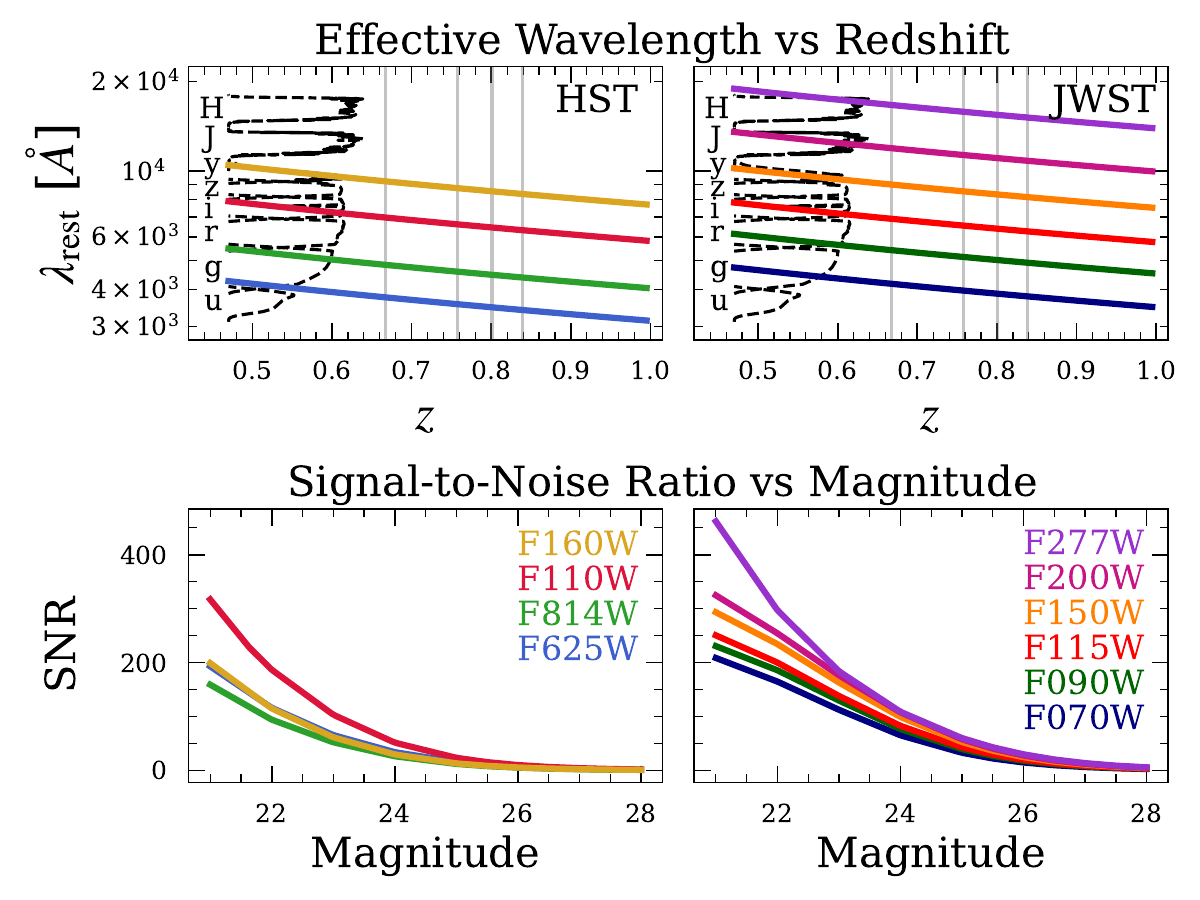}
    \caption{(Top) The effective wavelength of the \textit{HST} and \textit{JWST} filters considered in this work as a function of redshift. The gray vertical lines show the redshifts of the four base objects considered in this work. (Bottom) The signal-to-noise ratio as a function of magnitude for the \textit{HST} and \textit{JWST} filters, assuming an exposure time of 750 s for all filters. These curves are computed using the \textit{HST} and \textit{JWST} exposure time calculators, and can be scaled by exposure time.}
    \label{fig:effwave-redshift}
\end{figure}

When simulating follow-up observations from \textit{HST} and \textit{JWST}, it is important to understand 1) the observer-frame wavelength regime being probed by each filter, and 2) the SNR as a function of magnitude for a given exposure time. In the top panel of Figure \ref{fig:effwave-redshift}, we show the effective wavelength of four \textit{HST} filters and two \textit{JWST} filters as a function of wavelength, as compared to rest-frame $ugrizyJH$ filters. The solid gray vertical lines show the redshifts of the four systems considered in this work.

In the bottom panel of Figure \ref{fig:effwave-redshift}, we show the curves of SNR as a function of magnitude for a fixed exposure time. All curves are scaled to correspond to an exposure time of 750 s. We use these functions to produce realistic uncertainties on the space-based observations, under the assumption of some exposure time which corresponds to a desired target SNR to be achieved for an observation of the SN~Ia in question at peak. As the multiple images will have different brightnesses at the time of observation, these curves enable realistic uncertainty estimates at any point in the light curve for all images.

\label{sec:appendixB}

\section{Light Curve Model Differences}
\begin{figure}
    \centering
    \includegraphics[width=0.99\linewidth]{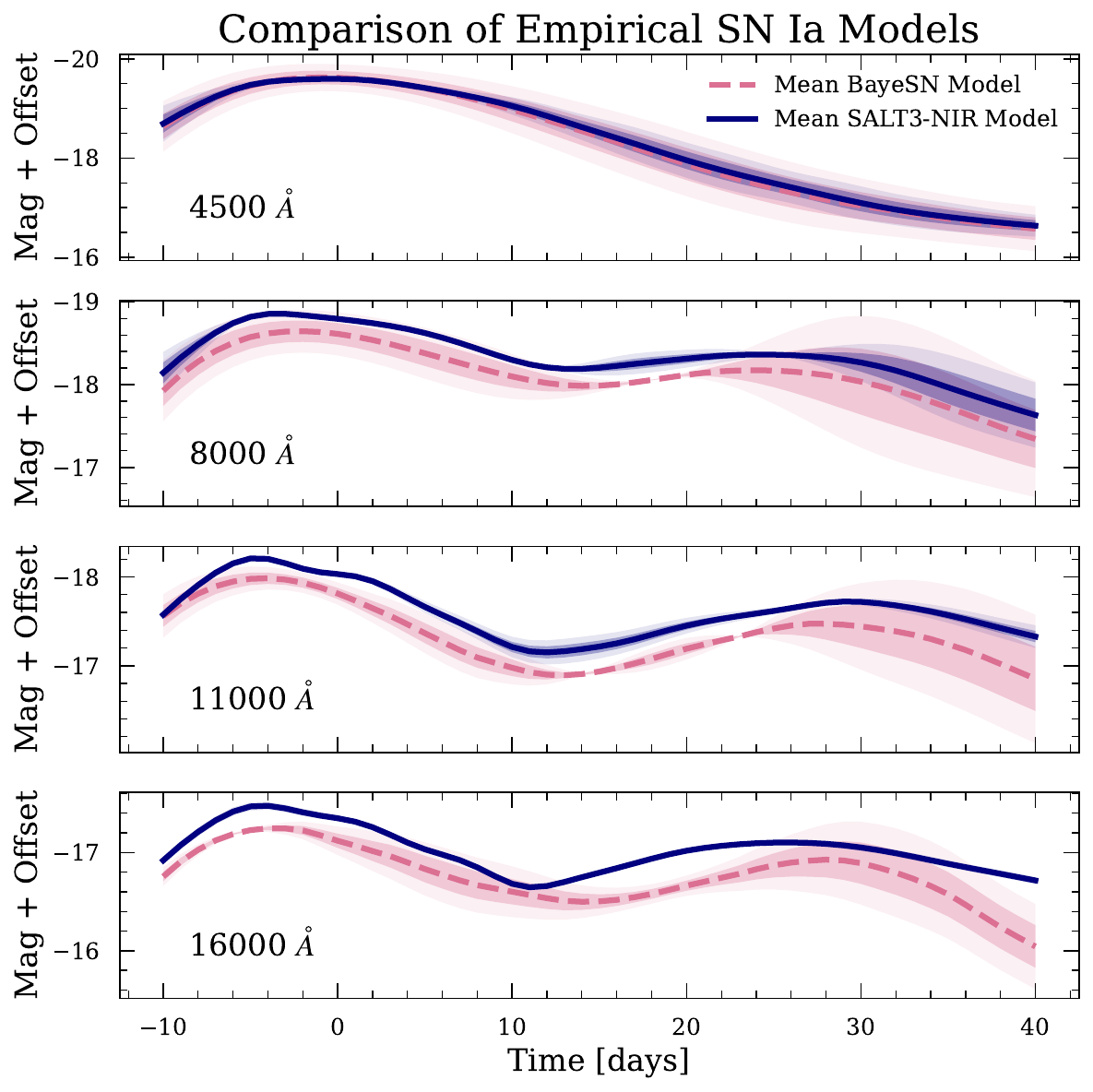}
    \caption{Comparison of the \textsc{BayeSN} (pink) and \texttt{SALT} (blue) empirical models for SNe~Ia from 4500\AA\ (B-band) to 16000\AA\ (H-band). We visualise the variation in each model by sampling values of the stretch parameter from a unit normal distribution. The mean light curve for each model is shown as a line, with the $1\sigma$ and $2\sigma$ deviations shown as dark and light shaded regions.}
    \label{fig:template-comp-Ia}
\end{figure}

One key finding of this work is the systematic uncertainty associated with template misspecification within the \textsc{Glimpse} framework. In Figure \ref{fig:template-comp-Ia}, we visualise the \textsc{BayeSN} and \texttt{SALT3-NIR} models in four wavelength regimes across the optical and NIR. We show both the mean light curve in each regime, as well as the variation arising from the stretch parameter, which is the primary driver of differences in light curve shape for SNe~Ia. There is strong agreement in the models in the blue region of the spectra (top panel). Owing to the large training sample available in this wavelength regime, the models are both able to capture well the full variation seen in SNe~Ia in this regime. In redder regions, however, the models show significant differences in both overall shape, as well as the extent of the variations. Limited observations of SNe~Ia in the NIR to date to train models with make it challenging to capture the observed variability in this wavelength regime.

This figure demonstrates how differences between the models may be irreconcilable in the NIR. As a result, the uncertainties on parameter estimates from \textsc{Glimpse} are driven up when fitting high-SNR NIR observations. This wavelength regime is the primary target of space telescope follow-up. As discussed in \S\ref{sec:discussion-mismatch-future}, the training of a new empirical SN~Ia model based on the hundreds of new NIR spectra from CSP-II will allow models like \textsc{Glimpse} to take better advantage of powerful NIR observations from \textit{HST}, \textit{JWST}, and other facilities.

We also highlight that the results of an alternative example of model misspecification are presented for the glSN~IIP system in \S\ref{sec:cc-sn}. The template differences in this case are up to 0.4 mag, similarly to the differences in the mean \textsc{BayeSN} and \texttt{SALT3-NIR} templates, though we note that these light curve templates are intended to describe the evolution of single objects and therefore don't have any flexibility to reconcile these differences. For this system, model misspecification leads to a median increase in the uncertainty on the time delay of 0.59 days for unresolved systems and 0.34 days for resolved systems under the assumption of six epochs of \textit{HST} data in the optical and NIR. This result is in line with the results for resolved glSN~Ia systems presented in \S\ref{sec:discussion-mismatch}, though slightly more significant owing to the more pronounced differences between simulation and fitting templates for the glSN~IIP.

As the range of normal SNe~Ia are generally well-described by at least one empirical SN~Ia, the systematic uncertainty associated with model misspecification is not expected to exceed the result of up to 0.25 days as presented in \S\ref{sec:discussion-mismatch}. On the other hand, CC~SNe show more diversity than SNe~Ia, so model misspecification may be more significant of a challenge for glSNe~CC. In the case of a lensed peculiar SNe, thorough tests of the impact of different assumed models on time-delay inference should be run to understand the extent of this systematic uncertainty if cosmology is going to be done with time delays from a template-based light-curve model for that system. The bias on the time delays will give the same level of bias on your inferred $H_{0}$, which can be on the order of the 5-10\% level for time delays of 10-20 days.
\label{sec:appendixC}

\bsp
\label{lastpage}
\end{document}